\documentclass[11pt]{article}

\usepackage{mathrsfs}
\usepackage{natbib,graphicx,setspace,lscape,longtable}
\usepackage{mathrsfs,amsmath,amsthm,amssymb,color}
\usepackage{natbib,epsfig,graphicx,pdfpages}
\usepackage{rotating}
\usepackage{subfigure}
\usepackage{enumerate}
\usepackage{graphicx}
\usepackage{algorithm}
\newcommand\sbullet[1][.5]{\mathbin{\vcenter{\hbox{\scalebox{#1}{$\bullet$}}}}}
\usepackage{url}
\usepackage[table]{xcolor}
\usepackage{colortbl}
\usepackage[noend]{algpseudocode}
\usepackage{algpseudocode}
\usepackage{tabularx}
\usepackage{setspace}
\usepackage[hidelinks,colorlinks=true,linkcolor=blue,citecolor=blue]{hyperref}
\bibpunct{(}{)}{;}{a}{,}{,}
\usepackage[compact]{titlesec}

\usepackage{cancel}
\makeatletter
\def\thm@space@setup{%
  \thm@preskip=0cm
  \thm@postskip=\thm@preskip 
}
\makeatother

\newtheorem{theorem}{Theorem}
\newtheoremstyle{exampstyle}
  {-0.1pt} 
  {-0.1pt} 
  {\itshape} 
  {} 
  {\bfseries} 
  {.} 
  {.5em} 
  {} 
\newtheorem{lemma}{Lemma}
\newtheorem{proposition}{Proposition}
\newtheorem{remark}{Remark}

\newtheorem{assumption}{{Assumption}}

\def\beq{\begin{equation}}
\def\eeq{\end{equation}}
\def\beqr{\begin{eqnarray}}
\def\eeqr{\end{eqnarray}}
\def\beqrs{\begin{eqnarray*}}
\def\eeqrs{\end{eqnarray*}}
\def\bet{\begin{theorem}}
\def\eet{\end{theorem}}
\def\bel{\begin{lemma}}
\def\eel{\end{lemma}}
\def\bep{\begin{proposition}}
\def\eep{\end{proposition}}
\def\bg{\begin{figure}[tbph]\begin{center}}
\def\eg{\end{center}\end{figure}}

\def\bc{\begin{center}}
\def\ec{\end{center}}

\def\wh{\widehat}

\def\cov{\mbox{Cov}}

\def\diag{\mbox{diag}}

\newcommand{\Var}{\textnormal{Var}}
\newcommand{\Cov}{\textnormal{Cov}}

\newcommand{\vc}{\textnormal{vec}}

\newcommand{\bA}{{\mathbf A}}
\newcommand{\bB}{{\mathbf B}}
\newcommand{\bF}{{\mathbf F}}
\newcommand{\bE}{{\mathbf E}}
\newcommand{\bG}{{\mathbf G}}
\newcommand{\bH}{{\mathbf H}}
\newcommand{\bI}{{\mathbf I}}
\newcommand{\bK}{{\mathbf K}}
\newcommand{\bL}{{\mathbf L}}
\newcommand{\bM}{{\mathbf M}}
\newcommand{\bN}{{\mathbf N}}
\newcommand{\bQ}{{\mathbf Q}}
\newcommand{\bP}{{\mathbf P}}
\newcommand{\bR}{{\mathbf R}}
\newcommand{\bS}{{\mathbf S}}

\newcommand{\bU}{{\mathbf U}}
\newcommand{\bV}{{\mathbf V}}
\newcommand{\bW}{{\mathbf W}}
\newcommand{\bX}{{\mathbf X}}
\newcommand{\bY}{{\mathbf Y}}
\newcommand{\bZ}{{\mathbf Z}}
\newcommand{\ba}{{\mathbf a}}
\newcommand{\bb}{{\mathbf b}}

 \newcommand{\bfc}{{\mathbf c}}
\newcommand{\be}{{\mathbf e}}
\newcommand{\bff}{{\mathbf f}}
 
\newcommand{\bh}{{\mathbf h}}

\newcommand{\bp}{{\mathbf p}}
\newcommand{\bq}{{\mathbf q}}
\newcommand{\br}{{\mathbf r}}

\newcommand{\bu}{{\mathbf u}}
\newcommand{\bv}{{\mathbf v}}
\newcommand{\bw}{{\mathbf w}}
\newcommand{\bx}{{\mathbf x}}
\newcommand{\by}{{\mathbf y}}
\newcommand{\bz}{{\mathbf z}}
\newcommand{\balpha} {\boldsymbol{\alpha}}
\newcommand{\bbeta}  {\boldsymbol{\beta}}
\newcommand{\bfeta}  {\boldsymbol{\eta}}

\newcommand{\bdelta} {\boldsymbol{\delta}}

\newcommand{\bOmega}{\boldsymbol{\Omega}}
\newcommand{\bomega}{\boldsymbol{\omega}}

\newcommand{\bSigma}{\boldsymbol{\Sigma}}

\newcommand{\bve}{\mbox{\boldmath$\varepsilon$}}

\newcommand{\bPhi} {\boldsymbol{\Phi}}
\newcommand{\bPsi} {\boldsymbol{\Psi}}

\newcommand{\bxi} {\boldsymbol{\xi}}

\newcommand{\bzeta} {\boldsymbol{\zeta}}
\newcommand{\bGamma} {\boldsymbol{\Gamma}}
\newcommand{\bLambda} {\boldsymbol{\Lambda}}
\newcommand{\bC}{{\mathbf C}}
\newcommand{\bD}{{\mathbf D}}

\newcommand{\ve}{{\varepsilon}}
\renewcommand{\epsilon}{{\ve}}
\renewcommand{\hat}{\widehat}

\newcommand{\tr}{\mbox{tr}}

\makeatletter
\def\BState{\State\hskip-\ALG@thistlm}
\makeatother

\def\JRSSB{{\sl Journal of the Royal Statistical Society}, {\bf B}}

\def\BKA{{\sl Biometrika}}
\def\JASA{{\sl Journal of the American Statistical Association}}

\numberwithin{equation}{section}

\usepackage[tmargin=1in, lmargin=1in, rmargin=1in, bmargin=1in]{geometry} 

\definecolor{darkgreen}{rgb}{0.0, 0.5, 0.0}
\definecolor{ashgrey}{rgb}{0.7, 0.75, 0.71}

\title{\bf Denoising and Multilinear Projected-Estimation of  High-Dimensional Matrix-Variate Factor Time Series}
\date{}
\author{ \small Zhaoxing Gao \\
\small School of Mathematical Sciences\\
	\small University of Electronic Science \& Technology of China\\
	\small China
\and \small Ruey S. Tsay\footnote{Corresponding author: \href{mailto:ruey.tsay@chicagobooth.edu}{ruey.tsay@chicagobooth.edu} (R.S. Tsay). Booth School of Business, University of Chicago, 5807 S. Woodlawn Avenue, Chicago, IL 60637, USA.}\\
\small Booth School of Business\\
	\small University of Chicago\\
	\small  USA
}
\begin{document}
\maketitle
\begin{abstract}
 This paper proposes a new multi-linear projection method for denoising and estimation of high-dimensional matrix-variate factor time series. 
It assumes that a $p_1\times p_2$ matrix-variate time series  
consists of a dynamically dependent, lower-dimensional matrix-variate factor process 
and a $p_1\times p_2$ matrix idiosyncratic series. In addition, the latter 
series assumes a matrix-variate factor structure such that its row and column covariances may have diverging/spiked eigenvalues to accommodate the case of low signal-to-noise ratio often 
encountered in applications. 
We use an iterative projection 
procedure to reduce the dimensions and noise effects in estimating front and back loading matrices 
and to obtain faster convergence rates than those of the traditional methods available in the literature. 
We further introduce 
 a two-way projected Principal Component Analysis to mitigate the diverging noise effects, and implement a high-dimensional white-noise testing procedure to estimate the dimension of the 
 matrix factor process.  Asymptotic properties of the proposed method are established if the dimensions and sample size go to infinity.  We also use simulations and real examples to assess the performance of the proposed method 
 in finite samples and to compare its forecasting ability with some existing ones in the literature.  
 The proposed method 
 fares well in out-of-sample forecasting. 
In a supplement, we demonstrate the efficacy of the proposed approach even when the idiosyncratic terms exhibit serial correlations with or without a diverging white noise effect.
\end{abstract}
\textbf{Keywords}:
Denoising, Multilinear Projection,  Factor Model, Matrix Time Series, Eigen-analysis

\newpage
\section{Introduction}

Large data sets are widely accessible nowadays. In many applications,  the data consists of many variables observed over time and form naturally a high-dimensional time series. For example, the returns of a large number of assets can be treated as a high-dimensional vector time series and play an important role in asset pricing, portfolio allocation, and risk management. Large panel time series data are also commonplace in economic, biological, and environmental studies; see, for instance, multiple macroeconomic variables of many countries and air pollution indexes from many monitoring stations.  
To analyse those 
large and high-dimensional data sets, various  dimension reduction methods have been proposed 
and extensively studied in the  
literature. Examples include the canonical correlation analysis (CCA) of \cite{BoxTiao_1977}, the principal component analysis (PCA) of \cite{StockWatson_2002}, and the scalar component model of \cite{TiaoTsay_1989}. The factor model approach can be found in \cite{BaiNg_Econometrica_2002}, \cite{StockWatson_2005}, \cite{forni2000,forni2005}, \cite{panyao2008}, \cite{LamYaoBathia_Biometrika_2011}, \cite{lamyao2012}, \cite{gaotsay2018a,gaotsay2020a,gaotsay2018b,gaotsay2021,gaotsay2021c}, among others.  
{The aforementioned 
vector time series methods may become inadequate, because 
 many current large-scale time series data 
are naturally represented not as a list or table of numbers, but as a multi-indexed array or tensor.}  An approach to analysing 
tensor-variate time series in the literature is to arrange the data into a high-dimensional vector time series, but 
such an approach overlooks the tensor structure of the data and often employs a large number of parameters in estimation as pointed out in  \cite{werner2008}. 

The matrix-variate time series is a sequence of second-order random tensors,  and studies of matrix time series form naturally  building-blocks for analysing higher-order tensor 
time series. See, for instance, Section 4 of \cite{gao2020}.
But direct analysis of matrix-variate time series is less studied in the literature; \cite{walden2002} considered the 
analysis in signal and image processing,  \cite{chenxiaoyang2020} proposed an autoregressive model for matrix-variate time series, and it was later extended to tensor-variate series by \cite{wanglianzhengli2020} and 
\cite{wangzhengli2021}.  
Similar to the case of vector time series, 
factor models have been proposed for 
modelling high-dimensional matrix time series. 
Roughly speaking, two different approaches have been 
employed in the literature to describe the 
low-dimensional factor structure. 
The first approach assumes that the matrix time series is driven by a low-dimensional matrix factor process, which is dynamically dependent, plus a matrix-variate idiosyncratic term, which is serially uncorrelated. See the factor model in \cite{wang2018} and its extension to the tensor case in \cite{Hanetal2020}. With domain or prior knowledge of the series under study,  \cite{chentsaychen2018} studied constrained matrix-variate factor models by imposing linear constraints on the loading matrices.  The second approach to matrix-variate factor modelling focuses on the cross-sectional factors in 
matrix data 
without using their dynamic information; 
see \cite{chenfan2023} and \cite{yu2022}, among others. 

\begin{figure}
\begin{center}
\subfigure[]{\includegraphics[width=0.45\textwidth]{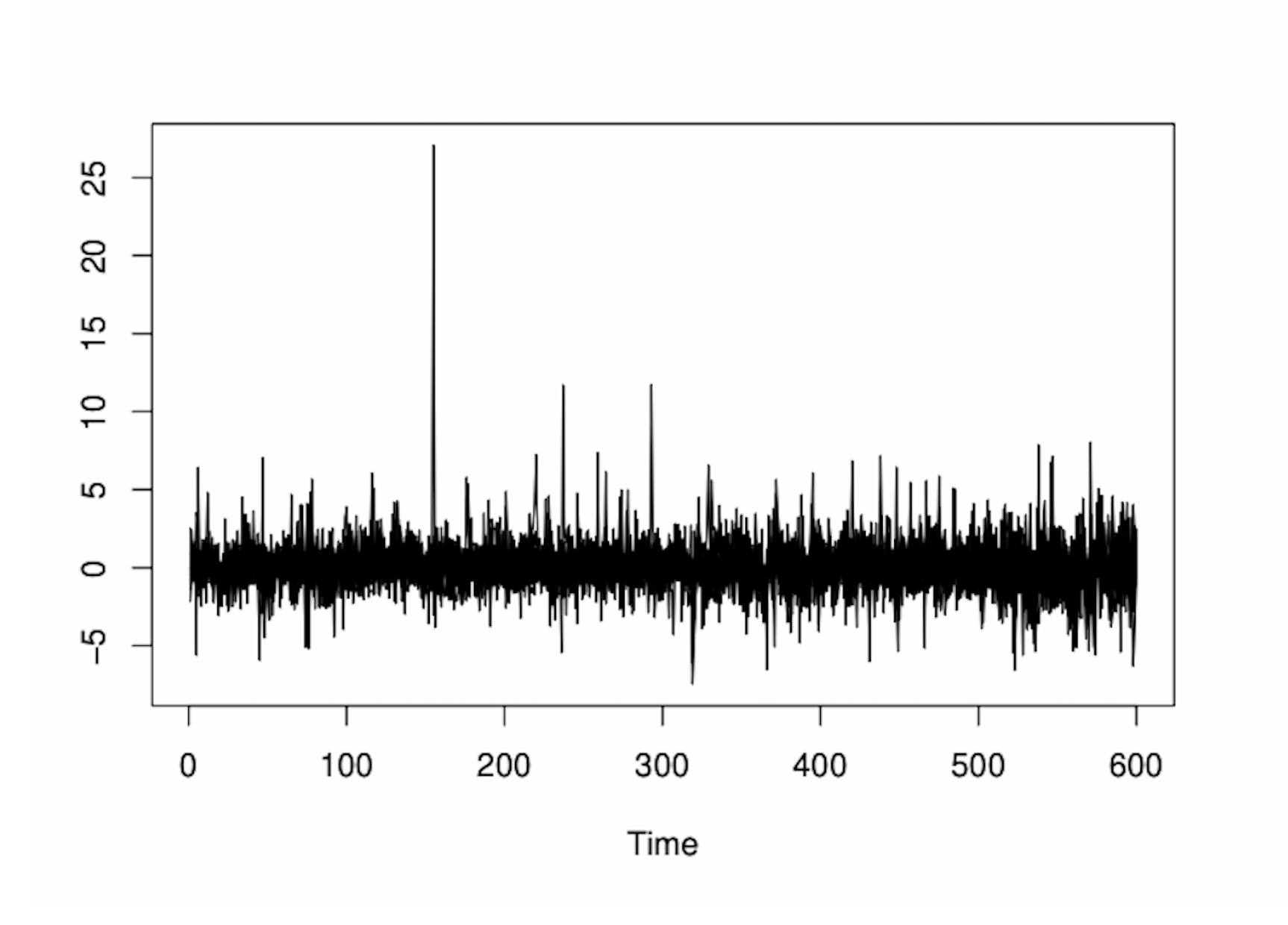}}
\subfigure[]{\includegraphics[width=0.45\textwidth]{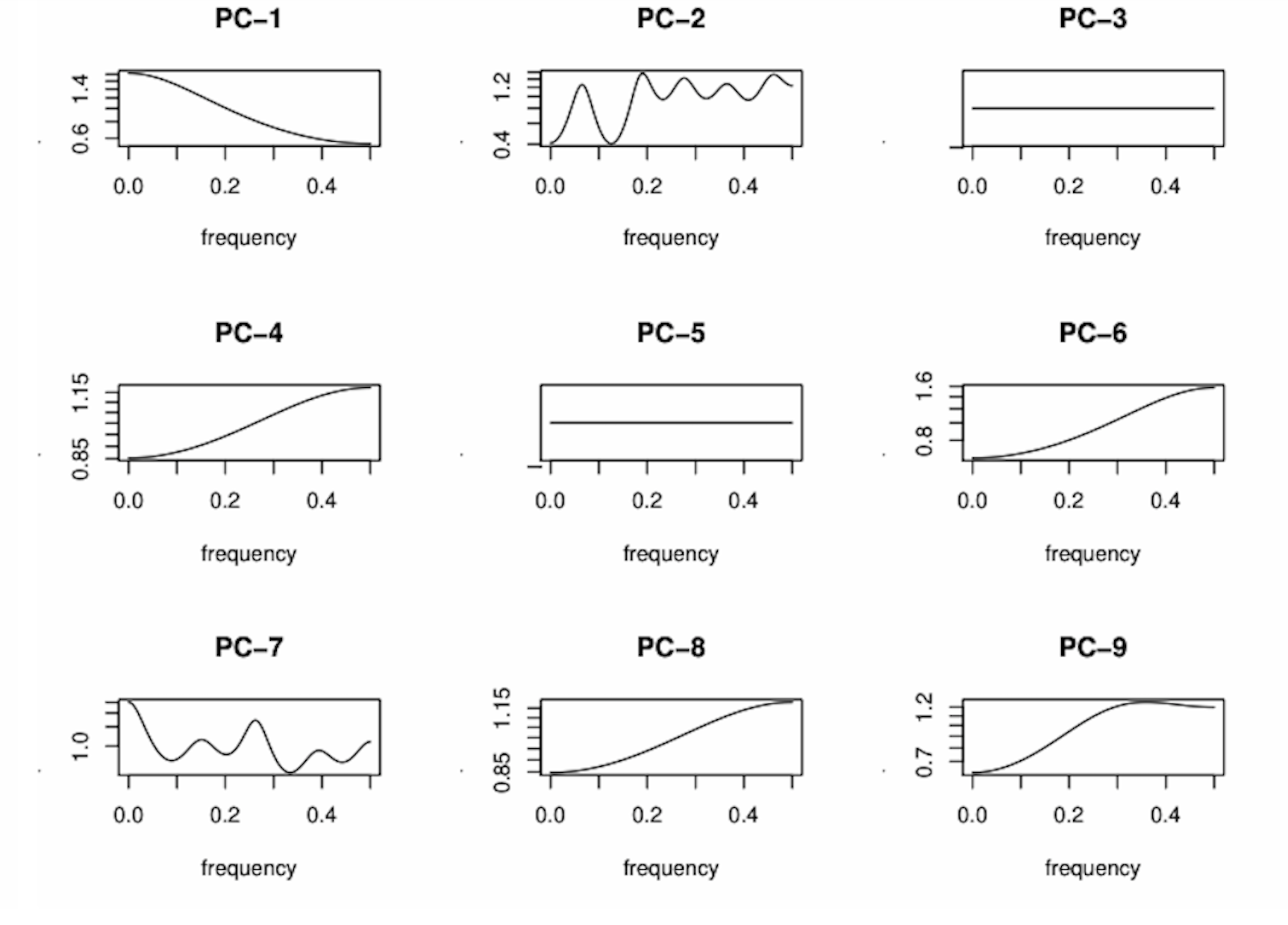}}
\caption{\small (a) Time plots of daily returns of 49 U.S. Industrial Portfolios with 600 observations from
July 13, 1988 to November 23, 1990; (b) The spectral densities of the first nine sample  Principal Components (PCs).  }\label{fig1}
\end{center}
\end{figure}

However, the aforementioned matrix factor models all assume that the covariance of the vectorized idiosyncratic terms is bounded in recovering the latent low-dimensional factors, whereas many empirical examples suggest that the noise effect may be prominent. 
To illustrate,  we plot the daily returns of 49 U.S. industrial portfolios from July 13, 1998 to November 23, 1990 in Figure \ref{fig1}(a). The data is accessible from Prof. Kenneth R. French's library. We apply the PCA to the data and show the spectral densities of the first nine empirical 
principal components in Figure~\ref{fig1}(b). Clearly, 
the third and the fifth components are essentially white noises, indicating that the noise effect of the data can be  prominent.  
See \cite{black1986} for further information. 
When prominent white noises exist, it becomes  
harder to detect the dynamically dependent 
factors.
To address this issue, \cite{gaotsay2021c} introduced a novel matrix-variate factor model from a transformation perspective, where the observed series is represented as a non-singular linear transformation of common factors and white noise series. An advantage of this approach is its ability to handle low signal-to-noise ratios, that is, the case of prominent noises.


This paper focuses on extracting dynamically dependent factors from matrix-variate time series under the framework {that 
the observed series is driven by a 
dynamically dependent, low-dimensional matrix factor process plus a matrix 
noise.} It contributes a further development in factor modelling of high-dimensional matrix-variate time series by proposing an alternative approach to \cite{gaotsay2021c}.  
Specifically, this paper 
provides an efficient alternative approach 
to studying the dynamical dependence of 
matrix time series. 
For a $p_1\times p_2$ dimensional time series $\bY_t=[\by_{1,t},...,\by_{p_2,t}]$,  we start with the factor model considered in \cite{wang2018},
\begin{equation}\label{ft:1}
\bY_t=\bL_1\bF_t\bR_1'+\bE_t,\quad t=1,2,...,T,
\end{equation}
where $\bF_t\in \mathbb{R}^{r_1\times r_2}$ is a latent  matrix-variate common factor process that captures all the dynamic dependence of $\bY_t$,   $\bL_1\in \mathbb{R}^{p_1\times r_1}$ and $\bR_1\in \mathbb{R}^{p_2\times r_2}$ are the front and back loading matrices,  respectively,  and $\bE_t$ is a $p_1\times p_2$ white noise idiosyncratic term. 
{More discussions on the assumption of $\bE_t$ is given in Section 2 and the {\it Online Supplement}.}  
In practice, one expects $r_1 << p_1$ and $r_2 << p_2$. 
In \cite{wang2018}, 
estimation of the front loading space is based on {the sum} of  lag-$k$ autocovariances between all pairs of $\by_{i,t}$ and $\by_{j,t-k}$, for 
$1 \leq i,j \leq p_2$, and for 
$k=1, \ldots, k_0$, where $k_0$ is a 
pre-specified positive integer. 
Similarly, there are $p_1^2$ pairs of autocovariances between the rows of $\bY_t$ in estimating the back loading space at a fixed lag, and the estimation is based on 
summing over lags $k=1, \ldots, k_0$. 
Consequently, the convergence rates of the estimators depend on the dimension $p_2$ (or $p_1$) when estimating the front (or back) loading space.  The resulting convergence rates can be slow when the row 
and column dimensions are high. In addition,  all the eigenvalues of $\cov(\vc(\bE_t))$ are assumed to be bounded when recovering the common dynamic factors. As demonstrated earlier, this condition may 
not hold in many 
empirical applications. This paper introduces a new approach that makes significant 
improvements by (a) speeding up the 
convergence rate in estimation  and (b) 
allowing the noise effects to be prominent.

To begin, we briefly illustrate our idea as follows. Suppose the factor dimensions $r_1$ and $r_2$ are known and let $\bC$ be a $p_2\times r_2$  orthonormal matrix. We project the data onto the directions of $\bC$ and construct a new time series $\bY_t\bC$, which is a $p_1\times r_2$ matrix.  This projection reduces the column dimension from $p_2$ to $r_2$ and should improve the convergence rate in estimating the front loading space via the method of \cite{wang2018} using the 
{projected} series $\bY_t\bC$.  The same argument also applies when estimating the back loading space.  Such ideas are {used in \cite{ye2005} and \cite{hung2012} to approximate a  matrix} by a low-rank one when the data are independent and identically distributed (i.i.d).  \cite{yu2022} applied the {same idea 
to matrix factor models} and found that the convergence rates of the estimated factors and loading 
matrix can be improved.  However,  {our study is 
different from those 
of the aforementioned papers.}  Firstly, we focus on serially or dynamically dependent time-series data while the methods in \cite{ye2005} and \cite{hung2012}  only deal with i.i.d. data.  Secondly,  the factors considered in our approach are dynamically dependent, and we adopt certain auto-covariances to estimate the factors and the corresponding loading matrices  whereas \cite{yu2022}  only studies the cross-sectional dependence of the matrix-variate data without taking {any} lagged information into account. Finally,  in contrast with the bounded eigenvalue assumption of $\cov(\vc(\bE_t))$ in \cite{wang2018} and other forms of weak cross-correlations among $\bE_t$ in \cite{chenfan2023} and \cite{yu2022},   we assume  that the idiosyncratic error term also admits a matrix-variate factor structure such that some eigenvalues of its row and column covariances can diverge as the dimension ($p_1,p_2$) of the system increases.

The goal of this paper is to propose a new iterative estimation method for high-dimensional matrix-variate time series of Equation (\ref{ft:1}).  {There are two reasons that we adopt the autocovariance-based method in \cite{wang2018} and \cite{gaotsay2021c} to extract the dynamically dependent factors rather than the cross-sectional PCA method in \cite{chenfan2023}.} First,  the cross-sectional PCA {cannot distinguish the factors from the  diverging noises} since both of them are associated with large eigenvalues.  Second, from a time-series modeling perspective, the dynamically dependent factors are more useful in forecasting, as explained in \cite{gaotsay2018b}.
Unlike the common orthonormal projection method of \cite{wang2018} and \cite{gaotsay2021c} in estimating the loading matrices,  we first project the observed data onto certain  row or column factor space, which reduces the dimension in one direction of the data when estimating the loading matrix of the other.  The procedure can be iterated until convergence.
 The resulting final estimators of the loading matrices turn out to have faster convergence rates than those obtained by traditional methods even when the noise effect is prominent.  We introduce 
 a two-way projected PCA  to  mitigate the diverging noise effects,  and adopt  high-dimensional white-noise testing procedures  to estimate the dimensions of the factor matrix.  Asymptotic properties of the proposed method are established as the dimensions and sample size approach infinity.  Simulated and real examples are used to assess the performance of the proposed method.  We also compared the proposed method with some existing ones in the literature concerning the forecasting ability of the estimated factors and found that the proposed approach 
 fares well in out-of-sample forecasting.
 

 The estimation framework employed in this paper is 
 different from those in \cite{wang2018} and \cite{gaotsay2021c}. Firstly, except for imposing 
 a  matrix-variate factor structure on the idiosyncratic noises, 
 the model considered  in this paper is the same as the additive matrix-variate factor model in \cite{wang2018}. On the other hand, the factor model in \cite{gaotsay2021c} is based on a two-way nonsingular matrix transformation, in which the dimension of the noise is lower than that of 
 the observed data.
 Secondly, we consider the existence of prominent noise effect in the data, which is similar to the setting in \cite{gaotsay2021c}, whereas  \cite{wang2018} only deals with the case in which 
 all factors are strong and the idiosyncratic noises 
 are weak with bounded eigenvalues in its covaraince. In fact, the model considered in \cite{wang2018} can be treated as a special case of ours by setting the strength parameter of the noises to one in Assumption \ref{asm4} of Section \ref{sec3}, and our procedure and results 
 continue to apply to their model as discussed in Remark \ref{rmk3} of Section \ref{sec3}. Thirdly, the convergence rates of all parameter estimates of the proposed procedure are higher than their counterparts in \cite{wang2018} and \cite{gaotsay2021c} regardless of the noise effect, and the asymptotic normalities of the estimated loadings, absent in \cite{wang2018} and \cite{gaotsay2021c}, are established, indicating that the proposed estimation method has  significant theoretical contribution. 
 
 We adopt the white noise assumption on the idiosyncratic terms primarily for comparison purposes with the existing methods, including those in \cite{wang2018} and \cite{gaotsay2021c}.  This assumption is also reasonable in many applications including  asset returns where the autocorrelations are often negligible 
 under the efficient market theory. 
 However, it is important to note that the proposed method remains feasible even if the noise $\bE_t$ exhibits serial correlations with  a diverging noise effect, as demonstrated and discussed in Section A.4 of the supplement. Notably, we establish and discuss the theoretical properties of the auto-covariance-based method introduced by \cite{lamyao2012} and further developed by \cite{wang2018}  in Section A.4 for both vector and matrix-valued factor models, even in the presence of serially correlated idiosyncratic terms. To the best of our knowledge, this is the first instance in the literature where the feasibility of the auto-covariance-based method by \cite{lamyao2012} is demonstrated in the context of serially correlated idiosyncratic terms. This represents a significant theoretical contribution. Furthermore,  the justifications for the proposed method when $\bE_t$ exhibits serial correlation and consists of a diverging common white noise component is given in the supplement, which serves as another valuable theoretical contribution.
 


The rest of the paper is organized as follows. We introduce the proposed model and 
estimation methodology in Section \ref{sec2} and 
study some theoretical properties of the proposed model and estimation method in Section \ref{sec3}. 
Numerical studies with both simulated and real data sets are given in Section \ref{sec4}, and 
Section \ref{sec5} provides some concluding remarks. A detailed modeling algorithm, some tables and figures, and
all technical proofs are given in an online Supplementary Material. Throughout the article,
 we use the following notation. For a vector
$\bu=(u_1,..., u_p)'\in R^p,$  $||\bu||_2 =\|\bu'\|_2= (\sum_{i=1}^{p} u_i^2)^{1/2} $
is the Euclidean norm, $\|\bu\|_\infty=\max_{1\leq i\leq p}|u_i|$ is the $\ell_\infty$-norm, where the superscript ${'}$ denotes 
the transpose of a vector or matrix.
$\bI_p$ denotes the 
$p\times p$ identity matrix. For a matrix $\bH=(h_{ij})$, $\|\bH\|_1=\max_j\sum_i|h_{ij}|$, $|\bH|_\infty=\max_{i,j}|h_{ij}|$,  $\|\bH
\|_F=\sqrt{\sum_{i,j}h_{ij}^2}$ is the Frobenius norm, $\|\bH
\|_2=\sqrt{\lambda_{\max} (\bH' \bH ) }$, where
$\lambda_{\max} (\cdot) $ denotes for the largest eigenvalue of a matrix, and $\|\bH\|_{\min}$ is the square root of the minimum non-zero eigenvalue of $\bH'\bH$.  We also use the notation $a\asymp b$ to denote that $a$ and $b$ have the same order.

\section{Models and Methodology}\label{sec2}

\subsection{Setting}

Let $\bY_t=[\by_{1,t},...,\by_{p_2,t}]\in \mathbb{R}^{p_1\times p_2}$ be a matrix-variate time series with $\by_{j,t}=(y_{1,j,t},...,y_{p_1, j,t})'\in \mathbb{R}^{p_1}$ and $E(\by_{j,t})={\bf 0}$, for $1\leq j\leq p_2$.  We consider the factor model in (\ref{ft:1}) with dynamically dependent factors. 

The loading matrices $\bL_1$ and $\bR_1$ are not uniquely defined because $c\bL_1$ and $\bR_1/c$, where $c\neq 0$,  also hold for Equation (\ref{ft:1}).  In addition,  when $\bL_1$ is given,  $\bR_1$ and $\bF_t$ are not uniquely identifiable since $(\bR_1, \bF_t)$ can be replaced by $(\bR_1\bH', \bF_t\bH^{-1})$ for a nonsingular matrix $\bH$ without altering Equation (\ref{ft:1}),  and a similar argument applies to $\bL_1$ and $\bF_t$ for a given $\bR_1$.  However,  the linear spaces  spanned by the columns of $\bL_1$ and $\bR_1$,  denoted by $\mathcal{M}(\bL_1)$ and $\mathcal{M}(\bR_1)$,  respectively,  can be uniquely determined.  

To proceed,  we decompose $\bL_1$ and $\bR_1$ as $\bL_1=\bA_1\bW_1$ and $\bR_1=\bP_1\bG_1$,  where $\bA_1$ and $\bP_1$ are semi-orthogonal matrices,  i.e.,  $\bA_1'\bA_1=\bI_{r_1}$ and $\bP_1'\bP_1=\bI_{r_2}$. Clearly, $\mathcal{M}(\bA_1)=\mathcal{M}(\bL_1)$ and $\mathcal{M}(\bP_1)=\mathcal{M}(\bR_1)$.  Let $\bX_t=\bW_1\bF_t\bG_1'$,  Model (\ref{ft:1}) can be rewritten as
\begin{equation}\label{ft:2}
\bY_t=\bA_1\bX_t\bP_1'+\bE_t,\quad  t=1,2,...,T.
\end{equation}
We focus on the case in which $\bX_t$ is dynamically dependent and captures all the prominent dynamic information of the data.  Consequently, the idiosyncratic error term 
$\bE_t$ can {be either a matrix-variate white noise  without dynamic dependence, 
as the settings considered in \cite{lamyao2012}, \cite{wang2018}, and \cite{gaotsay2018b, gaotsay2021c}, or a matrix noise with certain serial dependence discussed below.} In either case, the idiosyncratic term is accompanied by an additional diverging white noise effect that reflects the low signal-to-noise phenomenon {shown} in Figure~\ref{fig1}. 
Moreover, observing that the diverging white noise effect may exist in each row or column of $\bY_t$ (and hence $\bE_t$) for large $p_1$ {or} $p_2$, we assume that  the idiosyncratic term $\bE_t$ also admits an underlying factor structure:
\begin{equation}\label{ido:ft}
\bE_t=\bL_2\bxi_t\bR_2'+\bfeta_t,
\end{equation}
where $\bL_2\in R^{p_1\times k_1}$ and $\bR_2\in R^{p_2\times r_2}$ are the  loading matrices for the white noise term $\bxi_t\in R^{r_1\times r_2}$,  which characterizes the strength of the white noise effect.  
There are two sets of assumptions on $\bfeta_t\in R^{p_1\times p_2}$:
\\ ({\bf A1}). $\bfeta_t$ is a white noise sequence independent of $\bxi_t$ and $\Cov(\vc(\bfeta_t))$ is bounded.  \\
({\bf A2}). $\bfeta_t$ is weakly stationary  with serial correlations  independent of $\bxi_t$, and $\Cov(\vc(\bfeta_t))$ is bounded.\\
Thus, $\bE_t$ is a white noise term under assumption (A1), and a serially correlated sequence under (A2).  The bounded condition of covariance $\Cov(\vc(\bfeta_t))$ is natural since the prominent white noise effect is captured in $\bL_2\bxi_t\bR_2'$ under (A1), and the prominent dynamic information is captured in $\bL_1\bF_t\bR_1'$ under (A2).

For the purpose of comparisons with \cite{wang2018} and \cite{gaotsay2021c}, we adopt the framework in (A1) and illustrate the proposed method in the main text.  The feasibility of the proposed estimation method under (A2) is discussed in Section A.4 of the supplement. Note that the form of the idiosyncratic term in (\ref{ido:ft}) is general {because} each column and row of $\bE_t$ can consist of diverging noise effect as the dimensions $p_1$ and $p_2$ grow.  Moreover, the covariance of $\vc(\bE_t)$ is more general than the Kronecker form adopted in  \cite{dingcook2018} and \cite{chenxiaoyang2020},  among many others, and it reduces to a Kronecker structure if $\bfeta_t=0$ and $\Cov(\vc(\bxi_t))=\bI_{k_1k_2}$.  By the singular-value decomposition,  we assume 
\begin{equation}\label{svd:L2}
\bL_2=\bA_2\bD_2^{1/2}\bU_2'\,\,\text{and}\,\,\bR_2=\bP_2\bLambda_2^{1/2}\bV_2',
\end{equation}
where $\bA_2\in R^{p_1\times k_1}$, $\bU_2\in R^{k_1\times k_1}$, $\bP_2\in R^{p_2\times k_2}$, and $\bV_2\in R^{k_2\times k_2}$ are semi-orthogonal matrices, and $\bD_2$ and $\bLambda_2$ are diagonal with diverging diagonal entries.
{Under the above setting,  our goals are to estimate the front and back loading spaces $\mathcal{M}(\bA_1)$ and $\mathcal{M}(\bP_1)$,  to identify the order of the factor process $(r_1,r_2)$, and to recover the factor processes, allowing the noise effect to be prominent.}


\subsection{Estimation}


We begin by assuming that the factor orders $(r_1,r_2)$ are known 
  and develop the ideas at the population level. This is followed by discussions on sample estimation and methods to estimate the factor orders. 

\subsubsection{Common Orthonormal Projections of Projected Data}
The original loading matrices associated with Model (\ref{ft:2}) and (\ref{ido:ft}) are $\bL_1,\bR_1,\bL_2$, and $\bR_2$, and their equivalent semi-orthogonal versions are $\bA_1,\bP_1,\bA_2$, and $\bP_2$, respectively.
We denote the orthogonal complements of the latter  
by $\bB_1,\bQ_1,\bB_2,$ and $\bQ_2$, respectively.
Furthermore,  for $i=1, 2$, let  $\boldsymbol{\ell}_{i,j}$, $\br_{i,j}$, $\ba_{i,j}$, $\bb_{i,j}$, $\bp_{i,j}$ and $\bq_{i,j}$ be  the $j$-th columns of $\bL_i$, $\bR_i$, $\bA_i$, $\bB_i$, $\bP_i$ and $\bQ_i$, respectively, where the range of $j$ depends on the dimension of the corresponding matrix. 

Consider the method used in \cite{wang2018}, which is essentially a common orthonormal projection of  \cite{gaotsay2021c}. Let $\vc(\bY_t)=(\by_{1,t}',...,\by_{p_2,t}')'$ and $\bdelta_{t}=[\vc(\bY_{t-1})',...,\vc(\bY_{t-k_0})']'$ be the vector of past $k_0$ lagged values of $\bY_t$, where  $k_0$ is a prescribed positive integer.  We seek the direction $\ba$ such that it maximizes the covariance strength between all $\ba'\by_{i,t}$'s and the past lagged vector $\bdelta_t$, which 
{characterizes} the dynamic dependence of the columns on the past ones.  
Equivalently,  we solve the following optimization problem:
\begin{equation}\label{opm}
\max_{\ba\in\mathbb{R}^{p_1}}\sum_{i=1}^{p_2}\|\cov(\ba'\by_{i,t},\bdelta_t)\|_2^2,\quad\text{subject to}\quad \ba'\ba=1.
\end{equation}
Note that
\[\sum_{i=1}^{p_2}\|\cov(\ba'\by_{i,t},\bdelta_t)\|_2^2=\ba'\left[\sum_{k=1}^{k_0}\sum_{i=1}^{p_2}\sum_{j=1}^{p_2}\bSigma_{y,ij}(k)\bSigma_{y,ij}(k)'\right]\ba.  \]

\noindent Let $\bp_{1,i\sbullet}$  be the $i$-th  row vector of $\bP_1$. Define $\bSigma_{y,ij}(k)=\cov(\by_{i,t},\by_{j,t-k})$,  $\bSigma_{xp,ij}(k)=\cov(\bX_t\bp_{1,i\sbullet}',\bX_{t-k}\bp_{1,j\sbullet}')$. Then, $\ba$ is an eigenvector of the matrix

\begin{equation}\label{m1}
\bM_1=\sum_{k=1}^{k_0}\sum_{i=1}^{p_2}\sum_{j=1}^{p_2}\bSigma_{y,ij}(k)\bSigma_{y,ij}(k)'=\bA_1\left\{\sum_{k=1}^{k_0}\sum_{i=1}^{p_2}\sum_{j=1}^{p_2}\left[\bSigma_{xp,ij}(k)\bSigma_{xp,ij}(k)'\right]\right\}\bA_1'.
\end{equation}

We observe that $\bM_1\bB_1=\bf 0$, that is, the columns of $\bB_1$ are the eigenvectors associated with the zero eigenvalues of $\bM_1$, and the front factor loading space $\mathcal{M}(\bA_1)$ is spanned by the eigenvectors corresponding to the $r_1$ non-zero eigenvalues of $\bM_1$. Equivalently, the space spanned by the first $r_1$ solutions to the problem (\ref{opm}) are just the front factor loading space $\mathcal{M}(\bA_1)$. Note that $\bM_1$ in (\ref{m1}) is the same as equations (10)-(11) in \cite{wang2018} because  {the same factor assumption is used.} 
However, $\bM_1$ in (\ref{m1}) is derived from a common orthogonal projection procedure, providing a rational illustration for its use in time-series factor modelling.  {Note that the matrix $\bM_1$ used to estimate $\bA_1$ is fundamentally different from the $\wh\bM_1$ of \cite{yu2022} and the $\wh\bM_R$ of \cite{chenfan2023} because the covariances between all pairs of $(\by_{i,t},\by_{j,t-k})$, for $1\leq i,j\leq p_2$, are used in (\ref{m1}) while  only the covariances  $\Cov(\by_{i,t})$,   for each $1\leq i\leq p_2$, are used in  \cite{yu2022} and  \cite{chenfan2023}.}

The $r_2$ orthonormal directions of the columns of $\bP_1$ can be obtained by applying the same procedure on $\bY_t'$. We can similarly construct $\bM_2$ as $\bM_1$ in (\ref{m1})  based on $\bY_t'$ such that $\bM_2\bQ_1=\bf 0$, and therefore, $\mathcal{M}(\bP_1)$ is the space spanned by the first $r_2$ non-zero eigenvectors of $\bM_2$. 

We next illustrate our main idea using population statistics 
to develop new estimators of $\bA_1$ and $\bP_1$.  Suppose $\bA_1$ and $\bP_1$ are available from the aforementioned  method. Let $\bZ_t=\bY_t\bP_1$, that is, we project the data onto the directions of $\bP_1$ first. Note that $\bA_1$ and $\bP_1$ are semi-orthogonal matrices. It follows from Model (\ref{ft:2}) that
\begin{equation}\label{proj}
\bZ_t:=\bY_t\bP_1=\bA_1\bX_t\bP_1'\bP_1+\bE_t\bP_1=\bA_1\bX_t+\bE_t\bP_1,
\end{equation}
where the dimension $p_2$ has been reduced to $r_2$ in the new data $\bZ_t$,   which is a much smaller one.  Note that the convergence rates of the estimated loading matrices using the original data $\bY_t$ in \cite{wang2018} and \cite{gaotsay2021c} depend heavily on $p_1$ and $p_2$.  The purpose of the projection method in (\ref{proj}) is to increase the convergence rates or equivalently, to reduce the rates of the upper bounds in Theorem 1 of \cite{wang2018} and Theorem 2 of \cite{gaotsay2021c}. 

Consequently, similarly to $\bM_1$ in (\ref{m1}), we can construct
\begin{equation}\label{Mst}
\bM_1^*=\sum_{k=1}^{k_0}\sum_{i=1}^{r_2}\sum_{j=1}^{r_2}\bSigma_{z,ij}(k)\bSigma_{z,ij}(k)',
\end{equation}
which only involves $k_0r_2^2$ terms while there 
are $k_0p_2^2$ terms in (\ref{m1}).
By a similar argument as that in (\ref{m1})  the front factor loading space $\mathcal{M}(\bA_1)$ can also be estimated by the space spanned by the eigenvectors corresponding to the $r_1$ non-zero eigenvalues of $\bM_1^*$. Similarly, we can construct $\bM_2^*$ based on $\bY_t'\bA_1$ and re-estimate $\bP_1$ or the linear space spanned by its columns $\mathcal{M}(\bP_1)$.  In practice,  the true loading matrices $\bP_1$ and $\bA_1$ are not available, we will propose an iterative way to estimate them later.

\vspace{0.2in}
\subsubsection{Two-Way Projected Principal Component Analysis}

In this section, we introduce a 2-way projected PCA to mitigate the noise effect. A similar idea is used in \cite{gaotsay2021c} for a transformed-matrix factor model.  {Let $\bzeta_t=\bD_2^{1/2}\bU_2'\bxi_t\bV_2\bLambda_2^{1/2}$, it follows from  (\ref{ido:ft}) that $\bE_t=\bA_2\bzeta_t\bP_2'+\bfeta_t$.}
If $\bB_1$ and $\bQ_1$ are available, it follows from (\ref{ft:2}) that
\begin{equation}\label{wt}
\bB_1'\bY_t\bQ_1=\bB_1'\bE_t\bQ_1=\bB_1'\bA_2\bzeta_t\bP_2'\bQ_1+\bB_1'\bfeta_t\bQ_1,
\end{equation}
which is a $(p_1-r_1)\times(p_2-r_2)$ matrix-variate white noise process.  On the other hand, the structure of $\bE_t$ in (\ref{ido:ft}) implies that $\bB_2$ and $\bQ_2$ are the directions that capture weaker strength of the covariance of $\vc(\bE_t)$.  Therefore, we project the data $\bY_t$ onto these two spaces as follows,
\begin{equation}\label{pnoise}
\bB_2'\bY_t=\bB_2'\bA_1\bX_t\bP_1'+\cancel{\bB_2'\bA_2\bzeta_t\bP_2'}+\bB_2'\bfeta_t\,\,\text{and}\,\, \bY_t\bQ_2=\bA_1\bX_t\bP_1'\bQ_2+\cancel{\bA_2\bzeta_t\bP_2'\bQ_2}+\bfeta_t\bQ_2,
\end{equation}
where the diverging white noise effects are mitigated by $\bB_2$ and $\bQ_2$, respectively.  
{Since $\bzeta_t$, $\bfeta_t$ and $\bX_t$ are uncorrelated with each other,  the left projection directions} in the columns of $\bB_2$ for $\bY_t$ capture weaker dependence between $\bY_t$ and $\vc(\bB_1'\bY_t\bQ_1)$ from the left-hand-side, and the directions $\bQ_2$ capture weaker dependence between $\bY_t$ and $\vc(\bB_1'\bY_t\bQ_1)$ from the right-hand-side.

Let $\bOmega_{y_i}=\cov(\by_{i,t},\vc(\bY_t))$ and $\bOmega_{\zeta,i}=\cov(\bzeta_t\bp_{2,i\sbullet}',\vc(\bzeta_t))$, where $\bp_{2,i\sbullet}$ is the $i$th row vector of $\bP_2$.  It follows from  (\ref{ft:2}) and (\ref{wt}) that 
\begin{equation}\label{p:cov}
\Cov(\by_{i,t},\vc(\bB_1'\bY_t\bQ_1))=\bOmega_{y_i}(\bQ_1\otimes\bB_1)=\bA_2\bOmega_{\zeta,i}(\bP_2'\bQ_1\otimes\bA_2'\bB_1)+\cov(\bfeta_{i,t},\vc(\bB_1'\bfeta_t\bQ_1)).
\end{equation}
Therefore, we define
\begin{equation}\label{ppca}
\bS_1=\sum_{i=1}^{p_2}[\bOmega_{y_i}(\bQ_1\otimes\bB_1)][\bOmega_{y_i}(\bQ_1\otimes\bB_1)]',
\end{equation}
which is a positive semi-definite matrix and $\bB_2$ consists of the eigenvectors corresponding to the smallest $p-k_1$ eigenvalues of $\bS_1$ because $\bB_2'\bA_2={\bf 0}$.  Similarly, we can construct $\bS_2$ based on the transpose of $\bY_t\bQ_2$ in the second equation in (\ref{pnoise}) and obtain $\bQ_2$.  We observe that any subspace matrices $\bB_2^*=\bB_2\bC_1$ and $\bQ_2^*=\bQ_2\bC_2$ are still orthogonal to $\bA_2$ and $\bP_2$, respectively. To properly extract the factor process, we only need to find two matrices $\bC_1\in R^{(p_1-k_1)\times r_1}$ and $\bC_2\in R^{(p_2-k_2)\times r_2}$ such that $\bB_2^*\in R^{p_1\times r_1}$ and $\bQ_2^*\in R^{p_2\times r_2}$ are the two subspaces of $\bB_2$ and $\bQ_2$, respectively. It follows from (\ref{ft:2}) that
\begin{equation}\label{proj:2}
\bB_2^*{'}\bY_t\bQ_2^*=\bB_2^*{'}\bA_1\bX_t\bP_1'\bQ_2^*+\bB_2^*{'}\bE_t\bQ_2^*,
\end{equation}
where the diverging noise effect has been mitigated, and consequently,  if the noise in (\ref{proj:2}) is negligible and the square matrices $\bB_2^*{'}\bA_1$ and $\bP_1'\bQ_2^*$ are invertible, we can construct the factor process as
\begin{equation}\label{etr:ft}
\bX_t\approx (\bB_2^*{'}\bA_1)^{-1}\bB_2^*{'}\bY_t\bQ_2^*(\bP_1'\bQ_2^*)^{-1}.
\end{equation}
In theory, $\bB_2^*$ (and $\bQ_2^*$) can be any subspaces of $\bB_2$ (and $\bQ_2$), but we will propose a way to select them at the sample level in the next subsection.

\begin{remark}
(i) When the projection matrices $\bB_2^*$ and $\bQ_2^*$ are available,  (\ref{proj:2}) is a transformed version of matrix-variate factor models where the diverging noise effect has been eliminated. An alternative way to estimate the factors is to apply the proposed procedure below to the data $\bB_2^*{'}\bY_t\bQ_2^*$. However, the loading matrices  will be some transformed versions of $\bB_2^*{'}\bA_1$ and $\bQ_2^*{'}\bP_1$ instead of $\bA_1$ and $\bP_1$.  Moreover, the estimation errors of the loading matrices will be also depending on the errors in estimating $\bA_1,\bP_1,\bB_2$, and $\bQ_2$ in previous steps.  Therefore, we adopt the way in (\ref{etr:ft}) to extract the factors such that we can estimate the loading matrices corresponding to the original ones.\\
(ii) To illustrate the invertibilities of $\bB_2^*{'}\bA_1$ and $\bP_1'\bQ_2^*$ in (\ref{etr:ft}).  Note that $\bB_2$ is an orthogonal complement of $\bL_2$,  i.e., $\bB_2'\bL_2={\bf 0}$,  then there exists a  transformation matrix $\bK=[\bK_1',\bK_2']'$  with $\bK_1\in R^{k_1\times r_1}$, $\bK_2\in R^{(p_1-k_1)\times r_1}$, and $rank(\bK)=r_1$ such that $\bA_1=[\bL_2,\bB_2]\bK=\bL_2\bK_1+\bB_2\bK_2$. Then $\bB_2^*{'}\bA_1=\bC_1'\bK_2$ and we require $rank(\bK_2)=r_1$ for the invertibility of $\bB_2^*{'}\bA_1$. A sufficient (not necessary) condition is that the $r_1$-dimensional column space of $\bA_1$ is a subspace of the $(p_1-k_1)$-dimensional column space of $\bB_2$.  When this does not hold,  it is still reasonable to assume $rank(\bK_2)=r_1$ since the number of rows $p_1-k_1$ of $\bK_2$ is much larger than that ($k_1$) of $\bK_1$ in the full-rank matrix $\bK$ as the dimension $p_1$ diverges. The way to obtain $\wh\bB_2^*$ and $\wh\bQ_2^*$ discussed in Section~\ref{sec3} also guarantees that $(\wh\bB_2^*{'}\wh\bA_1)^{-1}$ and $(\wh\bQ_2^*{'}\wh\bP_1)^{-1}$ behave well in finite samples.
A more general condition is given in Assumption \ref{asm5}(ii) of Section~\ref{sec3} below to guarantee the invertibilities of $\bB_2^*{'}\bA_1$ and $\bP_1'\bQ_2^*$.
\end{remark}

\vspace{0.2in}
\subsubsection{Sample Estimation}

Given a realization $\{\bY_t:t=1,...,T\}$, the goal is to estimate $\bA_1$ and $\bP_1$, or equivalently $\mathcal{M}(\bA_1)$ and $\mathcal{M}(\bP_1)$, and 
the dimension $(r_1,r_2)$ of the factor matrix, and to recover the latent factor matrix process $\bX_t$. 
To begin, we need to choose some initial estimator $(\wh r_{1}^o, \wh r_{2}^o)$ of  the factor order $(r_1,r_2)$. 
This is carried out via the diagonal-path selection method in Section 2.5 of \cite{gaotsay2021c}.  
We briefly outline the method below and {provide details in the online supplement}.

{
First, 
construct a sample version 
of $\bM_1$ defined in (\ref{m1}) as
\begin{equation}\label{m1hat}
\wh\bM_1=\sum_{k=1}^{k_0}\sum_{i=1}^{p_2}\sum_{j=1}^{p_2}\wh\bSigma_{y,ij}(k)\wh\bSigma_{y,ij}(k)',
\end{equation}
where $\wh\bSigma_{y,ij}(k)$ is the lag-$k$ sample autocovariance between $\by_{i,t}$ and $\by_{j,t-k}$.  The estimation accuracy of autocovariances and, hence, $\wh\bM_1$ deteriorates as the lag $k$ increases.  Therefore, some caution is needed when selecting $k_0$ in real applications.  For stationary time series, a relatively small $k_0$ is sufficient to capture the dynamic information of the data.  One may also choose an optimal $k_0$ in terms of the out-of-sample forecasting performance of the extracted factors.  Simulation studies in Section~\ref{sec4} indicate that the results seem not sensitive to the choice of $k_0$.} 
Similarly we can apply the same procedure to $\{\bY_t',t=1,...,T\}$ and construct $\wh\bM_2$.   The idea of our method follows from Equation (\ref{wt}) that $\bB_1'\bY_t\bQ_1$ is a matrix-variate white noise process. Let $\wh \bGamma_1^o$ and $\wh\bGamma_2^o$ be the matrices of eigenvectors (in the decreasing order of the corresponding eigenvalues) of the sample matrix $\wh\bM_1$ in (\ref{m1hat}) and $\wh \bM_2$, respectively.  Define  $\wh\bN_t=\wh\bGamma_1^o{'}\bY_t\wh\bGamma_2^o$ and let $\wh\bN_t(i,j)\in\mathbb{R}^{(p_1-i+1)\times (p_2-j+1)}$ be the lower-right submatrix consisting of the $i$-th to the $p_1$-th rows and the $j$-th to the $p_2$-th columns of $\wh\bN_t$, and $\wh\bN_t^*(i,j)\in \mathbb{R}^{(i-1)\times (j-1)}$ be the upper-left  submatrix of $\wh\bN_t$. Our test procedure searches  
the  order $(i,j)$ such that $\wh\bN_t^*(i,j)$ consists of all the factors and the remaining 
elements of $\wh\bN_t$ are white noises. The estimate of $(r_1,r_2)$ is then $(i-1,j-1)$. The testing procedure is discussed  below, and the test statistic used depends on the dimension $p_1p_2$.

If the dimension $p_1p_2$ is small, implying that $\bY_t$ is a low dimensional matrix, we recommend using the well-known Ljung-Box statistic $Q_s(m)$ for multivariate time series, where $s$ and $m$ denote the dimension of the vector and the number of lags used. See, for example, \cite{hosking1980} and \cite{Tsay_2014}. Specifically, we first search the minimum of $r_1$ and $r_2$ along the diagonal of $\wh\bN_t$. Consider the null hypothesis
\[H_0(l): \vc(\wh\bN_t(l,l))\,\, \text{is a vector white noise}, \]
with type-I error $\alpha$. $H_0(l)$ is rejected if $Q_{d_l}(m)\geq \chi_{d_l^2 m,1-\alpha}^2$, where  $d_l=(p_1-l+1)(p_2-l+1)$ is the dimension of $\vc(\wh\bN_t(l,l))$ and $\chi_{d_l^2 m,1-\alpha}^2$ is the $(1-\alpha)$-th quantile of a chi-squared distribution with $d_l^2m$ degrees of freedom. We start with $l=1$. If $H_0(1)$ is rejected, we increase $l$ by 1 and repeat the testing 
procedure until we  cannot reject $H_0(l)$, and denote the resulting  
order as $l^*$. Two situations can happen. If $l^*=\min(p_1,p_2)$ and we still  reject $H_0(l^*)$, we fix one dimension (say $p_1$ when $p_1=l^*$), and test whether $\vc(\wh\bN_t(p_1,p_1+j))$ is white noise or not by starting with $j=1$ until we cannot reject $H_0$. If $l^*<\min(p_1,p_2)$, then we perform a back testing to determine the maximum order of the factor matrix. That is, we first test whether $\vc(\wh\bN_t(l^*-1+i,l^*-1))$ is a vector white noise starting with $i=1$. Increase $i$ by 1 
and repeat the testing procedure until we cannot reject $H_0$ at $i=i^*$. Second, we test whether $\vc(\wh\bN_t(l^*+i^*-2,l^*-1+j))$ is a vector white noise starting with $j=1$. Increase $j$ by 1 
and repeat the testing procedure until we reject $H_0$ at $j=j^*$. Then, 
{we have $\wh r_1^o=l^*+i^*-2$ and $\wh r_2^o=l^*+j^*-2$.} Finally, $\wh \bGamma_1^o=[\wh\bA_1^o,\wh\bB_1^o]$ and $\wh\bGamma_2^o=[\wh\bP_1^o,\wh\bQ_1^o]$, where $\wh\bA_1^o\in\mathbb{R}^{p_1\times \wh r_1^o}$ and $\wh\bP_1^o\in\mathbb{R}^{p_2\times \wh r_2^o}$.

For large $p_1$ and/or $p_2$, we use the same testing procedure, but replace the 
$Q_s(m)$ test statistics by high-dimensional white noise (HDWN) tests. This is so, because 
$Q_s(m)$ is no longer adequate.  We consider two HDWN test statistics
in this paper. The first test statistic is introduced by \cite{changyaozhou2017} and makes use of the maximum absolute auto- and cross-correlations of the component series. Specifically, let $\wh\bGamma_N(k)=[\wh\rho_{ij}(k)]_{1\leq i,j\leq d_l}$ be the lag-$k$ sample auto-correlation matrix of $\vc(\wh\bN_t(l,l))$, the test statistic $S_T$ is defined as
\[S_T=\max_{1\leq k\leq m}\max_{1\leq i,j\leq d_l}T^{1/2}|\wh\rho_{ij}(k)|,\]
and its limiting distribution under $H_0(l)$ can be approximated by that of the $L_\infty$-norm of a normal random vector, which can be simulated by a bootstrapping algorithm. The second  
HDWN test statistic is developed by \cite{Tsay_2018}.   Let $\wh\bGamma_{N,k}=[\wh\Gamma_{N,k}(i,j)]_{1\leq i,j\leq d_l}$ be the lag-$k$ sample rank auto-correlation matrix of an orthogonalized vector of $\vc(\wh\bN_t(l,l))$, where the orthogonalization can be done via PCA if $d_l<T$ and we only vectorize the top-left $\min(p_1,\sqrt{\epsilon T})$-by-$\min(p_2,\sqrt{\epsilon T})$ principal submatrix of $\wh\bN_t(l,l)$ for some $\epsilon\in(0,1)$ if $d_l\geq T$; See \cite{gaotsay2018b} for details. The test statistic is defined as
\[S(m)=\max\{\sqrt{T}|\wh\Gamma_{N,k}(i,j)|:1\leq i,j\leq d_l,1\leq k\leq m\},\]
and its limiting distribution under $H_0(l)$ is a function of the standard Gumbel distribution via 
the extreme value theory. 
The critical values and the rejection regions of the test statistics are available in closed form and can be found in \cite{Tsay_2018} or Section 2.3 in \cite{gaotsay2018b}.

Next, we discuss the proposed method based on the main idea discussed in Section 2.2.1.  The initial choices of $r_1$ and $r_2$ are given by $r_{1}^0=\wh r_1^o$ and $r_{2}^0=\wh r_2^o$, respectively. 
 Suppose $\bP_{0,1}\in R^{p_2\times r_{2}^0}$ is a non-random semi-orthogonal matrix, which can be arbitrary in theory as long as  $\bP_{0,1}'\bP_1\neq {\bf 0}$. In practice, we generate a sequence of orthonormal matrices and select $\bP_{0,1}$ to maximize the average singular values of  $\bP_{0,1}'\wh\bP_1^o$. We  project the data $\bY_t$ onto $\bP_{0,1}$ as $\wh\bZ_{0,t}=\bY_t\bP_{0,1}=[\wh\bz_{1,t}^0,...,\wh\bz_{r_{2}^0,t}^0]$,  and construct the sample version of $\bM_{0,1}^*$ in (\ref{Mst}) as 
\begin{equation}\label{mst:hat}
\wh\bM_{0,1}^*=\sum_{k=1}^{k_0}\sum_{i=1}^{r_2^0}\sum_{j=1}^{r_2^0}\wh\bSigma_{z_0,ij}(k)\wh\bSigma_{z_0,ij}(k)',
\end{equation}
where $\wh\bSigma_{z_0,ij}(k)$ is the sample covariance between $\wh\bz_{i,t}^0$ and $\wh\bz_{j,t-k}^0$.  A new estimator for $\bA_1$ is denoted by $\wh\bA_{0,1}$ whose columns  are the normalized eigenvectors corresponding to the $r_1^0$ largest eigenvalues of $\wh\bM_{0,1}^*$.  Consequently, we may obtain the null space of $\wh\bM_{0,1}^*$, denoted by $\mathcal{M}(\wh\bB_{0,1})$, where $\wh\bB_{0,1}$ consists of the eigenvectors corresponding to the $p_1-r_1^0$ smallest eigenvalues of $\wh\bM_{0,1}^*$.

Let $\wh\bW_{0,t}=\bY_t'\wh\bA_{0,1}=[\bw_{1,t}^0,...,\bw_{r_1^0,t}^0]$.  Applying the same procedure to $\{\wh\bW_{0,t}\}_{t=1}^T$, we can construct $\wh\bM_{0,2}^*$, and 
obtain the estimator $\wh\bP_{1,1}$ of $\bP_1$, and $\wh\bQ_{1,1}$ of $\bQ_1$ at the same time.  We then project the data $\bY_t$ onto $\wh\bP_{1,1}$ as $\wh\bZ_{1,t}=\bY_t\wh\bP_{1,1}$ and construct $\wh\bM_{1,1}^*$ to obtain a new estimator $\wh\bA_{1,1}$ for $\bA_1$.  With $\wh\bA_{1,1}$, we  apply the procedure to $\wh\bW_{1,t}=\bY_t'\wh\bA_{1,1}$ and repeat this process until the estimators converge. In this way, we obtain a sequence of estimators $\{\bP_{0,1},\wh\bP_{1,1},...,\wh\bP_{i,1}\}$ for $\bP_1$ and another sequence $\{\wh\bA_{0,1},\wh\bA_{1,1},...,\wh\bA_{i,1}\}$ for $\bA_1$, if we stop at the $i$-th step, where $\bP_{0,1}$ is the initial non-random semi-orthogonal matrix.

The above estimation depends on the initial choices of $r_1$ and $r_2$ in each iteration. But the theoretical proofs in the supplement for Section 3 suggest that the convergence rates of the first $\min(r_1, r_1^0)$ columns of $\wh\bA_{i,1}$ and the $\min(r_2,r_2^0)$ columns of $\wh\bP_{i,1}$ still hold as those in Section 3 when $r_1^0=r_1$ and $r_2^0=r_2$.  Suppose we stop at the $i$-th iteration and let $\wh \bGamma_{i,1}=[\wh\bA_{i,1},\wh\bB_{i,1}]$ and $\wh\bGamma_{i,2}=[\wh\bP_{i,1},\wh\bQ_{i,1}]$ be the matrices of eigenvectors (in the decreasing order of the corresponding eigenvalues) of the sample matrix $\wh\bM_{i,1}^*$ in (\ref{mst:hat}) and $\wh \bM_{i-1,2}^*$, respectively.  We apply the diagonal-path selection method mentioned above to the transformed series $\wh\bGamma_{i,1}'\bY_t\wh\bGamma_{i,2}$ and obtain the re-estimated factor order $(\wh r_1, \wh r_2)$. Then we re-partition the  transformation matrices as $\wh \bGamma_{i,1}=[\wh\bA_{1},\wh\bB_{1}]$ and $\wh\bGamma_{i,2}=[\wh\bP_{1},\wh\bQ_{1}]$, where $\wh\bA_1\in\mathbb{R}^{p_1\times \wh r_1}$ and $\wh\bP_1\in\mathbb{R}^{p_2\times \wh r_2}$.

  The initial choices of the  factor orders do not affect the convergence rates of the final estimators $\wh\bA_1$ and $\wh \bP_1$ so long as $r_1^0$ and $r_2^0$ are chosen as small integers.  In this paper, we suggest choosing $r_1^0=\wh r_1^o$ and $r_2^0=\wh r_2^o$, where $(\wh r_1^o,\wh r_2^o)$ are obtained by the diagonal-path selection method.
  The method is also briefly illustrated in Section A.1 of the supplement if we adopt the initial estimators $\wh\bA_1^o$ and $\wh\bP_2^o$ and their orthogonal complements.  Finally, the estimator $(\wh r_1, \wh r_2)$ of $(r_1,r_2)$ is  re-estimated based on the full eigenvector matrices $[\wh\bA_1,\wh\bB_1]$ and $[\wh\bP_1,\wh\bQ_1]$ as illustrated in Section A.1 of the supplement.

Once we have the estimators $\wh\bA_1$ and $\wh\bP_1$,  we consider methods for obtaining the estimators of $\bB_2$ and $\bQ_2$.  Let 
\begin{equation}\label{s1:hat}
\wh\bS_1=\sum_{i=1}^{p_2}[\wh\bOmega_{y_i}(\wh\bQ_1\otimes\wh\bB_1)][\wh\bOmega_{y_i}(\wh\bQ_1\otimes\wh\bB_1)]',
\end{equation}
where $\wh\bOmega_{y_i}$ is the sample estimator of $\bOmega_{y_i}$ defined in Section 2.3.  When the numbers of diverging noises $k_1$ and $k_2$ are known, letting $\wh\bB_2$ and $\wh\bQ_2$ be the sample estimators of $\bB_2$ and $\bQ_2$, respectively,  we suggest choosing  $\wh\bB_2^*=\wh\bB_2\bC_1$ and $\wh\bQ_2^*=\wh\bQ_2\bC_2$, where the columns of $\bC_1$ are chosen as the $\wh r_1$ eigenvectors of $\wh\bB_2'\wh\bA_1\wh\bA_1'\wh\bB_2$ corresponding to the $\wh r_1$ largest eigenvalues, and the columns of $\bC_2$ are the $\wh r_2$ eigenvectors of $\wh\bQ_2'\wh\bP_1\wh\bP_1'\wh\bQ_2$ corresponding to the largest $\wh r_2$ eigenvalues.  These choices guarantee that both $\wh\bB_2^*{'}\wh\bA_1$ and $\wh\bP_1'\wh\bQ_2^*$ behave well in empirical calculations. 
Finally, we recover the latent factor matrix as
\begin{equation}\label{rec:ft}
\wh\bX_t=(\wh\bB_2^*{'}\wh\bA_1)^{-1}\wh\bB_2^*{'}\bY_t\wh\bQ_2^*(\wh\bP_1'\wh\bQ_2^*)^{-1}. 
\end{equation}
When $k_1$ and $k_2$ are unknown,  we adopt the ratio-based technique to estimate them. Let $\wh\mu_{1,1}\geq ...\geq  \wh\mu_{1,p_1}$ and $\wh\mu_{2,1}\geq ...\geq  \wh\mu_{2,p_2}$ be the eigenvalues of $\wh\bS_1$ and $\wh\bS_2$, respectively.  Define
\begin{equation}\label{ratios}
\wh k_{1,0}=\arg\min_{1\leq j\leq R}\wh\mu_{1,j+1}/\wh\mu_{1,j}\,\,\text{and}\,\,\wh k_{2,0}=\arg\min_{1\leq j\leq R}\wh\mu_{2,j+1}/\wh\mu_{2,j}.
\end{equation}
We may choose $\wh k_1=\wh k_{1,0}$ and $\wh k_{2}=\wh k_{2,0}$. In fact, as discussed in Remark 2 of \cite{gaotsay2021c}, the proposed method still works even if $\wh k_{1,0}\leq \wh k_1\leq p_1-\wh r_1$ and $\wh k_{2,0}\leq \wh k_2\leq p_2-\wh r_2$ because $\wh\bB_2$ and $\wh\bQ_2$ are still orthogonal to the diverging components and their subspaces can still mitigate the diverging noises.

\section{Theoretical Properties}\label{sec3}
We present here the asymptotic theory for the estimation methods described in Section 2 when $T,p_1,p_2\rightarrow\infty$. For simplicity, we assume  $r_1$ and $r_2$ are known and fixed because the theoretical analysis suggests that the initial estimators $\wh r_1^o$ and $\wh r_2^o$ do not change the convergence rates of the final estimators of the loading spaces with known $r_1$ and $r_2$.  The consistency of the white noise tests in determining $r_1$ and $r_2$ of the factor process is shown thereafter. We also assume the numbers of diverging noise components $k_1$ and $k_2$ are known because (1) accurate estimations of them are not necessary as discussed at the end of Section 2.2.3 and (2) the consistency of the ratio-based method proposed at the end of Section 2.2.3 is straightforward and will be discussed at the end of this section.

We begin with some assumptions.

\begin{assumption}\label{asm1}
The process $\{\vc(\bY_t),\vc(\bF_t)\}$ is $\alpha$-mixing with the mixing coefficient satisfying the condition $\sum_{k=1}^\infty\alpha_p(k)^{1-2/\gamma}<\infty$ for some $\gamma>2$, where
\[\alpha_p(k)=\sup_{i}\sup_{A\in\mathcal{F}_{-\infty}^i,B\in \mathcal{F}_{i+k}^\infty}|P(A\cap B)-P(A)P(B)|,
\]
and $\mathcal{F}_i^j$ is the $\sigma$-field generated by $\{(\vc(\bY_t),\vc(\bF_t)):i\leq t\leq j\}$.
\end{assumption}
\begin{assumption}\label{asm2}
For any $i=1,...,r_1 r_2$ and $1\leq j\leq p_1p_2$, $E|f_{i,t}|^{2\gamma}<C_1$ and $E|\omega_{j,t}|^{2\gamma}<C_2$, where $f_{i,t}$ and $\omega_{j,t}$ are the $i$-th and $j$-th element of $\bff_t=\vc(\bF_t)$ and $\bomega_t=\vc(\bE_t)$, respectively, $C_1$ and $C_2$ are positive constants, and $\gamma$ is given in Assumption 1.
\end{assumption}



\begin{assumption}\label{asm3}
(i) There exists $\delta_1\in (0,1)$ such that $\|\bL_1\|_2^2\asymp p_1^{1-\delta_1}\asymp\|\bL_1\|_{\min}$ and $\|\bR_1\|_2^2\asymp p_2^{1-\delta_1}\asymp\|\bR_1\|_{\min}$; (ii) $\|\ba_{1,i}\|_2\asymp p_1^{-1/2}$ and $\|\bp_{1,j}\|_2\asymp p_2^{-1/2}$, for $1\leq i\leq p_1$ and $1\leq j\leq p_2$, where $\ba_{1,i}$ and $\bp_{1,j}$ are the $i$-th and the $j$-th row vectors of $\bA_1$ and $\bP_1$, respectively.
\end{assumption}

\begin{assumption}\label{asm4}
The diagonal matrices $\bD_2$ and $\bLambda_2$ in (\ref{svd:L2}) satisfy that  $\bD_2=\diag(d_1,...,d_{k_1})$ and $\bLambda_2=\diag(\gamma_{1},...,\gamma_{k_2})$
  with  $d_1\asymp...\asymp d_{k_1}\asymp p_1^{1-\delta_2}$ and $\gamma_1\asymp...\asymp\gamma_{k_2}\asymp p_2^{1-\delta_2}$ for some $\delta_2\in(0,1)$.
\end{assumption}

\begin{assumption}\label{asm5}
(i) For any $1\leq l_1\leq p_1$, $1\leq l_2\leq p_2$, $\bh\in \mathbb{R}^{l_1l_2}$, $\bH_1\in\mathbb{R}^{p_1\times l_1 }$ and $\bH_2\in \mathbb{R}^{p_2\times l_2}$ with $\|\bh\|_2=c<\infty$, $\bH_1'\bH_1=\bI_{l_1}$ and $\bH_2'\bH_2=\bI_{l_2}$, we assume $E|\bh'\vc(\bH_1'\bfeta_{t}\bH_2)|^{2\gamma}<\infty$; (ii)  $\sigma_{\min}(\bC_1'\bB_2{'}\bA_1)\geq C_3$ and $\sigma_{\min}(\bC_2'\bQ_2{'}\bP_1)\geq C_4$ for some constants $C_3, C_4>0$ and some semi-orthogonal matrices $\bC_1\in \mathbb{R}^{p_1\times r_1}$ and $\bC_2\in \mathbb{R}^{p_2\times r_2}$ satisfying $\bC_1'\bC_1=\bI_{r_1}$ and $\bC_2'\bC_2=\bI_{r_2}$, where $\sigma_{\min}$ denotes the minimum non-zero singular value of a matrix. 
\end{assumption}
Assumption \ref{asm1} controls the dynamic dependence of the time series under study. 
See \cite{gaoetal2017} for a theoretical justification for VAR series. 
Assumption \ref{asm2} is used to establish the convergence of the sample covariance matrices. Similar to \cite{fan2013} and \cite{gaotsay2021c},  Assumptions \ref{asm3}-\ref{asm4}  impose some strengths on the loading matrices $\bL_1$ and $\bR_1$ of Equation (\ref{ft:1}) and the white noise effect of $\bY_t$, which allow the factors to be either pervasive or slightly weak. 
Assumption~\ref{asm3}(ii) ensures that each component $y_{i,j,t}$ of $\bY_t$ has a finite variance under the normalized conditions $\bA_1'\bA_1=\bI_{r_1}$ and $\bP_1'\bP_1=\bI_{r_2}$. Note that 
$\delta_1=0$ corresponds to the strong or pervasive factors which are used in \cite{BaiNg_Econometrica_2002} and \cite{fan2013}. When $\delta_1>0$,  the corresponding factors are called weaker ones and we can link the convergence rates of the estimated factors explicitly to the strength of the factors.  Similar argument applied to the choice of $\delta_2$ in characterizing the strength of noises. Note that Assumption \ref{asm3} is equivalent to Condition 4 in \cite{wang2018}, as explained in \cite{gaotsay2021c}.
We restrict $\delta_1$ and $\delta_2$ in $(0,1)$ in Assumptions \ref{asm3}-\ref{asm4} and exclude the two end points to better illustrate the advantages of the  proposed method, but the theory still holds if we take the limits to either side.  Assumption \ref{asm5}(i) is mild and includes the  standard normal distribution as a special case, and Assumption 5(ii) is reasonable since $\bB_2^*$ is a subspace of $\bB_2$, $\bQ_2^*$ is a subspace of $\bQ_2$, and it implies  that $\bB_2^{*}{'}\bA_1$ and $\bQ_2^{*}{'}\bP_1$ are nonsingular.

As discussed in Section 2,  we will estimate $\bB_2$ or equivalently $\mathcal{M}(\bB_2)$, which is the subspace spanned by the eigenvectors associated with the $p_1-k_1$ smallest eigenvalues of $\bS_1$.  Assume $\wh{\bB}_2$ consists of the eigenvectors corresponding to the smallest $p_1-k_1$ eigenvalues of $\wh\bS_1$.  Under some conditions, we can show that $\mathcal{M}(\wh{\bB}_2)$ is consistent to $\mathcal{M}(\bB_2)$.  This is also the case in the literature on high-dimensional PCA with i.i.d. data. See, for example, \cite{shenetal2016} and the references therein. Therefore, the choice of $\wh\bB_2^*$ should be a subspace of $\wh{\bB}_2$. The choices of $\bC_1$ and $\bC_2$, and hence the estimates $\wh\bB_2^*=\wh{\bB}_2\bC_1$ and $\wh\bQ_2^*=\wh{\bQ}_2\bC_2$ will be discussed later.

Recall that $\wh\bA_1^o$ and $\wh\bP_1^o$ and their corresponding orthogonal complements $\wh\bB_1^o$ and $\wh\bQ_1^o$ are obtained by the traditional methods in \cite{wang2018} and \cite{gaotsay2021c}.  We further denote by  $\wh\bB_2^o$ and $\wh\bQ_2^o$ the estimators of the orthogonal matrices used to mitigate the diverging noise effects based on the projected PCA of \cite{gaotsay2021c} under their transformation model. Let $D(\bH_1,\bH_2)$ be a discrepancy measure between two $p\times r$ semi-orthogonal matrices $\bH_1$ and $\bH_2$, defined as
\begin{equation}
D({\bf H}_1,{\bf
H}_2)=\sqrt{1-\frac{1}{r}\textrm{tr}({\bf H}_1{\bf H}_1'{\bf
H}_2{\bf H}_2')}.\label{eq:D}
\end{equation}
We first present the consistency results of these estimators.
\begin{proposition}\label{prop1}
Suppose Assumptions 1--5 hold and $r_1$ and $r_2$ are known and fixed.  As $T\rightarrow\infty$, if $p_1^{\delta_1}p_2^{\delta_1}T^{-1/2}=o(1)$, then 
\[D({\wh\bA}_1^o,{\bA}_1)=O_p(p_1^{\delta_1}p_2^{\delta_1}T^{-1/2})\,\,\text{and}\,\,D({\wh\bP}_1^o,{\bP}_1)=O_p(p_1^{\delta_1}p_2^{\delta_1}T^{-1/2}),\]
and the above results also hold for $D({\wh\bB}_1^o,{\bB}_1)$ and $D({\wh\bQ}_1^o,{\bQ}_1)$.
Furthermore,
\[
D(\wh\bB_2^o,\bB_2)
=O_p(p_1^{\delta_2}p_2^{3\delta_2/2}T^{-1/2}+p_1^{\delta_1}p_2^{\delta_1+\delta_2}T^{-1/2}),\text{and}\,\,D(\wh\bQ_2^o,\bQ_2)=O_p(p_1^{3\delta_2/2}p_2^{\delta_2}T^{-1/2}+p_1^{\delta_1+\delta_2}p_2^{\delta_1}T^{-1/2}).
\]
\end{proposition}

\begin{remark}{
The convergence rates of the factor loadings are the same as those in Theorem 1 of \cite{wang2018} and Theorem 2 of \cite{gaotsay2021c}, even though the latter only uses a 
single parameter to describe the common strength of the loadings. 
The parameter $\delta_2$ is used to characterize the strength of the diverging noise effects as that of \cite{gaotsay2021c}. }
\end{remark}

\begin{theorem}\label{tm1}
Suppose Assumptions 1--5 hold and $r_1$ and $r_2$ are known and fixed. As $T\rightarrow\infty$, if $p_1\asymp p_2$, $p_1^{\delta_1-\delta_2}p_2^{\delta_1-\delta_2}T^{-1/2}=o(1)$, then 
\begin{equation}\label{da1:err}
D({\wh\bA}_1,{\bA}_1)=\left\{\begin{array}{ll}
O_p(T^{-1/2}),&\text{if}\,\, \delta_1\leq \delta_2,\delta_2\leq1/2,\\
O_p(p_1^{\delta_1-\delta_2}p_2^{\delta_1-\delta_2}T^{-1/2}),&\text{if}\,\, \delta_1>\delta_2,\delta_2\leq 1/2,\\
O_p(T^{-1/2}),&\text{if}\,\, \delta_1\leq 1/2,\delta_2> 1/2,\\
O_p(p_1^{\delta_1}p_2^{\delta_1-1}T^{-1/2}),&\text{if}\,\, \delta_1> 1/2,\delta_2> 1/2,
\end{array}\right.
\end{equation}
and 
\begin{equation}\label{dp1:err}
D({\wh\bP}_1,{\bP}_1)=\left\{\begin{array}{ll}
O_p(T^{-1/2}),&\text{if}\,\, \delta_1\leq \delta_2,\delta_2\leq1/2,\\
O_p(p_1^{\delta_1-\delta_2}p_2^{\delta_1-\delta_2}T^{-1/2}),&\text{if}\,\, \delta_1>\delta_2,\delta_2\leq 1/2,\\
O_p(T^{-1/2}),&\text{if}\,\, \delta_1\leq 1/2,\delta_2> 1/2,\\
O_p(p_2^{\delta_1}p_1^{\delta_1-1}T^{-1/2}),&\text{if}\,\, \delta_1> 1/2,\delta_2> 1/2,
\end{array}\right.
\end{equation}
and the first result above also holds for $D({\wh\bB}_1,{\bB}_1)$ and the second for $D({\wh\bQ}_1,{\bQ}_1)$.
Furthermore, if $p_1^{\delta_2}p_2^{\delta_2}T^{-1/2}=o(1)$, 
\begin{align}\label{db2:th}
D(\wh\bB_2,\bB_2)=&O_p\left(p_1^{\delta_2}p_2^{\delta_2}T^{-1/2}+p_1^{-1+2\delta_2}p_2^{-1/2+2\delta_2}T^{-1/2}\right.\\
&\left.+(1+p_1^{-1+\delta_2}p_2^{-1/2+\delta_2}+p_1^{-2+2\delta_2}p_2^{-1+2\delta_2})(D({\wh\bQ}_1,{\bQ}_1)+D({\wh\bB}_1,{\bB}_1))\right),\notag
\end{align}
and
\begin{align}\label{dq2:th}
D(\wh\bQ_2,\bQ_2)=&O_p\left(p_2^{\delta_2}p_1^{\delta_2}T^{-1/2}+p_2^{-1+2\delta_2}p_1^{-1/2+2\delta_2}T^{-1/2}\right.\\
&\left.+(1+p_2^{-1+\delta_2}p_1^{-1/2+\delta_2}+p_2^{-2+2\delta_2}p_1^{-1+2\delta_2})(D({\wh\bQ}_1,{\bQ}_1)+D({\wh\bB}_1,{\bB}_1))\right).\notag
\end{align}
\end{theorem}

\begin{remark}\label{rmk3}
(i) 
The conditions for Theorem~\ref{tm1} are equivalent to those in \cite{wang2018}, except for the ones on the diverging noises in Assumptions~\ref{asm4}-\ref{asm5}. Under Assumptions 3-4 that $\delta_1,\delta_2\in (0,1)$,  the convergence rates of the estimated factor loading spaces in Theorem \ref{tm1} are faster than those in Proposition \ref{prop1} and \cite{wang2018}.  Specifically,  the stochastic bound between the estimated front loading space and the true one is reduced by an order of $p_1^{\delta_1}p_2^{\delta_1}$ if $\{\delta_1\leq\delta_2,\delta_2\leq 1/2\}$ or $\{\delta_1\leq 1,\delta_2>1/2\}$,  an order of $p_1^{\delta_2}p_2^{\delta_2}$ if $\{\delta_2>\delta_2,\delta_2\leq 1/2\}$, and an order of $p_1^{-1}$ if $\{\delta_1>1/2,\delta_2>1/2\}$. If $p_1\asymp p_2$, the convergence rates of the back loading space in (\ref{dp1:err}) are the same as those in (\ref{da1:err}), and they are all faster than their counterparts in Proposition \ref{prop1} and \cite{wang2018}.  When the factors are strong ones (i.e., $\delta_1=0$), the convergence rates are the same as those in \cite{wang2018}. \\
(ii) In particular,  Model (\ref{ft:2}) with noise term in (\ref{ido:ft}) reduces to the one in \cite{wang2018} if there is no diverging noise effect, i.e., $\delta_2=1$ in Assumption~\ref{asm4}, and we only require (A1) and Assumptions \ref{asm1}--\ref{asm3} for (\ref{da1:err}) and (\ref{dp1:err}) to hold, which are equivalent to the Conditions 1--4 in \cite{wang2018}, where the distinct eigenvalue condition 
is not a necessary one.  Therefore, with the same set of conditions and $\delta_2=1$,  the stochastic bounds in (\ref{da1:err}) are reduced by a factor of $p_1^{\delta_1}p_2^{\delta_1}$ or $p_2$ for the front loading, and $p_1^{\delta_1}p_2^{\delta_1}$ or $p_1$ for the back loading in (\ref{dp1:err}) compared with those in \cite{wang2018}. This marks a significant improvement.  \\
(iii) If $\delta_1=0$ and $\delta_2=1$, the factors are strong and the noises are weak.  The non-asymptotic rate of the estimated loadings is $O_p(T^{-1/2})$, which is independent of the dimensions $p_1$ and $p_2$.  Under the framework that $p_1\asymp p_2\asymp p$, the non-asymptotic rate in Corollary 3.1 of \cite{yu2022} is  $O_p((Tp)^{-1/2}+(p_1p_2)^{-1})$.  This difference can be explained in two-fold.  First,  the estimation method of this paper is different from that in \cite{yu2022}.  Specifically, the auto-covariance-based method in this paper makes use of the dependence information between all pairs of the columns and rows of the data $\bY_t$, whereas the method of \cite{yu2022} treats the matrix  $\bY_t$ as a whole and only considers cross-sectional covariance between two matrices. Therefore, the essential length/sample size in \cite{yu2022} is of order $Tp$ while it is  $T$ in our procedure. Second,  the white noise assumption used in this paper is different from those in \cite{yu2022},  and hence, the auto-covariance of the white noise terms would produce a factor of $T^{-1/2}$. If the noises are not white,  Step 1 after (A.4.15) in the supplement yields a rate of $O_p(T^{-1/2}+(p_1p_2)^{-1})$, which is the same as that in \cite{yu2022} if the sample size $T$ is replaced by $Tp$ of \cite{yu2022}.\\
(iv) Due to the improvement achieved in the first step, the convergence rates of the estimated $\wh\bB_2$ and $\wh\bQ_2$((A.3.22) and (A.3.23) in the supplement), which are used to mitigate the prominent noise effect, are also improved and become faster under many situations. For example, it is reasonable to assume that $p_1\asymp p_2\asymp p$ and $\delta_1\asymp \delta_2\asymp \delta$ for a dimension parameter $p$ and a strength parameter $\delta\in (0,1)$, then the upper bounds for $D({\wh\bB}_2^o,{\bB}_2)$ and $D(\wh\bQ_2^o,\bQ_2)$ are all of order $p^{3\delta}T^{-1/2}$ in Proposition \ref{prop1}
whereas their counterparts in (\ref{db2:th}) and (\ref{dq2:th}) above are of order $\max(p^{2\delta}T^{-1/2},p^{4\delta-3/2}T^{-1/2})$, which is smaller.  We omit further details for other cases.
\end{remark}
{Given $\wh{\bB}_2$ and $\wh\bQ_2$, one can choose $\wh\bB_2^*$ and $\wh\bQ_2^*$ as follows. Let $\wh\bC_1=(\wh\bfc_{1,1},...,\wh\bfc_{1,r_1})\in \mathbb{R}^{(p_1-k_1)\times r_1}$ and $\wh\bC_2=(\wh\bfc_{2,1},...,\wh\bfc_{2,r_2})\in \mathbb{R}^{(p_2-k_2)\times r_2}$, 
where $\wh \bfc_{1,i}$ is the vector associated with the $i$-th largest eigenvalue of $\wh{\bB}_2'\wh\bA_1\wh\bA_1'\wh{\bB}_2$ and $\wh \bfc_{2,j}$ is the vector associated with the $j$-th largest eigenvalue of $\wh{\bQ}_2{'}\wh\bP_1\wh\bP_1'\wh{\bQ}_2$.  We may define $\wh\bB_2^*=\wh{\bB}_2\wh\bC_1$ and $\wh\bQ_2^*=\wh{\bQ}_2\wh\bC_2$, which guarantee that the matrices $(\wh\bB_2^*{'}\wh\bA_1)^{-1}$ and $(\wh\bQ_2^*{'}\wh\bP_1)^{-1}$ behave well in recovering the factor $\wh\bX_t$,  and  the diverging part of the noise covariance matrix can be eliminated, as shown in Theorems \ref{tm20} and \ref{tm2} below.  Using the eigenvalue ratios in (\ref{ratios}), consistency of the obtained $\wh k_{1,0}$ and $\wh k_{2,0}$ to their true values $k_1$ and $k_2$ can be established by a similar argument as that in \cite{lamyao2012}, \cite{ahn2013}, and \cite{gaotsay2021}, among many others.  On the other hand, there are many ways to choose the numbers of components $\wh k_1$ and $\wh k_2$ in Assumption \ref{asm4} so long as $\wh k_1\geq \wh k_{1,0}$, $\wh k_2\geq \wh k_{2,0}$, $p_1-\wh k_1>r_1$ and $p_2-\wh k_2>r_2$.   We do not repeat the details here and adopt the true $k_1$ and $k_2$ in the following theorems stating the convergence rates of the extracted common factors and components.  }
\begin{theorem}\label{tm20}
Under the Assumptions in Theorem \ref{tm1}, we have
\begin{align*}\label{axt:h}
\frac{1}{\sqrt{p_1p_2}}\|\wh\bX_t-\bH_L\bX_t\bH_R'\|_2=&O_p\left(D(\wh\bB_2,\bB_2)D(\wh\bQ_2,\bQ_2)+p_2^{-1/2}D(\wh\bB_2,\bB_2)+p_1^{-1/2}D(\wh\bQ_2,\bQ_2)+(p_1p_2)^{-1/2}\right),
\end{align*}
where $\bH_L=(\wh\bB_2^*{'}\wh\bA_1)^{-1}\wh\bB_2^*{'}\bA_1$ and $\bH_R=(\wh\bQ_2^*{'}\wh\bP_1)^{-1}\wh\bQ_2^*{'}\bP_1$, as shown in the  supplement.
\end{theorem}

\begin{theorem}\label{tm2}{
Let the Assumptions in Theorem \ref{tm1} hold.\\
(i) The estimated common component satisfies
\begin{align*}
(p_1p_2)^{-1/2}\|\wh\bA_1\wh\bX_t\wh\bP_1'-\bA_1\bX_t\bP_1'\|_2=&O_p\left(D(\wh\bB_2,\bB_2)D(\wh\bQ_2,\bQ_2)+p_2^{-1/2}D(\wh\bB_2,\bB_2)+p_1^{-1/2}D(\wh\bQ_2,\bQ_2)\right.\notag\\
&\left.+(p_1p_2)^{-\delta_1/2}D(\wh\bA_1,\bA_1)+(p_1p_2)^{-\delta_1/2}D(\wh\bP_1,\bP_1)+(p_1p_2)^{-1/2}\right).
\end{align*}
(ii) If $\delta_1=\delta_2=0$, the $(i,j)$-th element of the common component satisfies
\begin{equation}\label{axp:pt}
\left|(\wh\bA_1\wh\bX_t\wh\bP_1')_{i,j}-(\bA_1\bX_t\bP_1')_{i,j}\right|\leq O_p\left(\sqrt{\frac{p_1}{T}}+\sqrt{\frac{p_2}{T}}+\frac{\sqrt{p_1p_2}}{T}+\frac{1}{\sqrt{p_1p_2}}\right),
\end{equation} 
for $1\leq i\leq p_1,1\leq j\leq p_2$, as  $p_1,p_2,T\rightarrow\infty$.
}
\end{theorem}

\begin{remark}
(i) A similar result is given in Theorem 3 of \cite{LamYaoBathia_Biometrika_2011}, Theorem 3 of \cite{wang2018}, Theorem 5 of \cite{gaotsay2018b} and Theorem 3 of \cite{gaotsay2021c} in extracting the factor processes. If $\delta_1=\delta_2=0$, i.e. the factors and the noise terms are all strong, the  error rate of the estimated factors in Theorem~\ref{tm20} is $O_p(T^{-1}+(p_1T)^{-1/2}+(p_2T)^{-1/2}+(p_1p_2)^{-1/2})$,  which is only slightly larger than the averaging rate $O_p(T^{-1}+N^{-1/2})$ in Proposition 1 of \cite{BaiNg2023}, where $N$ is the dimension therein. The averaging error rate of all the common components in Theorem \ref{tm2} is $O_p((p_1p_2)^{-1/2}+T^{-1/2})$, which is the same as that of a single component specified in Theorem 3 of \cite{Bai_Econometrica_2003} in dealing with the traditional approximate factor model.\\
(ii) As discussed in Remark 2,  all the distances between the estimated eigen-spaces and the true ones are smaller than those obtained by the traditional methods in \cite{wang2018} and \cite{gaotsay2021c}, implying that the stochastic bound in Theorem \ref{tm2}(i) is smaller than the ones stated in Theorem 3 of \cite{gaotsay2021c} under the assumption that the noise effect is prominent.\\
(iii) A  pointwise convergence rate of $|(\wh\bA_1\wh\bX_t\wh\bP_1')_{i,j}-(\bA_1\bX_t\bP_1')_{i,j}|$ for general $\delta_1$ and $\delta_2$  is provided in the proof of Theorem~\ref{tm2} of the supplement.  We only present the rate for the special case when $\delta_1=\delta_2=0$ for simplicity. It can be seen that the pointwise consistency can be obtained if $p_1=o(T)$ and $p_2=o(T)$.  Under additional condition in Assumption~\ref{asm7} below, we can further improve the rates in (\ref{axp:pt}) to $O_p(T^{-1/2}+(p_1p_2)^{-1/2})$, which is in line with the result in Theorem 3 of \cite{Bai_Econometrica_2003}.  We omit the details, 
but point it out in Remark A.3.1 of the supplement. 
\end{remark}
Next, we study the consistency of the white noise tests described in Section 2. The  conditions depend on the test statistic used. We only consider the two test statistics $S_T$ and $S(m)$ of Section 2.2.3.
and  present the consistency when $p_1$ and $p_2$ are large since the case of small $p_1$ and $p_2$ is trivial. {A random vector  $\bx_t$ is sub-Gaussian if there exists a constant $C>0$ such that $P(|\bv'(\bx_t-E\bx_t)|>x)\leq C\exp(-Cx^2)$ for any constant vector $\bv$ with $\|\bv\|_2=1$.}
We need an additional assumption.
\begin{assumption}
 $\vc({\bF_t})$,  $\vc(\bxi_{t})$, and $\vc(\bfeta_t)$ are sub-Gaussian random vectors.
\end{assumption}

\begin{theorem}\label{tm3}
Assume that Assumptions 1--6 hold.\\
(i) If $p_1^{1-\delta_1}p_2^{1-\delta_1}T^{-2}+p_1^{1-\delta_2}p_2^{1-\delta_2}T^{-1}=o(1)$ on the event of $\{\delta_1\leq \delta_2,\delta_2\leq 1/2\}$ or $\{\delta_1\leq 1/2,\delta_2>1/2\}$, and $p_1^{1+3\delta_1-4\delta_2}p_2^{1+3\delta_1-4\delta_2}T^{-2}+p_1^{1+2\delta_1-3\delta_2}p_2^{1+2\delta_1-3\delta_2}T^{-1}=o(1)$ on  $\{\delta_1>\delta_2,\delta_2\leq 1/2\}$, and $p_1^{3\delta_1-1}p_2^{3\delta_1-1}T^{-2}+p_1^{1+2\delta_1-\delta_2}p_2^{2\delta_1-\delta_2-1}T^{-1}+p_1^{2\delta_1-\delta_2-1}p_2^{1+2\delta_2-\delta_2}T^{-1}=o(1)$ on $\{\delta_1> 1/2,\delta_2> 1/2\}$, then the test statistic $S_T$  can consistently estimate $r_1$ and $r_2$, i.e. $P(\wh r_1=r_1,\wh r_2=r_2)\rightarrow 1$ as $T\rightarrow\infty$.\\
(ii)  If $p_1^{(1-\delta_1)/2}p_2^{(1-\delta_1)/2}T^{-1}\sqrt{\log(T)}+p_1^{1/2}p_2^{1/2}T^{-1/2}\sqrt{\log(Tp_1p_2)}=o(1)$ on the event of $\{\delta_1\leq \delta_2,\delta_2\leq 1/2\}$ or $\{\delta_1\leq 1/2,\delta_2>1/2\}$, and $p_1^{1/2+3\delta_1/2-2\delta_2}p_2^{1/2+3\delta_1/2-2\delta_2}T^{-1}\sqrt{\log(T)}+p_1^{1/2+\delta_1-\delta_2}p_2^{1/2+\delta_1-\delta_2}T^{-1/2}$ $\sqrt{\log(Tp_1p_2)}=o(1)$ on  $\{\delta_1>\delta_2,\delta_2\leq 1/2\}$, and $p_1^{3\delta_1/2-1/2}p_2^{3\delta_1/2-1/2}T^{-1}\sqrt{\log(T)}+p_1^{\delta_1+1/2}p_2^{\delta_1-1/2}T^{-1/2}$ $\sqrt{\log(Tp_1p_2)}=o(1)$ on $\{\delta_1> 1/2,\delta_2> 1/2\}$,  then 
the test statistic $S(m)$ can consistently estimate $r_1$ and $r_2$.
\end{theorem}

By Theorem~\ref{tm3}, we see that the requirements for $p_1$ and $p_2$ are weaker than their counterparts in Theorem 4 of \cite{gaotsay2021c}, but we do not repeat them here to save space.

{Finally,  we state the asymptotic normality of the estimated loading vectors with some additional assumptions for the special case of $\delta_1=\delta_2=0$. The limiting distributions can also be established for other choices of $\delta_1$ and $\delta_2$ following the proof in the supplement. Recall that the $i$-th column (or row) vector of the matrix $\wh\bP_1$ is denoted as $\wh\bp_{1,\sbullet i}$ (or $\wh\bp_{1,i\sbullet}$). Similar notation also applies to $\wh\bA_1$.
\begin{assumption}\label{asm7}
Let $\bbeta_{1,\sbullet i}$ and $\balpha_{1,\sbullet j}$ be the limits of $\wh\bp_{1,\sbullet i}$ and $\wh\ba_{1,\sbullet j}$, respectively.
(i) For $1\leq l \leq p_1$,  
\[\frac{\sqrt{p_1T}}{(p_1p_2)^2}\bA_1'\sum_{k=1}^{k_0}\sum_{i=1}^{r_2}\sum_{j=1}^{r_2}\frac{1}{T}\sum_{t=k+1}^T\bY_t\bbeta_{1,\sbullet i}\bbeta_{1,\sbullet j}'\bY_{t-k}'\frac{1}{T}\sum_{t=k+1}^T\bY_{t-k}\bbeta_{1,\sbullet j}\bbeta_{1,\sbullet i}'\be_{l\sbullet,t}\rightarrow_d N({\bf 0},\bSigma_1).\]
(ii) For $1\leq l \leq p_2$,  
\[\frac{\sqrt{p_2T}}{(p_1p_2)^2}\bP_1'\sum_{k=1}^{k_0}\sum_{i=1}^{r_2}\sum_{j=1}^{r_2}\frac{1}{T}\sum_{t=k+1}^T\bY_t'\balpha_{1,\sbullet i}\balpha_{1,\sbullet j}'\bY_{t-k}\frac{1}{T}\sum_{t=k+1}^T\bY_{t-k}'\balpha_{1,\sbullet j}\balpha_{1,\sbullet i}'\be_{\sbullet l,t}\rightarrow_d N({\bf 0},\bSigma_2).\]
\end{assumption}

\begin{theorem}\label{thm5}
Suppose Assumptions \ref{asm1}--\ref{asm7} and the conditions in Theorems \ref{tm1}--\ref{tm3} hold.  If $\delta_1=\delta_2=0$, then
\[\sqrt{p_1T}(\wh\ba_{1,l\sbullet}-\bH_{1,T}\ba_{1,l\sbullet})\rightarrow_d N({\bf 0},\bV_1^{-1}\bH_1\bSigma_1\bH_1'\bV_1^{-1}),\quad 1\leq l\leq p_1,\]
and
\[\sqrt{p_2T}(\wh\bp_{1,l\sbullet}-\bH_{2,T}\bp_{1,l\sbullet})\rightarrow_d N({\bf 0},\bV_2^{-1}\bH_2\bSigma_2\bH_2'\bV_2^{-1}),\quad 1\leq l\leq p_2,\]
where $\bH_{1,T}$ and $\bH_{2,T}$ are some rotation matrices with limits $\bH_1$ and $\bH_2$, respectively,  and $\bV_1$ and $\bV_2$ are the limits of diagonal matrices specified in the supplement.
\end{theorem}
Note that the convergence rate of the loading vectors is actually $\sqrt{T}$ in Theorem~\ref{thm5} because we assumed $\bA_1'\bA_1=\bI_{r_1}$ and $\bP_1'\bP_1=\bI_{r_2}$, and the factors $\sqrt{p_1}$ and $\sqrt{p_2}$ are associated with the scaled loading vectors. This is different from the rate in \cite{chenfan2023} and \cite{yu2022}, where $\bA_1'\bA_1/p_1=\bI_{r_1}$ and $\bP_1'\bP_1/p_2=\bI_{r_2}$ are assumed.  In addition, we can treat the sample size as $Tp_2$ when estimating the front loading and $Tp_1$ in estimating the other in the procedure of \cite{chenfan2023} and \cite{yu2022}, and, therefore, the convergence rate of the loading vectors therein is still of order the square-root of the sample size.

}
\section{Numerical Properties}\label{sec4}
\subsection{Simulation}

We illustrate the finite-sample properties of the proposed methodology under different choices of $p_1$ and $p_2$. 
As the dimensions of $\wh\bA_1$  and $\bA_1$ are not necessarily the same,
and  $\bL_1$ is not an orthogonal matrix in general, we first extend the discrepancy measure
in Equation (\ref{eq:D}) 
to a more general form below. Let $\bH_i$ be a
$p\times h_i$ matrix with rank$(\bH_i) = h_i$, and $\bP_i =
\bH_i(\bH_i'\bH_i)^{-1} \bH_i'$, for $i=1,2$. Define
\begin{equation}\label{dmeasure}
\bar{D}(\bH_1,\bH_2)=\sqrt{1-
\frac{1}{\max{(h_1,h_2)}}\textrm{tr}(\bP_1\bP_2)}.
\end{equation}
Then $\bar{D} \in [0,1]$. Furthermore,
$\bar{D}(\bH_1,\bH_2)=0$ if and only if
either $\mathcal{M}(\bH_1)\subset \mathcal{M}(\bH_2)$ or
$\mathcal{M}(\bH_2)\subset \mathcal{M}(\bH_1)$, and  it is 1 if and only if
$\mathcal{M}(\bH_1) \perp \mathcal{M}(\bH_2)$.
When $h_1 = h_2=h$ and $\bH_i'\bH_i= \bI_r$,
$\bar{D}(\bH_1,\bH_2)
$ reduces to that in Equation (\ref{eq:D}). 
We only present the simulation results for $k_0=2$ in Equation (\ref{m1hat}) to save space because other choices of $k_0$ produce similar patterns.  {See the simulation results for $k_0=1,3$, and $4$ in Tables~A.8-A.10 of the Supplementary Materials.}

{\noindent \bf Example 1.} Consider model (\ref{ft:1}) with common factors following the equation 
\[\bF_t=\bPhi\bF_{t-1}\bPsi'+\bN_t,\]
where $\bN_t$ is a matrix-variate white noise process with independent entries, $\bPhi\in \mathbb{R}^{r_1\times r_1}$ and $\bPsi\in \mathbb{R}^{r_2\times r_2}$ are two diagonal coefficient matrices.  The true dimension of the matrix factor is $(r_1,r_2)=(2,3)$, the orders of the diverging noise components are $(k_1,k_2)=(1,2)$ as defined in Assumption 4, the dimensions used are $(p_1,p_2)=(7,7)$, $(10,15)$, $(20,20)$, and $(20,30)$, and the sample sizes are $T=300$, $500$, $1000$, $1500$, $3000$. We consider three scenarios for the factor strength $\delta_1$ and $\delta_2$: $(\delta_1,\delta_2)=(0,0)$, $(0.2,0.4)$ and $(0.6,0.2)$. Similar results are obtained 
for other settings entertained, we omit the details to save space.  
For each scenario mentioned above, we first set the seed to \texttt{1234} and the elements of $\bL_1$, $\bR_1$ $\bL_2$ and $\bR_2$  are drawn independently from $U(-2,-1)\cup U(1,2)$. We then divide $\bL_1$ ($\bR_1$) and $\bL_2$ ($\bR_2$) by $p_1^{\delta_1/2}$ ($p_2^{\delta_1/2}$) and $p_1^{\delta_1/2}$ ($p_2^{\delta_1/2}$), respectively, to satisfy Assumptions \ref{asm3}-\ref{asm4}.  For simplicity, we generate $\vc(\bxi_t)\sim N(0,\bI_{k_1k_2})$ and $\vc(\bfeta_t)\sim N(0,\bI_{p_1p_2})$ in Model (\ref{ido:ft}).  $\bPhi$ and $\bPsi$ are diagonal matrices with their diagonal elements drawn independently from $U(0.5,0.9)$,  and $\vc(\bN_t)\sim N(0,\bI_{r_1r_2})$. We use 500 replications in each experiment.

We first study the performance of estimating the dimension of the matrix-variate factors.  Let $\wh \lambda_{1,1}\geq...\geq \wh \lambda_{1,p_1}$ be the eigenvalues of $\wh\bM_1$ in (\ref{m1hat}) and $\wh \lambda_{2,1}\geq...\geq \wh \lambda_{2,p_2}$ be the eigenvalues of $\wh\bM_2$ using the observed data.  \cite{wang2018} proposed the following way to determine the factor order:
\[\wh r_i^{WLC}=\arg\min_{1\leq j\leq p_i/2}\wh\lambda_{i,j+1}/\wh\lambda_{i,j},\quad i=1,2.\]
We compare the proposed method with those in \cite{wang2018}, \cite{gaotsay2021c}, and \cite{yu2022} to determine the order of the factor process. These estimators are denoted by $(\wh r_1, \wh r_2)$,  $(\wh r_1^{WLC},\wh r_2^{WLC})$, $(\wh r_1^o,\wh r_2^o)$, and $(\wh r_1^{YHKZ}, \wh r_2^{YHKZ})$, respectively. For simplicity, we only report the results of the test statistic $S(m)$ with $m=10$, defined in Section 2, and the results for the other test are similar.  The initial projection matrix $\bP_{0,1}$ is simply obtained by applying singular-value decomposition to $\bR_1$ and its left singular vectors are used as $\bP_{0,1}$ in the experiment, and the results are similar if we generate a sequence of orthonormal matrices as described in Section~A.1 of the supplement 
and choose  $\bP_{0,1}$  such that $\bP_{0,1}'\wh\bP_{1,0}^{o}$ has largest eigenvalues on an average basis. See the simulation results in Tables~A.11-A.13 of the supplement.  When $p_1p_2>T$, we only keep the upper $\epsilon\sqrt{T}$ row- and column-transformed series of $\wh\bGamma_1^{o}{'}\bY_t\wh\bGamma_2^{o}$ and $\wh\bGamma_1'\bY_t\wh\bGamma_2$ with $\epsilon=0.9$ in the testing. Similar results are obtained for other choices of $\epsilon$, but we do not report them here.  
The testing results are given in Tables~A.1-A.3 of the supplement where we consider the scenarios when the strengths of the factors and the noises are both strong ($\delta_1=\delta_2=0$),  the strength of the factors is stronger than that of the noises ($\delta_1=0.2,\delta_2=0.4$)  and the strength of the factors is weaker than that of the noises ($\delta_1=0.6,\delta_2=0.2$). From the tables,  when both the factors and the noises are strong ($\delta_1=\delta_2=0$) or the factors are stronger than the noises ($\delta_1=0.2,\delta_2=0.4$),  we see that for each setting of fixed $(p_1,p_2)$, the performance of the white noise test procedure in \cite{gaotsay2021c} and the proposed one in Section 2 is satisfactory in the context of large sample sizes, and generally the proposed one fares slightly better than the one in \cite{gaotsay2021c}.  The ratio-based method is satisfactory only when the factors are stronger than the noises (e.g. $\delta_1=0.2$ and $\delta_2=0.4$). 
The only exception is the case of $p_1=10$ and $p_2=7$ when the factors and the noises are both strong. This is understandable since the ratio method was developed for strong factors.  When the factors are weaker than the noises ($\delta_1=0.6,\delta_2=0.2$),  we see that the proposed method works well for all configurations of the dimensions and the performance improves as the sample size increases.  The method in \cite{yu2022} does not provide satisfactory performance in all scenarios when the noise effect is prominent.
Overall, our proposed method produces satisfactory results for all scenarios.



Next, consider the accuracy in estimating loading matrices. The mean of $\bar{D}(\wh\bA_1,\bL_1)$  (denoted by $MD(\wh\bA_1,\bL_1$) and $\bar{D}(\wh\bP_1,\bR_1)$ (denoted by $MD(\wh\bP_1,\bR_1$) as well as those produced by the initial estimators $\wh\bA_1^o$ and $\wh \bP_1^o$  and the estimators $\wh\bA_1^{YHKZ}$ and $\wh\bP_1^{YHKZ}$ of \cite{yu2022} are shown in Tables~A.4-A.6 of the supplement under the scenarios mentioned before.  The empirical standard errors are given in parentheses.  Note that the standard errors can be larger than the corresponding means because they are the standard errors of the 500 discrepancies in each configuration, not those for the 
average discrepancies. From Tables~A.4-A.6, we see that the estimation accuracy of the loading matrix using the proposed and the \cite{gaotsay2021c} methods is better than that by \cite{yu2022}, and improves as the sample size increases even for moderately large $p_1p_2$, which is in line with the results in \cite{gaotsay2021c} and our asymptotic theory.  We also see that the proposed method can improve the estimation accuracy of the estimated loading matrices in general.




We next measure the estimation accuracy of the estimated common components by
\begin{equation}\label{DX}
d(\wh\bA_1\wh\bX_t\wh\bP_1',\bL_1\bF\bR_1')=\frac{1}{T\sqrt{p_1p_2}}\sum_{t=1}^n\|\wh\bA_1\wh\bX_t\wh\bP_1'-\bL_1\bF_t\bR_1'\|_2.
\end{equation}
We can similarly define ${d}(\wh\bA_1^o\wh\bX_t^{o}\wh\bP_1^o{'},\bL_1\bF\bR_1')$ and ${d}(\wh\bA_1^{YHKZ}\wh\bX_t^{YHKZ}\wh\bP_1^{YHKZ}{'},\bL_1\bF\bR_1')$.  For simplicity, define $\mathcal{\wh S}_{0,t}=\wh\bA_1^0\wh\bX_t^o\wh\bP_1^o{'}$, $\mathcal{\wh S}_{t}=\wh\bA_1\wh\bX_t\wh\bP_1{'}$,   $\mathcal{\wh S}_{YHKZ,t}=\wh\bA_1^{YHKZ}\wh\bX_t^{YHKZ}\wh\bP_1^{YHKZ}{'}$,  and $\mathcal{ S}_t=\bL_1\bF_t\bR_1'$. We only compare the proposed method with the ones of \cite{gaotsay2021c} and \cite{yu2022} because the comparison between the methods of \cite{gaotsay2021c} and \cite{wang2018} are studied in Section 4 of \cite{gaotsay2021c}.  The results are shown in Tables~A.7, from which we see that, for fixed $(p_1,p_2)$, the estimation accuracy by the proposed and \cite{gaotsay2021c} methods improves as the sample size increases, and our method performs slightly better than the one of \cite{gaotsay2021c} but much better than the one of \cite{yu2022} in most cases. When the sample size is small (e.g., 
$T=300$), the estimation errors of YHKZ are smaller than those of our proposed method for 
$(\delta_1,\delta_2)=(0.6,0.2)$ and $(p_1,p_2)=(10,15)$ and $(20,20)$. This occurs because the factor is significantly weaker than the noise in these cases, impacting estimation accuracy at smaller sample sizes. However, as 
$T$ increases, our method begins to outperform YHKZ, aligning with the results of Theorem \ref{tm2} in Section 3.
 Overall, under the prominent noise assumption, the proposed method outperforms the existing ones in the literature.

Finally, under the setting as above, 
we evaluate the asymptotic normality of the estimated loadings in Theorem \ref{thm5} when $(p_1,p_2)=(7,7)$, $(\delta_1,\delta_2)=(0,0)$,  and $T=500$.  Figure A.1 in the supplement plots the histograms of the first and second coordinates of $\sqrt{p_1T}(\wh\ba_{1,1\sbullet}-\bH_{1,T}\ba_{1,1\sbullet})$ via 2000 replications, where the standard errors of the normal curves are estimated by the sample versions using the residuals as that in Section 5 of \cite{Bai_Econometrica_2003}. From Figure A.1, we see that the estimators 
{behave closely to normal},  which is in agreement with our asymptotic theory.



\subsection{Real Data Analysis}
In this section, we {apply the proposed method to} 
two real examples. The first one focuses on the Fama-French return series formed by 
 market capitalization and investment levels,
and the second one involves a climate dataset with monthly measurements of Molecular Hydrogen (H$_2$) across North America.

{\noindent\bf Example 2.}  Consider the Fama-French return series studied in \cite{gaotsay2021c}. The data contain monthly returns of 100 portfolios in 
a $10$ by $10$ matrix formed by ten levels of market capitalization (Size, in rows, from small to large) and ten levels of investment (Inv, in columns, from low to high). Both size and investment are factors for  average stock returns used in \cite{famafrench2015}. The return series spans from July 1963 to December 2019 and consists of 678 monthly observations for each return process.  The data and relevant  
information are available at 
\url{http://mba.tuck.dartmouth.edu/pages/faculty/ken.french/data_library.html}.  Following the same procedure of Example 2 in \cite{gaotsay2021c}, we adjust each return series by subtracting the corresponding risk-free asset returns, available from the same website. Missing values were imputed by a simple exponential smoothing method. Time plots of the adjusted $10\times 10$ series are shown in Figure~4 of \cite{gaotsay2021c} with $p_1=p_2=10$ and $T=678$.

The method of \cite{gaotsay2021c} and that of \cite{wang2018} specify $(\wh r_1^o,\wh r_2^o)=(2,2)$ and $(\wh r_1^{WLC},\wh r_2^{WLC})=(1,1)$, respectively. Using the initial estimators $\wh r_1^o$ and $\wh r_2^o$, we further apply the proposed method and found that $(\wh r_1,\wh r_2)=(1,2)$,  implying that a $1\times 2$ matrix-variate latent factor process is detected.  
 The estimated front and back loading matrices after being multiplied by $30$ are reported in Table~\ref{loading}, which have similar implications as those in \cite{gaotsay2021c}.    
First, for Size, it seems that the 10 rows of the portfolios can be divided into two or three groups, and the dependence on the factors decreases as the Size increases from S1 to S10.  Second, for Investment, all the portfolios have similar dependence on the first column of the factor matrix, and the dependence on the second columns seems to have four groups; the lowest investment portfolio (corresponding to Inv1) seems to depend heavily on the second row of the factors,  the 2nd to the 6th and the 9th investment portfolios depend heavier on the first column of the factors than on the second.  The dependence of the 7th, 8th, and the 10th have similar strength on the two factor columns, which is slightly different from the finding in \cite{gaotsay2021c}.  Note that the signs of the first coefficients of the size loading and the investment loading are the same, implying that each return series shares a co-movement with respect to the $[1,1]$-factor series. This is understandable since we can treat this common factor as representing the market factor in  
the capital asset pricing model (CAPM) of \cite{sharpe1964}. The product of the first coefficients of the size loading and the investment loading can be treated as a market beta, even though 
the starting point of our approach is different from that of CAPM. The usefulness of the 
detected market factor, however, deserves a further investigation. Not surprisingly, similar results are also found in \cite{gaotsay2021c}.

\begin{table}
\caption{Fama-French return series: Size and Investment (Inv) loading matrices after being multiplied  by $30$. The two-dimensional loading vectors are ordered via sizes (S1--S10) and Investment (Inv1--Inv10) from small  to large and from low to high, respectively.}
\label{loading}
\begin{center}
\begin{tabular}{ccccccccccc}
\hline
Size Factor & S1 & S2 & S3 & S4 & S5 & S6 & S7 & S8 & S9 & S10 \\ \hline
Row 1 & \cellcolor[gray]{0.5}{-18} & \cellcolor[gray]{0.6}{-12} &
\cellcolor[gray]{0.6}{-11} & \cellcolor[gray]{0.6}{-9} &
\cellcolor[gray]{0.6}{-8}
& \cellcolor[gray]{0.6}{-7} &\cellcolor[gray]{0.6}{ -7} & \cellcolor[gray]{0.9}{-5}& \cellcolor[gray]{0.9}{-5}&\cellcolor[gray]{0.9}  {-3} \\
\hline
\hline
Inv Factor & Inv1 & Inv2 & Inv3 & Inv4 & Inv5 & Inv6 & Inv7 & Inv8 & Inv9 & Inv10 \\ \hline
Column 1 & {\cellcolor[gray]{0.8} 11}  & {\cellcolor[gray]{0.8} 10}
& {\cellcolor[gray]{0.8} 9} & {\cellcolor[gray]{0.8} 8}
& {\cellcolor[gray]{0.8} 8} & {\cellcolor[gray]{0.8} 8} &  {\cellcolor[gray]{0.8}9 }&  {\cellcolor[gray]{0.8}9} &  {\cellcolor[gray]{0.8}10} & {\cellcolor[gray]{0.8} 12 }\\
Column 2 & {\cellcolor[gray]{0.4} 23} &  {\cellcolor[gray]{0.9} 3} &  {\cellcolor[gray]{0.9}1}  & {\cellcolor[gray]{0.9}1} & { \cellcolor[gray]{0.9}0}
& {\cellcolor[gray]{0.9}-1} & {\cellcolor[gray]{0.5} -14}
& {\cellcolor[gray]{0.8} -7} & { \cellcolor[gray]{0.9}1}
& {\cellcolor[gray]{0.8} -10}\\
\hline
\end{tabular}
\end{center}

\end{table}

To obtain the extracted factors, by the two-way projected PCA of Section 2, we first examine the eigenvalues of the sample covariance matrices $\wh \bS_1$ and $\wh\bS_2$.  From Figure~A.2 of the 
supplement, we see that the largest eigenvalues of $\wh\bS_1$ and $\wh\bS_2$ are much larger than the others. Therefore, we choose $\wh k_1=\wh k_2=1$, and the recovered matrix-variate factors are shown in the upper panel of Figure~A.3 and their corresponding spectrum 
densities in the lower panel. From Figure~A.3, we see that the $[1,1]$-factor process has a flat spectrum which implies that it is a white noise by itself.  However, this does not violate our assumptions as we seek a matrix-variate factor process that captures most dynamic information of the data. It is possible that some component series in the factor matrix is not auto-correlated,  especially for asset return series, but the entire matrix-variate factor process is dynamically dependent. 



Next we examine and compare the forecasting performance of the extracted factors via the proposed method (denoted by proposal) and those by \cite{gaotsay2021c}(denoted by GT), \cite{wang2018} (denoted by WLC), \cite{chenfan2023} (denoted by CF) with $\alpha=-1$ therein, and \cite{yu2022} (denoted by YHKZ). 
We estimate the models using the data in the time span $[1,\tau]$ with $\tau=558,...,678-h$ for the $h$-step ahead forecasts, i.e., we use  returns of the last five years for out-of-sample forecasts, where
 we employ a simple AR(1) model for each detected common factor to 
 produce forecasts.   We also fit a scalar AR(1) (denoted by SAR) model to each individual return series as a benchmark approach.
The following two criteria are used to measure the forecast errors:
\begin{equation}\label{fef}
\text{FE}_F(h)=\frac{1}{120-h+1}\sum_{\tau=558}^{678-h}\frac{1}{\sqrt{p_1p_2}}\|\wh\bY_{\tau+h}-\bY_{\tau+h}\|_F,
\end{equation}
and
\begin{equation}\label{fe2}
\text{FE}_2(h)=\frac{1}{120-h+1}\sum_{\tau=558}^{678-h}\frac{1}{\sqrt{p_1p_2}}\|\wh\bY_{\tau+h}-\bY_{\tau+h}\|_2,
\end{equation}
where $p_1=p_2=10$. Table~\ref{Table4} reports the 1-step to 4-step ahead forecast errors of Equations (\ref{fef}) and (\ref{fe2}) for the five methods GT,  WLC,  CF,  YHKZ, and SAR. The smallest forecast error of each step is shown in boldface. From the table, we see that the proposed method is capable of producing accurate forecasts and most of the associated forecast errors based on the extracted factors by the method are smaller than those based on the factors extracted by {other methods}. 
In particular, the proposed method produces the smallest errors in 1-step ahead predictions, which is  useful because {many practitioners are interested in shorter-term forecasting in financial applications.}  Moreover, the performance of the contemporaneous factors extracted by CF and YHKZ cannot beat  that of SAR in 1-step ahead predictions, implying that the dynamically dependent factors considered in this paper can be more useful in out-of-sample forecasting.
The difference in forecasting errors between the six methods used in 
Table~\ref{Table4} is small. But it is generally not easy to produce accurate forecasts in 
asset returns and the improvements by the proposed method could have important 
implications to practitioners, especially over the five-year horizon.


\begin{table}[ht]\scriptsize
 \caption{The 1-step to 4-step ahead out-of-sample forecast errors of various methods for Example 2.  Proposal denotes the proposed method of the paper, GT denotes the method in \cite{gaotsay2021c},  and WLC, CF,  and YHKZ denote the forecasting errors based on the extracted factors by the method in \cite{wang2018}, \cite{chenfan2023}, and \cite{yu2022}, respectively.  SAR denotes a scalar AR model to each individual series.
 Boldface numbers denote the smallest error for a given forecast horizon.}
          \label{Table4}
\begin{center}
\begin{tabular}{cccccccccccccc}
\hline
&\multicolumn{6}{c}{FE$_{F}(h)$} &&\multicolumn{6}{c}{FE$_{2}(h)$}\\
\cline{2-7}\cline{9-14}
Step-$h$&Proposal&GT&WLC&CF&YHKZ&SAR&&Proposal&GT&WLC&CF&YHKZ&SAR\\
\cline{1-7}\cline{9-14}
1&{\bf 4.27}&4.31&4.35&4.35&4.35&4.34&&{\bf 3.74}&3.78&3.81&3.82&3.82&3.80\\
2&{\bf 4.26}&{\bf 4.26}&4.27&4.27&4.27&4.28&& {3.72}&{\bf 3.71}&3.74&3.74&3.74&3.74\\
3&{\bf 4.232}&4.24&{4.233}&4.233&4.233&4.24&&{\bf 3.68}&3.69&3.69&3.69&3.69&3.69\\
4&{\bf 4.27}&4.28&{4.27}&4.27&4.27&4.28&&{\bf 3.713}&3.74&{3.732}&3.732&3.732&3.74\\
\hline
\end{tabular}
          \end{center}
\end{table}
{\noindent\bf Example 3.}  Climate change is one of the most critical socio-technological issues mankind faces in the new century. In this application, we apply the proposed method to a climate data set 
of North America.  The data consist of  monthly measurements of Molecular Hydrogen (H$_2$) over 13 years from January 1990 to December 2002 on a $2.5\times 2.5$ degree grid that covers most of the United States, and can be downloaded at NOAA (\url{https://gml.noaa.gov/dv/ftpdata.html}). See also \cite{Lozano2009}. The locations of the measurements are shown  in Figure~A.4.  As one of the greenhouse gases,  the increase in H$_2$ concentration is likely to have an impact on the tropospheric hydroxyl radicals, which in turn provides indirect impact on other greenhouse lifetimes and the photochemical production of ozone. Therefore, it is worthy  studying H$_{2}$ as stated in \cite{Hauglustaine2002}. 
We take a lag-12 difference of the $8\times 14$ matrix-variate data to remove any possible seasonality or seasonal trends in the series. The resulting series are shown in Figure~A.5 with $p_1=8$, $p_2=14$ and $T=144$.
  


We  applied the white-noise tests and the ratio-based method to identify the dimension of the matrix-variate factors and found that $(\wh r_1^o, \wh r_2^o)=(2,2)$ and $(\wh r_1^{WLC},\wh r_2^{WLC})=(1,1)$. We then applied the proposed procedure and also found that $(\wh r_1,\wh r_2)=(2,2)$.  The estimated front and back loading matrices after being multiplied by $10$ are reported in Table~\ref{Table5}. For latitude loading matrices, locations can be roughly divided into two groups. All latitudes show similar dependence on the first row factor, but the middle areas rely less on the second row factors compared to the North or South.
For longitude factors, locations can be roughly divided into three groups. All the locations have similar dependence on the first column factor but those of the West areas depend more on the second column factor, and the West and the East locations all have heavier dependence on the second column factor than those in the middle.




\begin{table}
\caption{Latitude (LAT, South to North) and Longitude (LON, West to East)  loading matrices after being multiplied  by $10$ in Example 3.}
\label{Table5}
\small
\begin{center}
\begin{tabularx}{0.62\linewidth}{ccccccccc}
\hline
LAT Factor & 1 & 2 & 3 & 4 & 5 & 6 & 7&8 \\ \hline
Row 1 & \cellcolor[gray]{0.5}{-3.8} & \cellcolor[gray]{0.5}{-3.8} &
\cellcolor[gray]{0.5}{-3.7} & \cellcolor[gray]{0.5}{-3.6} &
\cellcolor[gray]{0.5}{-3.5}
&\cellcolor[gray]{0.5} {-3.4} &\cellcolor[gray]{0.5}{ -3.3}  &\cellcolor[gray]{0.5}{ -3.2} \\
Row 2 & \cellcolor[gray]{0.5}{-4.0} &  \cellcolor[gray]{0.5}{-3.7} &
\cellcolor[gray]{0.5}{-2.9} & \cellcolor[gray]{0.8}{-1.3} &
\cellcolor[gray]{0.8}{0.7}
& \cellcolor[gray]{0.5}{2.7} &\cellcolor[gray]{0.5}{ 4.5}&\cellcolor[gray]{0.5}{ 5.7}   \\
\hline
\end{tabularx}
\end{center}
\begin{center}
\small
\begin{tabularx}{0.95\linewidth}{ccccccccccccccc}
\hline
LON Factor &1  & 2 & 3 & 4 & 5 & 6 & 7 &8&9&10&11&12&13&14\\ \hline
Column 1 & \cellcolor[gray]{0.8}{2.9}  & \cellcolor[gray]{0.8}{ 2.9}
& {\cellcolor[gray]{0.8} 2.8} & {\cellcolor[gray]{0.8} 2.7}
& {\cellcolor[gray]{0.8} 2.7} & \cellcolor[gray]{0.8}{2.6} & \cellcolor[gray]{0.8} {2.6}& \cellcolor[gray]{0.8} {2.6}& \cellcolor[gray]{0.8} {2.6}&\cellcolor[gray]{0.8}  {2.6}& \cellcolor[gray]{0.8} {2.6}& \cellcolor[gray]{0.8}{2.6}& \cellcolor[gray]{0.8} {2.6}& \cellcolor[gray]{0.8} {2.6}\\
Column 2 &\cellcolor[gray]{0.5} {-6.0} & \cellcolor[gray]{0.5} { -4.6} &   \cellcolor[gray]{0.8}{-3.1}  &  \cellcolor[gray]{0.8}{-1.6} & { -0.3}
&{0.7} & \cellcolor[gray]{0.8}{1.5}& \cellcolor[gray]{0.8}{1.9}&\cellcolor[gray]{0.8} {2.2}&\cellcolor[gray]{0.8} {2.3}&\cellcolor[gray]{0.8} {2.3}& \cellcolor[gray]{0.8}{2.2}& \cellcolor[gray]{0.8}{1.9}&\cellcolor[gray]{0.8} {1.6}\\
\hline
\end{tabularx}
\end{center}
\end{table}

Next, we employed the two-way projected PCA to extract the factors, and the eigenvalues of the sample covariance matrices $\wh\bS_1$ and $\wh\bS_2$ are reported in Figure~A.6,  from which we see that we may choose $\wh k_1=5$ and $\wh k_2=1$.  The recovered matrix-variate factors are shown in Figure~A.7, which can be used for out-of-sample forecasting or to study the dynamic 
dependence of the original processes.  In this particular instance, the 
proposed method reduces the dimensions from $112$ to $4$,  marking a substantial reduction.



We also examined and compared the forecasting performance of the extracted factors using the same methods as those 
of Example 2. 
We use data of the last $44$ months for out-of-sample testing, and employ the forecast errors in (\ref{fef})--(\ref{fe2}), where $p_1=8$ and $p_2=14$,  the number of rolling-windows is $44$, and the summation is summing from $\tau=100$ to $144-h$.  From Table~\ref{Table6}, we see that the proposed method is capable of producing more accurate forecasts. It produces 
the smallest forecast errors in 1-step, 4-step and 6-step ahead forecasts, and its 2-step and 3-step ahead forecast errors are only slightly greater than those of GT. 
Furthermore, these two methods fare better than the ones in WLC, CF, and YHKZ,  which don't effectively remove noise when recovering common factors. In addition, all methods perform better than the benchmark SAR method. Therefore, the proposed method provides another useful approach for practitioners who are interested in out-of-sample forecasting 
of matrix-variate time series.

\begin{table}\scriptsize
 \caption{The  out-of-sample forecast errors of various methods for Example 3. Proposal denotes the proposed method in the paper, GT-ES denotes the method in \cite{gaotsay2021c},  and WLC,  CF,  and YHKZ denote the forecasting errors based on the extracted factors by the method in \cite{wang2018}, \cite{chenfan2023}, and \cite{yu2022}, respectively. SAR denotes a scalar AR model to each individual series.
 Boldface numbers denote the smallest error for a given forecast horizon.}
          \label{Table6}
\begin{center}
 \setlength{\abovecaptionskip}{0pt}
\setlength{\belowcaptionskip}{3pt}

\begin{tabular}{cccccccccccccc}
\hline
&\multicolumn{6}{c}{FE$_{F}(h)$} &&\multicolumn{6}{c}{FE$_{2}(h)$}\\
\cline{2-7}\cline{9-14}
Step-$h$&Proposal&GT&WLC&CF&YHKZ&SAR&&Proposal&GT&WLC&CF&YHKZ&SAR\\
\cline{1-7}\cline{9-14}
1&{\bf 8.55}&8.59&8.63&8.70&8.70&8.79&&{\bf 8.31}&8.34&8.43&8.49&8.49&8.55\\
2&{8.43}&{\bf 8.38}&8.61&8.72&8.72&8.86&& {8.22}&{\bf 8.19}&8.40&8.52&8.52&8.67\\
3&{8.92}&{\bf 8.91}&9.07&9.15&9.15&9.18&&8.70&{\bf 8.69}&8.88&8.96&8.96&8.98\\
4&{\bf 9.08}&{ 9.08}&9.18&9.24&9.24&9.20&&{\bf 8.87}&{8.87}&8.98&9.03&9.03&9.00\\
6&{\bf 8.61}&{8.62}&8.64&8.69&8.69&8.68&&{\bf 8.39}&{8.40}&8.43&8.47&8.47&8.47\\
\hline
\end{tabular}
          \end{center}
\end{table}

\section{Concluding Remarks}\label{sec5}
This article introduced a new factor procedure for modeling  high-dimensional matrix-variate time series. The new approach involves multiple projections of  the observed data onto certain  row or column factor spaces, and the estimation of the loading matrices 
{was carried out} in an iterative way. The advantage of such a procedure is that the convergence rates of the estimated loading matrices are faster than those 
of the traditional methods such as \cite{wang2018} or \cite{gaotsay2021c}. The proposed iterative method is easy to implement and the simulation results show that it can estimate the loading matrices and extract the factors more accurately compared with the existing ones. The empirical results suggest that the proposed procedure can effectively extract the number of common factors from complex data, and the extracted factors could be useful in out-of-sample forecasting.

\section*{Supplementary Material}
The supplementary material includes a detailed modelling algorithm for the proposed procedure in Section 2, some tables and figures from the numerical studies in Section 4, and all technical proofs of the theorems presented in the paper.
 



\newpage
\renewcommand*{\thefootnote}{\fnsymbol{footnote}}
\begin{center}
{\bf \Large Supplement to ``Denoising and Multilinear Projected-Estimation of  High-Dimensional Matrix-Variate Factor Time Series"}\\
{$^1$Zhaoxing Gao and $^2$Ruey S. Tsay\footnote{Corresponding author: \href{mailto:ruey.tsay@chicagobooth.edu}{ruey.tsay@chicagobooth.edu} (R.S. Tsay). Booth School of Business, University of Chicago, 5807 S. Woodlawn Avenue, Chicago, IL 60637, USA.}\\
$^1$School of Mathematical Sciences, University of Electronic Science \& Technology of China\\
$^2$Booth School of Business, University of Chicago}
\end{center}

\vskip 0.5 cm

\setcounter{page}{1}
\setcounter{figure}{0}
\setcounter{section}{0}
\setcounter{theorem}{0}
\makeatletter 
\renewcommand{\thefigure}{A.\arabic{figure}}
\renewcommand{\thetheorem}{A.\arabic{theorem}}
\setcounter{table}{0}
\makeatletter 
\renewcommand{\thetable}{A.\arabic{table}}
\renewcommand{\thesection}{A.\arabic{section}}
	\renewcommand{\thesubsection}{\thesection.\arabic{subsection}}

\newcommand{\bPi} {\boldsymbol{\Pi}}
\newtheorem{condition}{Condition}[section]

This supplementary material consists of a detailed modeling algorithm for the proposed procedure in Section 2, 
some tables and figures of the numerical 
studies in Section 4, and all technical proofs of the theoretical statements in the main article.


\section{Modeling Algorithm}\label{sec224}
We summarize the pseudo-code of the  iterative modeling procedure of Section 2 in Algorithm \ref{FM} below.   For a given $\wh \bP_{i,1}$, we form the matrix $\wh \bM_{i,1}^*$ and construct $\wh \bA_{i,1}$ by computing the eigenvectors of the matrix $\wh \bM_{i,1}^*$. With the computed $\wh\bA_{i,1}$, we can update the back loading as $\wh\bP_{i+1,1}$ by computing the eigenvectors of the matrix $\wh\bM_{i,2}^*$. The procedure can be repeated until convergence.
In theory,  the initial orthonormal matrix $\bP_{0,1}$ can be arbitrary so long as $\bP_{0,1}'\bP_1\neq {\bf 0}$.  In practice, we may generate a sequence of orthonormal matrices and choose the one $\bP_{0,1}$ such that $\bP_{0,1}'\wh\bP_1^o$ has largest singular values on an average basis.  

The convergence criterion can be established via a discrepancy measure between two matrices.  
We adopt the one used by
\cite{panyao2008}: for two $p\times r$ semi-orthogonal 
matrices ${\bf H}_1$ and ${\bf H}_2$ satisfying the condition ${\bf
H}_1'{\bf H}_1={\bf H}_2'{\bf H}_2=\bI_{r}$, the difference
between the two linear spaces $\mathcal{M}({\bf H}_1)$ and
$\mathcal{M}({\bf H}_2)$ is measured by
\begin{equation}
D({\bf H}_1,{\bf
H}_2)=\sqrt{1-\frac{1}{r}\textrm{tr}({\bf H}_1{\bf H}_1'{\bf
H}_2{\bf H}_2')}.\label{eq:D1}
\end{equation}
Note that $D({\bf H}_1,{\bf
H}_2) \in [0,1].$ 
It is equal to $0$ if and only if
$\mathcal{M}({\bf H}_1)=\mathcal{M}({\bf H}_2)$, and to $1$ if and
only if $\mathcal{M}({\bf H}_1)\perp \mathcal{M}({\bf H}_2)$. By Lemma A1(i) in \cite{panyao2008}, $D(\cdot,\cdot)$ is a well-defined distance measure on some quotient space of matrices.  Therefore, we can choose a small threshold $\eta>0$, and the convergence of the algorithm is determined by checking whether the following inequalities hold:
\begin{equation}\label{cri}
D(\wh\bA_{i,1},\wh\bA_{i-1,1})<\eta\,\,\text{and}\,\,D(\wh\bP_{i+1,1},\wh\bP_{i,1})<\eta,
\end{equation}
and $\wh\bA_{i,1}$ and $\wh\bP_{i+1,1}$ are the chosen estimators since we stop at the $i$-th step in Algorithm \ref{FM}.
Obviously, the convergence is guaranteed together with the criterion in (\ref{cri}) if model (2.1). 
holds.  On the other hand,  as discussed in Section 3 and shown in the proofs,  we may stop at $i=2$ since one iteration is often good enough and the convergence rates are faster than those in \cite{wang2018} and \cite{gaotsay2021c}.

We mention that the consistency of the estimators obtained by the algorithm is established under model (2.1). 
However, theoretical analysis suggests that the convergence rates may not be improved if we increase the number of iterations. Therefore, we may prescribe a maximum number of iterations $s_0$ such that the iteration stops if (\ref{cri}) is satisfied or $i=s_0$.  The simulation results in Section 4 suggest that $s_0=2$ is sufficient to produce accurate estimators.

\begin{algorithm}[ht]\small
\caption{Iterative Estimation of the Matrix-Variate Factor Models}\label{FM}
{\bf Input:} Data $\{\bY_1,...,\bY_T\}$\\
{\bf Output:} $\wh r_1$, $\wh r_2$, $\wh\bA_1$, $\wh\bP_1$, and \{$\wh\bX_1, \ldots,\wh\bX_T$\}
\begin{algorithmic}[1]
\State{Construct $\wh\bM_1$ and $\wh\bM_2$, and obtain the matrices $\wh\bGamma_1^0$ and $\wh\bGamma_2^0$}
\State{Obtain the initial factor order estimates $r_1^0=\wh r_1^o$ and $r_2^0=\wh r_2^o$ based on $\wh\bGamma_1^o{'}\bY_t\wh\bGamma_2^o$}
\State{Generate an initial orthonormal matrix $\bP_{0,1}\in R^{p_2\times r_2^0}$, set $\wh\bP_{0,1}=\bP_{0,1}$ and $i\gets 0$}
 \While{Not Convergent}
      \State  {form the matrix $\wh\bM_{i,1}^*$ based on the data $\bY_t\wh\bP_{i,1}$}
      \State     compute $\wh\bA_{i,1}$ consisting of the leading $r_1^0$ eigenvectors of $\wh\bM_{i,1}^*$
       \State{form the matrix $\wh\bM_{i,2}^*$ based on $\bY_t'\wh\bA_{i,1}$}
       \State{compute $\wh\bP_{i+1,1}$ consisting of the leading $r_2^0$ eigenvectors of $\wh\bM_{i,2}^*$}
       \State{$i\gets i+1$}
\EndWhile
\State{End{\bf while}}
\State{$i\gets i-1$}
\State{Compute the eigenvector matrices $\wh\bGamma_{i,1}$ and $\wh\bGamma_{i,2}$ based on $\wh\bM_{i,1}^*$ and $\wh\bM_{i,2}^*$, respectively}
\State{Obtain the factor order estimates $\wh r_1$ and $\wh r_2$ based on $\wh\bGamma_{i,1}'\bY_t\wh\bGamma_{i,2}$ using the HDWN test}
\State{Obtain $\wh\bA_1$ consisting of the top $\wh r_1$ eigenvectors of $\wh\bGamma_{i,1}$}
\State{Obtain $\wh\bB_1$ consisting of the top $\wh r_2$ eigenvectors of $\wh\bGamma_{i,2}$}

\State{Form the matrices $\wh\bS_1$ and $\wh\bS_2$}
\State{Obtain the $\wh k_1$ and $\wh k_2$}
\State{Obtain $\wh\bB_2$ consisting of the last $p_1-\wh k_1$ eigenvectors of $\wh\bS_1$}
\State{Obtain $\wh\bQ_2$ consisting of the last $p_2-\wh k_2$ eigenvectors of $\wh\bS_2$}
\State{Compute $\bC_1$ consisting of the top $\wh r_1$ eigenvectors of $\wh\bB_2'\bA_1\bA_1'\bB_2$}
\State{Compute $\bC_2$ consisting of the top $\wh r_2$ eigenvectors of $\wh\bQ_2'\bP_1\bP_1'\bQ_2$}
\State{$\wh\bB_2^*\gets\wh\bB_2\bC_1$}
\State{$\wh\bQ_2^*\gets\wh\bQ_2\bC_2$}
\State{Compute $\wh\bX_t\gets (\wh\bB_2^*{'}\wh\bA_1)^{-1}\wh\bB_2^*{'}\bY_t\wh\bQ_2^*(\wh\bP_1'\wh\bQ_2^*)^{-1} $}
\State{END}
\end{algorithmic}
\end{algorithm}

\section{Some Tables and Figures}
\begin{figure}[ht]
\begin{center}
{\includegraphics[width=0.8\textwidth]{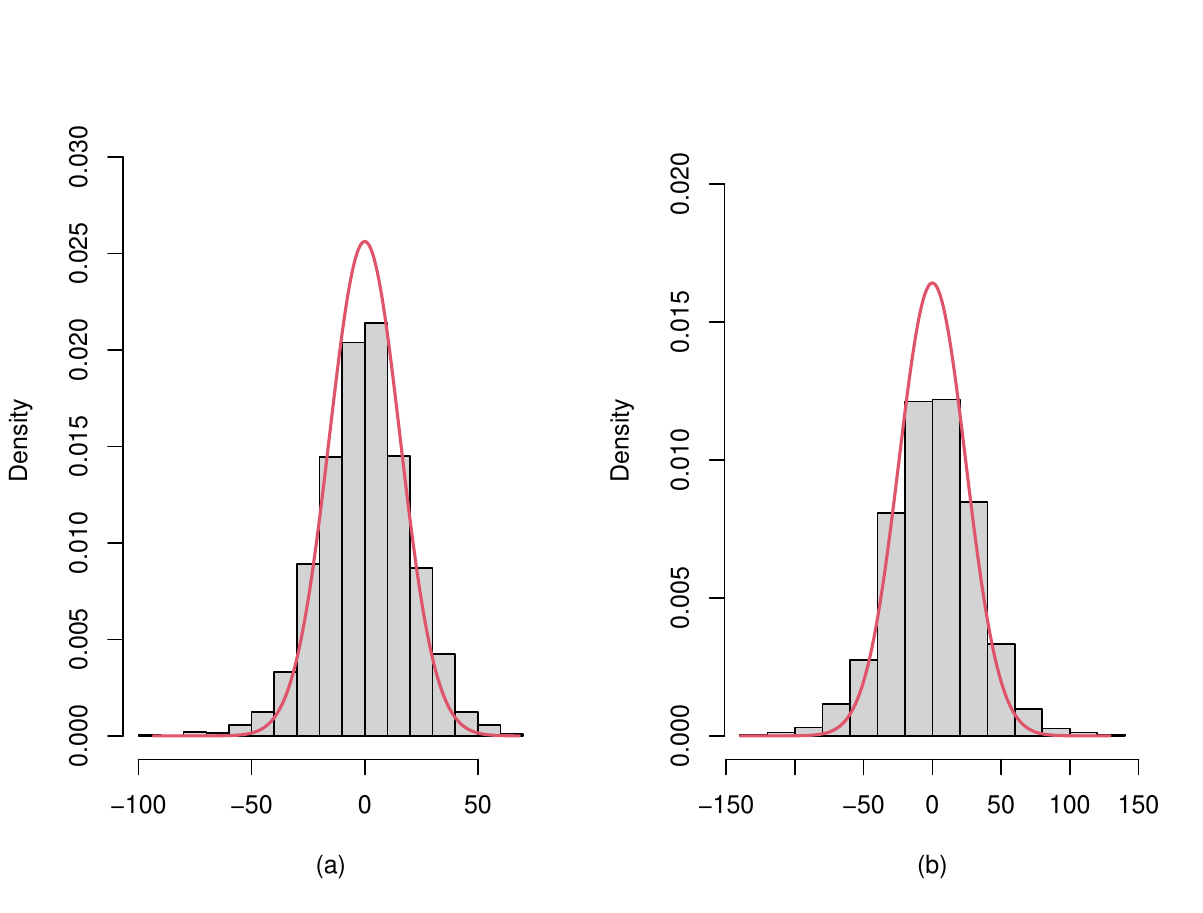}}
\caption{(a) Histograms of the first coordinate of $\sqrt{p_1T}(\wh\ba_{1,1\sbullet}-\bH_{1,T}\ba_{1,1\sbullet})$; (b) Histograms of the second coordinate of $\sqrt{p_1T}(\wh\ba_{1,1\sbullet}-\bH_{1,T}\ba_{1,1\sbullet})$, where we set $(p_1,p_2)=(7,7)$, $(\delta_1,\delta_2)=(0,0)$, $T=500$, and $k_0=2$.  The settings of the parameters are the same as those in Example 1. The curves represent the densities of normal distributions where the standard errors are estimated by the sample versions using the residuals as that in Section 5 of \cite{Bai_Econometrica_2003}.  2000 replications are used in the experiments. }\label{fig:inf}
\end{center}
\end{figure}

\begin{figure}
\begin{center}
{\includegraphics[width=0.7\textwidth]{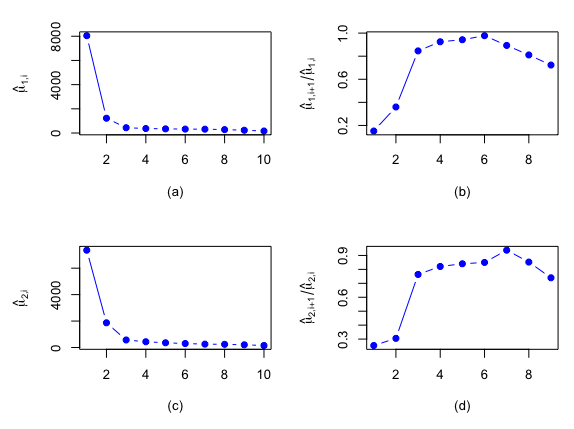}}
\caption{(a) The 10 eigenvalues of $\wh\bS_1$; (b) Plot of the ratios of consecutive eigenvalues of $\wh\bS_1$; (c) The 10 eigenvalues of $\wh\bS_2$; (d) Plot of the ratios of consecutive 
eigenvalues of $\wh\bS_2$ in Example 2.}\label{fig4}
\end{center}
\end{figure}

\begin{figure}
\begin{center}
{\includegraphics[width=0.7\textwidth]{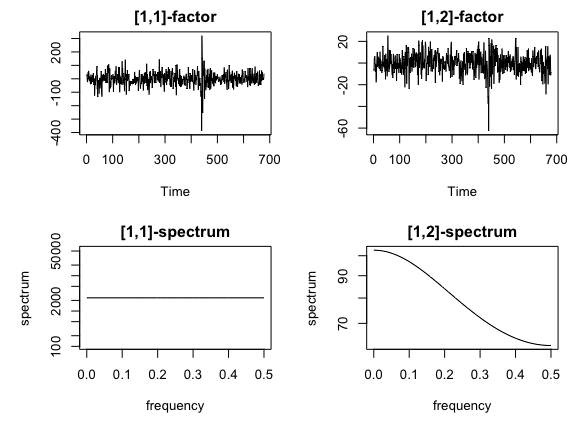}}
\caption{(a) The time series plots of the extracted $1\times 2$ common factors; (b) the corresponding spectrum densities of the factor processes}.\label{fig5}
\end{center}
\end{figure}

\begin{figure}
\begin{center}
{\includegraphics[width=0.8\textwidth]{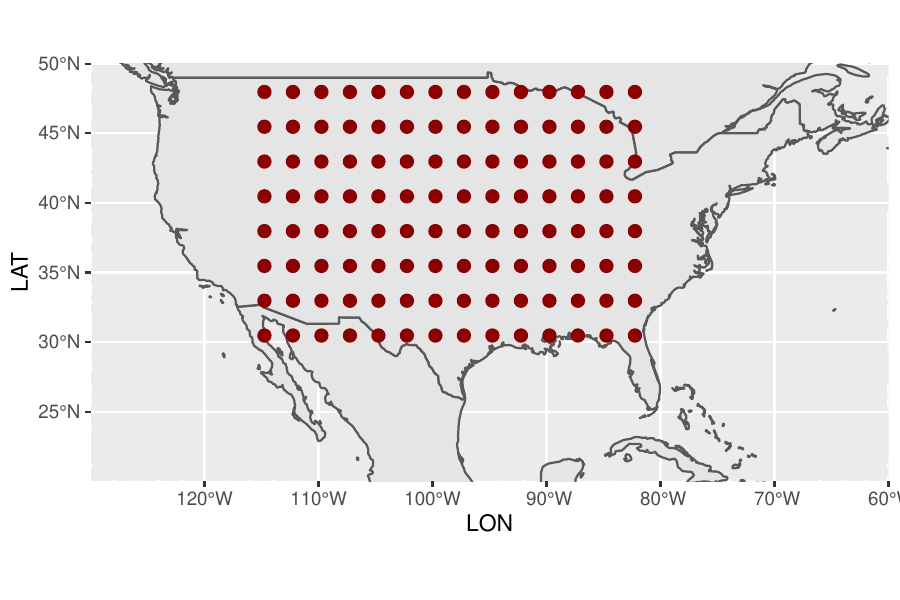}}
\caption{Locations (latitude vs. longitude) of the $8\times14$ grid that covers most of the United States of Example
3.}\label{fig77}
\end{center}
\end{figure}

\begin{figure}
\begin{center}
{\includegraphics[width=0.8\textwidth]{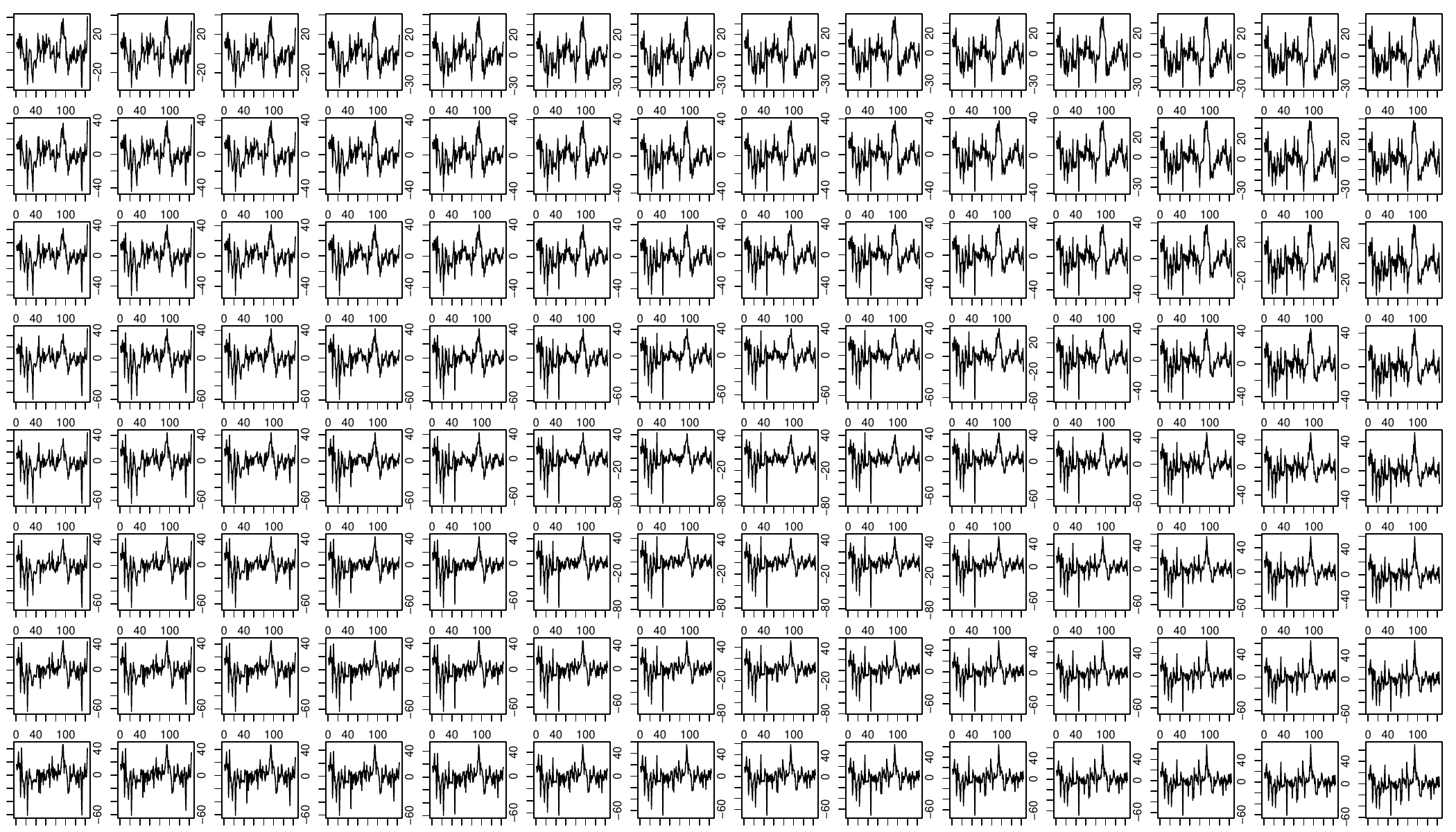}}
\caption{Time plots of the $8\times14$ measurements of the Molecular Hydrogen of Example
3.}\label{fig78}
\end{center}
\end{figure}

\begin{figure}
\begin{center}
{\includegraphics[width=0.7\textwidth]{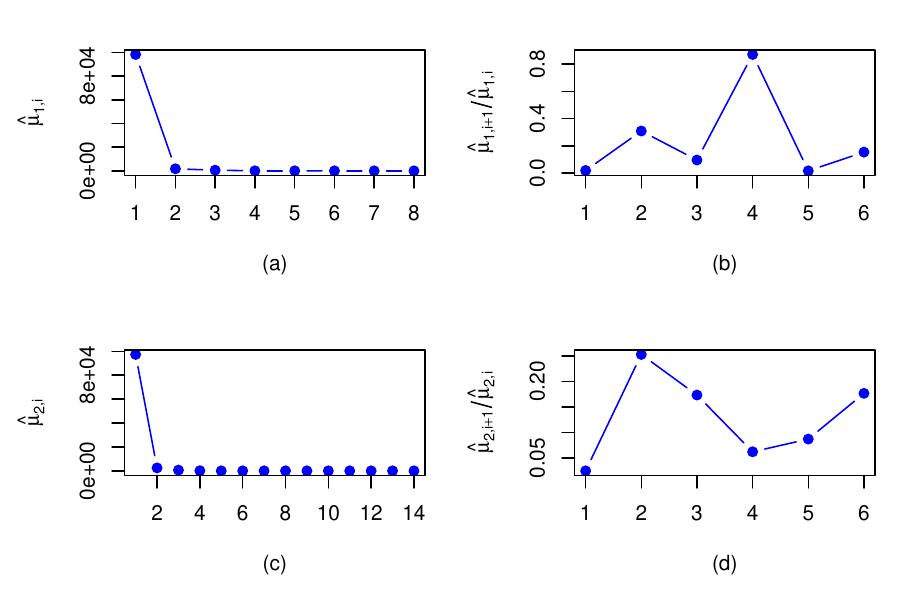}}
\caption{(a) The 8 eigenvalues of $\wh\bS_1$; (b) Plot of the ratios of consecutive eigenvalues of $\wh\bS_1$; (c) The 14 eigenvalues of $\wh\bS_2$; (d) Plot of the ratios of consecutive 
eigenvalues of $\wh\bS_2$.}\label{fig6}
\end{center}
\end{figure}

\begin{figure}
\begin{center}
\subfigure[]{\includegraphics[width=0.44\textwidth]{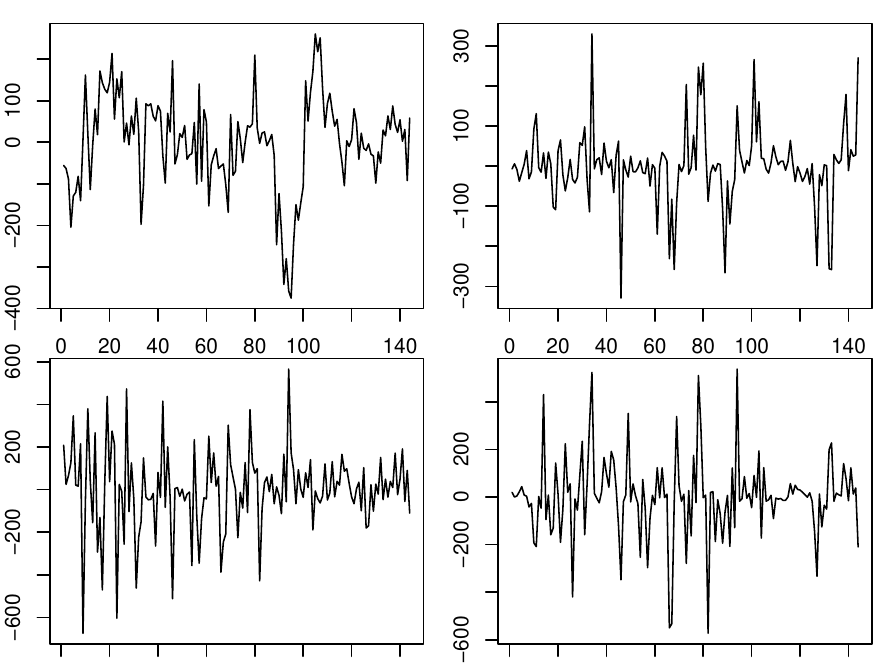}}
\subfigure[]{\includegraphics[width=0.44\textwidth]{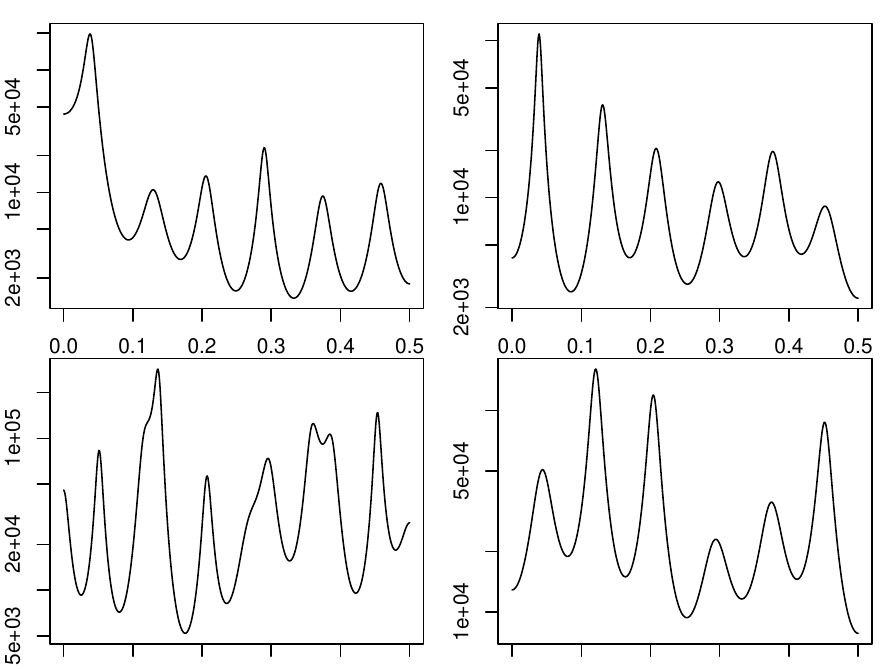}}
\caption{(a) The time series plots of the extracted $2\times 2$ common factors; (b) The spectrum of the factors in (a) of Example 3.}\label{fig7}
\end{center}
\end{figure}

In this section, we present some tables and figures used in the main article.  For $k_0=2$, Tables~\ref{Table-a1}-\ref{Table-a3} report the empirical probabilities of the testing results on determining the factor orders for Example 1 with $(r_1,r_2)=(2,3)$ and $(k_1,k_2)=(1,2)$ under  different strengths of factors and noises that $(\delta_1,\delta_2)=(0,0)$,  $(0.2,0.4)$, and $(0.6,0.2)$, respectively.  Tables~\ref{Table-a4}-\ref{Table-a6} present the estimation accuracy of loading matrices  using our proposed method and the traditional ones for $(\delta_1,\delta_2)=(0,0)$, $(0.2,0.4)$, and $(0.6,0.2)$ in Example 1, respectively.  Table~\ref{Table-a7}    reports the  distance between the extracted common components and the true ones when $(r_1,r_2)=(2,3)$ and  $(k_1,k_2)=(1,2)$ in Example 1, using our proposed method and the existing ones in \cite{gaotsay2021c} and \cite{yu2022}.  For $k_0=1,3,$ and $4$, Tables~\ref{Table-a8}-\ref{Table-a10} reports the empirical probabilities of the testing results on determining the factor orders,   the estimation accuracy of loading matrices  using our proposed method and the traditional ones, and  the  distance between the extracted common components and the true ones, respectively,  where $(\delta_1,\delta_2)=(0,0)$ and $(p_1,p_2)=(20,20)$, and the other settings are the same as those in Example 1. In Tables~\ref{Table-a11}-\ref{Table-a13}, we follow the modeling algorithm in Section 2.2.4 and randomly generate 10 orthonormal matrices, and choose the one as the initial projection matrix $\bP_{0,1}$ such that $\bP_{0,1}'\wh\bP_{1}^{o}$ has largest singular values on an average basis.  Figure~\ref{fig:inf} plots histograms of the first and the second coordinates of $\sqrt{p_1T}(\wh\ba_{1,1\sbullet}-\bH_{1,T}\ba_{1,1\sbullet})$, where the estimated normal curves are plotted according to the limiting distributions in Theorem~5.  Figure~\ref{fig4} shows the plots of eigenvalues of the covariances $\wh\bS_1$ and $\wh\bS_2$, and their associated ratios of Example 2.  Figure~\ref{fig5} plots the extracted factors and their associated spectrum densities in Example 2.  Figure~\ref{fig77} displays the locations of the measurements of H$_2$ in Example 3. 
Figure~\ref{fig78} shows the time series plots of the series in Example 3. Figure~\ref{fig6} shows the plots of eigenvalues of the covariances $\wh\bS_1$ and $\wh\bS_2$, and their associated ratios of Example 3.  Figure~{\ref{fig7}} plots the extracted factors and their associated spectrum densities in Example 3.


\begin{table}[ht]
 \caption{Empirical probabilities (EP) of determining the factor orders by different methods for Example 1 with $(r_1,r_2)=(2,3)$ and $(k_1,k_2)=(1,2)$, where $(p_1,p_2)$ and $T$ are the dimension and the sample size, respectively. $\delta_1$ and $\delta_2$ are the strength parameters of the factors and the errors, respectively. $500$ iterations are used. In the table, $\hat{r}, \hat{r}^0, 
 \hat{r}^{WLC}$, and $\wh r^{YHKZ}$ denote the proposed, 
\cite{gaotsay2021c},  \cite{wang2018}, and \cite{yu2022} method, respectively.}
          \label{Table-a1}
\begin{center}
 \setlength{\abovecaptionskip}{0pt}
\setlength{\belowcaptionskip}{3pt}

\begin{tabular}{c|ccc|ccccc}
\hline
 & &&&\multicolumn{5}{c}{$T$}\\
 \cline{5-9}
$(\delta_1, \delta_2)$ &$(p_1,p_2)$&$p_1p_2$&EP&$300$&$500$&$1000$&$1500$&$3000$\\
\hline
 (0,0) &$(7,7)$&49&$P(\wh r_1=r_1,\wh r_2=r_2)$&0.684&0.944&0.980&0.976&0.966\\
 &&&$P(\wh r_1^0=r_1,\wh r_2^0=r_2)$&0.612&0.914&0.976&0.978&0.954\\
  &&&$P(\wh r_1^{WLC}=r_1,\wh r_2^{WLC}=r_2)$&0.004&0.018&0.120&0.298&0.648\\
  &&&$P(\wh r_1^{YHKZ}=r_1,\wh r_2^{YHKZ}=r_2)$&0&0&0&0&0\\
  \cline{2-9}
 &$(10,15)$&150&$P(\wh r_1=r_1,\wh r_2=r_2)$&0.332&0.406&0.422&0.458&0.546\\
  &&&$P(\wh r_1^0=r_1,\wh r_2^0=r_2)$&0.268&0.330&0.394&0.410&0.536\\
  &&&$P(\wh r_1^{WLC}=r_1,\wh r_2^{WLC}=r_2)$&0.034&0.140&0.490&0.756&0.966\\
    &&&$P(\wh r_1^{YHKZ}=r_1,\wh r_2^{YHKZ}=r_2)$&0&0&0&0&0\\
    \cline{2-9}

&$(20,20)$&400&$P(\wh r_1=r_1,\wh r_2=r_2)$&0.456&0.700&0.856&0.876&0.940\\
 &&&$P(\wh r_1^0=r_1,\wh r_2^0=r_2)$&0.408&0.654&0.828&0.856&0.938\\
  &&&$P(\wh r_1^{WLC}=r_1,\wh r_2^{WLC}=r_2)$&0.002&0.008&0.046&0.116&0.472\\
    &&&$P(\wh r_1^{YHKZ}=r_1,\wh r_2^{YHKZ}=r_2)$&0&0&0&0&0\\
    \cline{2-9}
&$(20,30)$&600&$P(\wh r_1=r_1,\wh r_2=r_2)$&0.368&0.572&0.758&0.782&0.882\\
 &&&$P(\wh r_1^0=r_1,\wh r_2^0=r_2)$&0.214&0.406&0.668&0.708&0.858\\
  &&&$P(\wh r_1^{WLC}=r_1,\wh r_2^{WLC}=r_2)$&0&0&0.006&0.030&0.192\\
    &&&$P(\wh r_1^{YHKZ}=r_1,\wh r_2^{YHKZ}=r_2)$&0&0&0&0&0\\
\hline
\end{tabular}
          \end{center}
\end{table}

\begin{table}[ht]
 \caption{Empirical probabilities (EP) of determining the factor orders by different methods for Example 1 with $(r_1,r_2)=(2,3)$ and $(k_1,k_2)=(1,2)$, where $(p_1,p_2)$ and $T$ are the dimension and the sample size, respectively. $\delta_1$ and $\delta_2$ are the strength parameters of the factors and the errors, respectively. $500$ iterations are used. In the table, $\hat{r}, \hat{r}^0, 
 \hat{r}^{WLC}$, and $\wh r^{YHKZ}$ denote the proposed, 
 \cite{gaotsay2021c},  \cite{wang2018}, and \cite{yu2022} method,  respectively.}
          \label{Table-a2}
\begin{center}
 \setlength{\abovecaptionskip}{0pt}
\setlength{\belowcaptionskip}{3pt}

\begin{tabular}{c|ccc|ccccc}
\hline
 & &&&\multicolumn{5}{c}{$T$}\\
 \cline{5-9}
$(\delta_1, \delta_2)$ &$(p_1,p_2)$&$p_1p_2$&EP&$300$&$500$&$1000$&$1500$&$3000$\\
\hline
 (0.2,0.4) &$(7,7)$&49&$P(\wh r_1=r_1,\wh r_2=r_2)$&0.772&0.960&0.982&0.976&0.964\\
 &&&$P(\wh r_1^0=r_1,\wh r_2^0=r_2)$&0.760&0.950&0.982&0.978&0.964\\
  &&&$P(\wh r_1^{WLC}=r_1,\wh r_2^{WLC}=r_2)$&0.702&0.890&0.988&0.998&1\\
  &&&$P(\wh r_1^{YHKZ}=r_1,\wh r_2^{YHKZ}=r_2)$&0&0&0&0&0\\
  \cline{2-9}
 &$(10,15)$&150&$P(\wh r_1=r_1,\wh r_2=r_2)$&0.910&0.974&0.966&0.972&0.954\\
  &&&$P(\wh r_1^0=r_1,\wh r_2^0=r_2)$&0.910&0.972&0.966&0.968&0.952\\
  &&&$P(\wh r_1^{WLC}=r_1,\wh r_2^{WLC}=r_2)$&0.974&0.994&1&1&1\\
    &&&$P(\wh r_1^{YHKZ}=r_1,\wh r_2^{YHKZ}=r_2)$&0.054&0.004&0&0&0\\
    \cline{2-9}

&$(20,20)$&400&$P(\wh r_1=r_1,\wh r_2=r_2)$&0.922&0.978&0.978&0.976&0.976\\
 &&&$P(\wh r_1^0=r_1,\wh r_2^0=r_2)$&0.922&0.982&0.976&0.980&0.976\\
  &&&$P(\wh r_1^{WLC}=r_1,\wh r_2^{WLC}=r_2)$&0.986&0.998&1&1&1\\
    &&&$P(\wh r_1^{YHKZ}=r_1,\wh r_2^{YHKZ}=r_2)$&0&0&0&0&0\\
    \cline{2-9}
&$(20,30)$&600&$P(\wh r_1=r_1,\wh r_2=r_2)$&0.904&0.980&0.984&0.972&0.976\\
 &&&$P(\wh r_1^0=r_1,\wh r_2^0=r_2)$&0.902&0.974&0.984&0.970&0.978\\
  &&&$P(\wh r_1^{WLC}=r_1,\wh r_2^{WLC}=r_2)$&0.954&0.998&1&1&1\\
    &&&$P(\wh r_1^{YHKZ}=r_1,\wh r_2^{YHKZ}=r_2)$&0&0&0&0&0\\
\hline
\end{tabular}
          \end{center}
\end{table}


\begin{table}[ht]
 \caption{Empirical probabilities (EP) of determining the factor orders by different methods for Example 1 with $(r_1,r_2)=(2,3)$ and $(k_1,k_2)=(1,2)$, where $(p_1,p_2)$ and $T$ are the dimension and the sample size, respectively. $\delta_1$ and $\delta_2$ are the strength parameters of the factors and the errors, respectively. $500$ iterations are used. In the table,  $\hat{r}, \hat{r}^0, 
 \hat{r}^{WLC}$, and $\wh r^{YHKZ}$ denote the proposed, 
 \cite{gaotsay2021c},  \cite{wang2018}, and \cite{yu2022} method, respectively.}
          \label{Table-a3}
\begin{center}
 \setlength{\abovecaptionskip}{0pt}
\setlength{\belowcaptionskip}{3pt}

\begin{tabular}{c|ccc|ccccc}
\hline
 & &&&\multicolumn{5}{c}{$T$}\\
 \cline{5-9}
$(\delta_1, \delta_2)$ &$(p_1,p_2)$&$p_1p_2$&EP&$300$&$500$&$1000$&$1500$&$3000$\\
\hline
 (0.6,0.2) &$(7,7)$&49&$P(\wh r_1=r_1,\wh r_2=r_2)$&0.508&0.638&0.940&0.968&0.960\\
 &&&$P(\wh r_1^0=r_1,\wh r_2^0=r_2)$&0.168&0.256&0.654&0.838&0.956\\
  &&&$P(\wh r_1^{WLC}=r_1,\wh r_2^{WLC}=r_2)$&0&0&0&0&0\\
  &&&$P(\wh r_1^{YHKZ}=r_1,\wh r_2^{YHKZ}=r_2)$&0&0&0&0&0\\
  \cline{2-9}
 &$(10,15)$&150&$P(\wh r_1=r_1,\wh r_2=r_2)$&0.306&0.426&0.642&0.778&0.860\\
  &&&$P(\wh r_1^0=r_1,\wh r_2^0=r_2)$&0.072&0.036&0.066&0.142&0.390\\
  &&&$P(\wh r_1^{WLC}=r_1,\wh r_2^{WLC}=r_2)$&0&0&0.002&0&0.048\\
    &&&$P(\wh r_1^{YHKZ}=r_1,\wh r_2^{YHKZ}=r_2)$&0.054&0.004&0&0&0\\
    \cline{2-9}

&$(20,20)$&400&$P(\wh r_1=r_1,\wh r_2=r_2)$&0.016&0.050&0.332&0.628&0.968\\
 &&&$P(\wh r_1^0=r_1,\wh r_2^0=r_2)$&0.012&0&0.064&0.212&0.750\\
  &&&$P(\wh r_1^{WLC}=r_1,\wh r_2^{WLC}=r_2)$&0&0&0&0&0\\
    &&&$P(\wh r_1^{YHKZ}=r_1,\wh r_2^{YHKZ}=r_2)$&0&0&0&0&0\\
    \cline{2-9}
&$(20,30)$&600&$P(\wh r_1=r_1,\wh r_2=r_2)$&0.020&0.004&0.002&0.048&0.206\\
 &&&$P(\wh r_1^0=r_1,\wh r_2^0=r_2)$&0.004&0.022&0&0&0.036\\
  &&&$P(\wh r_1^{WLC}=r_1,\wh r_2^{WLC}=r_2)$&0&0&0&0&0\\
    &&&$P(\wh r_1^{YHKZ}=r_1,\wh r_2^{YHKZ}=r_2)$&0&0&0&0&0\\
\hline
\end{tabular}
          \end{center}
\end{table}


\begin{table}[ht]
 \caption{The estimation {discrepancy} of loading matrices defined in (4.1) when $(r_1,r_2)=(2,3)$ and  $(k_1,k_2)=(1,2)$ 
 in Example 1.  $(\wh\bA_1,\wh\bP_1)$,  $(\wh\bA_1^{o},\wh \bP_1^{o})$,  and $(\wh\bA_1^{YHKZ},\wh\bP_1^{YHKZ})$ are obtained by the proposed,  \cite{gaotsay2021c}, and \cite{yu2022} method, respectively.
 The sample sizes used are $T=300, 500, 1000, 1500, 3000$. Standard errors are given in the parentheses and 500 iterations are used. }
          \label{Table-a4}
          \scriptsize{
\begin{center}
 \setlength{\abovecaptionskip}{0pt}
\setlength{\belowcaptionskip}{3pt}

\begin{tabular}{cccccccc}
\hline
&&&\multicolumn{5}{c}{$T$}\\
\cline{4-8}
$(\delta_1,\delta_2)$&$(p_1,p_2)$&Method&$300$&$500$&$1000$&$1500$&$3000$\\
\hline
$(0,0)$& $(7,7)$&$MD(\wh\bA_1,\bL_1)$&0.066(0.161)&0.028(0.073)&0.023(0.075)&0.019(0.074)&0.021(0.084)\\
&&$MD(\wh\bA_1^o,\bL_1)$& 0.087(0.189)& 0.040(0.105)& 0.024(0.075)& 0.018(0.062)& 0.024(0.094)\\
&&$MD(\wh\bA_1^{YHKZ},\bL_1)$&0.577(5e-6)&0.577(3e-6)&0.577(1e-6)&0.577(8e-7)&0.577(4e-7)\\
&&$MD(\wh\bP_1,\bR_1)$&0.154(0.243)&0.032(0.108)&0.018(0.031)&0.010(0.049)&0.008(0.049)\\
&&$MD(\wh\bP_1^o,\bR_1)$&0.190(0.246)&0.047(0.118)&0.014(0.044)&0.012(0.049)&0.013(0.066)\\
&&$MD(\wh\bP_1^{YHKZ},\bR_1)$&0.600(0.103)&0.589(0.079)&0.586(0.037)&0.587(0.018)&0.585(0.001)\\
\cline{2-8}
& $(10,15)$&$MD(\wh\bA_1,\bL_1)$&0.044(0.117)&0.029(0.086)&0.017(0.059)&0.019(0.076)&0.020(0.087)\\
&&$MD(\wh\bA_1^o,\bL_1)$& 0.057(0.143)& 0.039(0.110)& 0.019(0.067)& 0.019(0.076)& 0.020(0.088)\\
&&$MD(\wh\bA_1^{YHKZ},\bL_1)$&0.577(1e-6)&0.577(8e-7)&0.577(4e-7)&0.577(2e-7)&0.577(1e-7)\\
&&$MD(\wh\bP_1,\bR_1)$&0.362(0.247)&0.309(0.250)&0.302(0.251)&0.276(0.252)&0.223(0.246)\\
&&$MD(\wh\bP_1^o,\bR_1)$&0.391(0.237)&0.346(0.247)&0.313(0.248)&0.301(0.251)&0.229(0.248)\\
&&$MD(\wh\bP_1^{YHKZ},\bR_1)$&0.632(2e-6)&0.632(1e-6)&0.632(6e-6)&0.632(4e-6)&0.632(2e-7)\\
\cline{2-8}
& $(20,20)$&$MD(\wh\bA_1,\bL_1)$&0.075(0.174)&0.062(0.157)&0.042(0.132)&0.040(0.131)&0.018(0.084)\\
&&$MD(\wh\bA_1^o,\bL_1)$& 0.095(0.119)& 0.066(0.164)& 0.048(0.141)& 0.039(0.129)& 0.018(0.080)\\
&&$MD(\wh\bA_1^{YHKZ},\bL_1)$&0.577(1e-6)&0.577(6e-7)&0.577(3e-7)&0.577(2e-7)&0.577(1e-7)\\
&&$MD(\wh\bP_1,\bR_1)$&0.272(0.232)&0.134(0.223)&0.057(0.144)&0.044(0.126)&0.025(0.094)\\
&&$MD(\wh\bP_1^o,\bR_1)$&0.277(0.269)&0.154(0.232)&0.064(0.151)&0.055(0.142)&0.028(0.099)\\
&&$MD(\wh\bP_1^{YHKZ},\bR_1)$&0.632(1e-6)&0.632(6e-7)&0.632(3e-7)&0.632(2e-7)&0.632(1e-7)\\
\cline{2-8}
& $(20,30)$&$MD(\wh\bA_1,\bL_1)$&0.010(0.216)&0.025(0.092)&0.020(0.084)&0.015(0.072)&0.010(0.057)\\
&&$MD(\wh\bA_1^o,\bL_1)$& 0.203(0.276)& 0.095(0.201)& 0.026(0.100)& 0.027(0.108)& 0.012(0.067)\\
&&$MD(\wh\bA_1,\bL_1)$&0.577(6e-7)&0.577(4e-7)&0.577(2e-7)&0.577(1e-7)&0.577(6e-8)\\
&&$MD(\wh\bP_1,\bR_1)$&0.320(0.269)&0.225(0.257)&0.121(0.210)&0.110(0.200)&0.061(0.156)\\
&&$MD(\wh\bP_1^o,\bR_1)$&0.307(0.258)&0.251(0.254)&0.163(0.233)&0.137(0.218)&0.071(0.168)\\
&&$MD(\wh\bP_1^{YHKZ},\bR_1)$&0.632(8e-7)&0.632(9e-7)&0.632(2e-7)&0.632(2e-7)&0.632(9e-8)\\
\hline
\end{tabular}
          \end{center}}
\end{table}

\begin{table}[ht]
 \caption{The estimation {discrepancy} of loading matrices defined in (4.1) when $(r_1,r_2)=(2,3)$ and  $(k_1,k_2)=(1,2)$ 
 in Example 1.  $(\wh\bA_1,\wh\bP_1)$,  $(\wh\bA_1^{o},\wh \bP_1^{o})$,  and $(\wh\bA_1^{YHKZ},\wh\bP_1^{YHKZ})$ are obtained by the proposed,  \cite{gaotsay2021c},  and \cite{yu2022} method, respectively.
 The sample sizes used are $T=300, 500, 1000, 1500, 3000$. Standard errors are given in the parentheses and 500 iterations are used. }
          \label{Table-a5}
          \scriptsize{
\begin{center}
 \setlength{\abovecaptionskip}{0pt}
\setlength{\belowcaptionskip}{3pt}

\begin{tabular}{cccccccc}
\hline
&&&\multicolumn{5}{c}{$T$}\\
\cline{4-8}
$(\delta_1,\delta_2)$&$(p_1,p_2)$&Method&$300$&$500$&$1000$&$1500$&$3000$\\
\hline
$(0.2,0.4)$& $(7,7)$&$MD(\wh\bA_1,\bL_1)$&0.030(0.079)&0.021(0.056)&0.018(0.066)&0.014(0.056)&0.022(0.094)\\
&&$MD(\wh\bA_1^o,\bL_1)$& 0.033(0.089)& 0.023(0.061)& 0.019(0.066)& 0.015(0.056)& 0.021(0.091)\\
&&$MD(\wh\bA_1^{YHKZ},\bL_1)$&0.577(8e-6)&0.577(5e-6)&0.577(2e-6)&0.577(1e-6)&0.577(8e-7)\\
&&$MD(\wh\bP_1,\bR_1)$&0.133(0.228)&0.028(0.095)&0.010(0.031)&0.013(0.058)&0.008(0.044)\\
&&$MD(\wh\bP_1^o,\bR_1)$&0.141(0.230)&0.035(0.104)&0.011(0.031)&0.013(0.054)&0.010(0.049)\\
&&$MD(\wh\bP_1^{YHKZ},\bR_1)$&0.085(0.096)&0.071(0.051)&0.065(0.025)&0.065(0.004)&0.064(0.003)\\
\cline{2-8}
& $(10,15)$&$MD(\wh\bA_1,\bL_1)$&0.022(0.066)&0.016(0.051)&0.017(0.072)&0.011(0.051)&0.016(0.080)\\
&&$MD(\wh\bA_1^o,\bL_1)$& 0.024(0.073)& 0.019(0.062)& 0.018(0.072)& 0.012(0.051)& 0.018(0.084)\\
&&$MD(\wh\bA_1^{YHKZ},\bL_1)$&0.536(0.145)&0.569(0.068)&0.577(1e-6)&0.577(7e-7)&0.577(4e-7)\\
&&$MD(\wh\bP_1,\bR_1)$&0.070(0.014)&0.030(0.064)&0.024(0.064)&0.022(0.068)&0.021(0.078)\\
&&$MD(\wh\bP_1^o,\bR_1)$&0.070(0.143)&0.030(0.060)&0.025(0.065)&0.024(0.075)&0.022(0.078)\\
&&$MD(\wh\bP_1^{YHKZ},\bR_1)$&0.565(0.190)&0.611(0.111)&0.632(2e-6)&0.632(1e-6)&0.632(6e-7)\\
\cline{2-8}
& $(20,20)$&$MD(\wh\bA_1,\bL_1)$&0.021(0.061)&0.017(0.056)&0.012(0.044)&0.011(0.051)&0.008(0.044)\\
&&$MD(\wh\bA_1^o,\bL_1)$& 0.020(0.056)& 0.018(0.056)& 0.012(0.044)& 0.012(0.051)& 0.008(0.044)\\
&&$MD(\wh\bA_1^{YHKZ},\bL_1)$&0.577(4e-6)&0.577(2e-6)&0.577(1e-6)&0.577(7e-7)&0.577(4e-7)\\
&&$MD(\wh\bP_1,\bR_1)$&0.055(0.139)&0.019(0.053)&0.018(0.062)&0.015(0.062)&0.014(0.066)\\
&&$MD(\wh\bP_1^o,\bR_1)$&0.056(0.140)&0.018(0.043)&0.019(0.065)&0.014(0.054)&0.015(0.066)\\
&&$MD(\wh\bP_1^{YHKZ},\bR_1)$&0.453(0.280)&0.516(0.242)&0.598(0.142)&0.625(0.067)&0.632(4e-7)\\
\cline{2-8}
& $(20,30)$&$MD(\wh\bA_1,\bL_1)$&0.025(0.092)&0.013(0.051)&0.013(0.062)&0.013(0.067)&0.011(0.067)\\
&&$MD(\wh\bA_1^o,\bL_1)$& 0.022(0.078)& 0.018(0.071)& 0.022(0.057)& 0.014(0.072)& 0.012(0.067)\\
&&$MD(\wh\bA_1,\bL_1)$&0.577(2e-6)&0.577(2e-6)&0.577(7e-7)&0.577(5e-7)&0.577(2e-7)\\
&&$MD(\wh\bP_1,\bR_1)$&0.059(0.146)&0.019(0.053)&0.011(0.031)&0.014(0.058)&0.010(0.049)\\
&&$MD(\wh\bP_1^o,\bR_1)$&0.064(0.151)&0.019(0.048)&0.013(0.038)&0.015(0.058)&0.010(0.044)\\
&&$MD(\wh\bP_1^{YHKZ},\bR_1)$&0.632(3e-6)&0.632(2e-6)&0.632(9e-7)&0.632(6e-7)&0.632(3e-7)\\
\hline
\end{tabular}
          \end{center}}
\end{table}

\begin{table}[ht]
 \caption{The estimation {discrepancy} of loading matrices defined in (4.1) when $(r_1,r_2)=(2,3)$ and  $(k_1,k_2)=(1,2)$ 
 in Example 1.  $(\wh\bA_1,\wh\bP_1)$,  $(\wh\bA_1^{o},\wh \bP_1^{o})$,  and $(\wh\bA_1^{YHKZ},\wh\bP_1^{YHKZ})$ are obtained by the proposed,  \cite{gaotsay2021c},  and \cite{yu2022} method, respectively.
 The sample sizes used are $T=300, 500, 1000, 1500, 3000$. Standard errors are given in the parentheses and 500 iterations are used. }
          \label{Table-a6}
          \scriptsize{
\begin{center}
 \setlength{\abovecaptionskip}{0pt}
\setlength{\belowcaptionskip}{3pt}

\begin{tabular}{cccccccc}
\hline
&&&\multicolumn{5}{c}{$T$}\\
\cline{4-8}
$(\delta_1,\delta_2)$&$(p_1,p_2)$&Method&$300$&$500$&$1000$&$1500$&$3000$\\
\hline
$(0.6,0.2)$& $(7,7)$&$MD(\wh\bA_1,\bL_1)$&0.237(0.225)&0.123(0.167)&0.047(0.071)&0.033(0.062)&0.035(0.096)\\
&&$MD(\wh\bA_1^o,\bL_1)$& 0.396(0.217)& 0.282(0.237)& 0.096(0.133)& 0.055(0.089)& 0.039(0.092)\\
&&$MD(\wh\bA_1^{YHKZ},\bL_1)$&0.578(2e-4)&0.578(1e-4)&0.577(6e-5)&0.577(4e-5)&0.577(2e-5)\\
&&$MD(\wh\bP_1,\bR_1)$&0.299(0.254)&0.212(0.243)&0.045(0.106)&0.026(0.072)&0.016(0.049)\\
&&$MD(\wh\bP_1^o,\bR_1)$&0.481(0.143)&0.379(0.193)&0.206(0.204)&0.114(0.155)&0.036(0.060)\\
&&$MD(\wh\bP_1^{YHKZ},\bR_1)$&0.893(0.048)&0.897(0.044)&0.903(0.039)&0.909(0.032)&0.916(0.021)\\
\cline{2-8}
& $(10,15)$&$MD(\wh\bA_1,\bL_1)$&0.279(0.218)&0.167(0.172)&0.074(0.092)&0.047(0.057)&0.031(0.066)\\
&&$MD(\wh\bA_1^o,\bL_1)$& 0.382(0.206)& 0.267(0.222)& 0.110(0.142)& 0.061(0.080)& 0.040(0.082)\\
&&$MD(\wh\bA_1^{YHKZ},\bL_1)$&0.577(3e-5)&0.577(2e-5)&0.577(9e-6)&0.577(6e-6)&0.577(3e-6)\\
&&$MD(\wh\bP_1,\bR_1)$&0.440(0.189)&0.355(0.227)&0.227(0.221)&0.153(0.189)&0.097(0.156)\\
&&$MD(\wh\bP_1^o,\bR_1)$&0.543(0.088)&0.541(0.113)&0.530(0.144)&0.485(0.184)&0.328(0.235)\\
&&$MD(\wh\bP_1^{YHKZ},\bR_1)$&0.633(5e-5)&0.633(3e-5)&0.632(1e-5)&0.632(9e-6)&0.632(4e-6)\\
\cline{2-8}
& $(20,20)$&$MD(\wh\bA_1,\bL_1)$&0.545(0.128)&0.471(0.204)&0.207(0.238)&0.084(0.143)&0.030(0.051)\\
&&$MD(\wh\bA_1^o,\bL_1)$& 0.583(0.033)& 0.567(0.076)& 0.440(0.227)& 0.276(0.256)& 0.057(0.108)\\
&&$MD(\wh\bA_1^{YHKZ},\bL_1)$&0.578(4e-5)&0.577(3e-5)&0.577(1e-5)&0.577(8e-6)&0.577(4e-6)\\
&&$MD(\wh\bP_1,\bR_1)$&0.589(0.076)&0.538(0.125)&0.367(0.229)&0.215(0.232)&0.039(0.076)\\
&&$MD(\wh\bP_1^o,\bR_1)$&0.621(0.042)&0.586(0.078)&0.481(0.152)&0.386(0.200)&0.145(0.193)\\
&&$MD(\wh\bP_1^{YHKZ},\bR_1)$&0.633(4e-5)&0.633(2e-5)&0.633(1e-5)&0.632(8e-6)&0.632(4e-6)\\
\cline{2-8}
& $(20,30)$&$MD(\wh\bA_1,\bL_1)$&0.579(0.044)&0.572(0.066)&0.463(0.216)&0.320(0.267)&0.073(0.159)\\
&&$MD(\wh\bA_1^o,\bL_1)$& 0.580(0.012)& 0.580(0.014)& 0.578(0.010)& 0.578(0.026)& 0.541(0.134)\\
&&$MD(\wh\bA_1,\bL_1)$&0.577(3e-5)&0.577(2e-5)&0.577(9e-6)&0.577(7e-6)&0.577(3e-6)\\
&&$MD(\wh\bP_1,\bR_1)$&0.619(0.048)&0.600(0.075)&0.541(0.081)&0.503(0.120)&0.424(0.204)\\
&&$MD(\wh\bP_1^o,\bR_1)$&0.633(0.018)&0.627(0.029)&0.555(0.098)&0.452(0.143)&0.248(0.152)\\
&&$MD(\wh\bP_1^{YHKZ},\bR_1)$&0.632(4e-5)&0.632(2e-5)&0.632(1e-5)&0.632(8e-6)&0.632(4e-6)\\
\hline
\end{tabular}
          \end{center}}
\end{table}

\begin{table}
 \caption{The  distance between the extracted common components and the true ones defined in (4.2) when $(r_1,r_2)=(2,3)$ and  $(k_1,k_2)=(1,2)$ 
 in Example 1.   $\mathcal{\wh S}_{t}$, $\mathcal{\wh S}_{0,t}$, and $\mathcal{\wh S}_{t}^{YHKZ}$ denote the extracted common components by the proposed, \cite{wang2018}, and \cite{yu2022} method, respectively.
 The sample sizes used are $T=300, 500, 1000, 1500, 3000$. Standard errors are given in the parentheses and 500 iterations are used. }
          \label{Table-a7}
\scriptsize{
\begin{center}
 \setlength{\abovecaptionskip}{0pt}
\setlength{\belowcaptionskip}{3pt}

\begin{tabular}{cccccccc}
\hline
&&&\multicolumn{5}{c}{$T$}\\
\cline{4-8}
$(\delta_1,\delta_2)$&$(p_1,p_2)$&Method&$300$&$500$&$1000$&$1500$&$3000$\\
\hline
$(0,0)$& $(7,7)$&$d(\mathcal{\wh S}_{t},\mathcal{S}_t)$&1.140(1.24)&0.504(0.76)&0.501(3.07)&0.360(0.196)&0.378(0.26)\\
&&$d(\mathcal{\wh S}_{0,t},\mathcal{S}_t)$& 3.373(18.71)& 0.900(4.22)& 0.745(6.71)& 0.396(0.55)& 0.473(1.37)\\
&&$d(\mathcal{\wh S}_{t}^{YHKZ},\mathcal{S}_t)$&3.040(0.09)&3.033(0.07)&3.041(0.05)&3.039(0.04)&3.038(0.03)\\
\cline{2-8}
& $(10,15)$&$d(\mathcal{\wh S}_{t},\mathcal{S}_t)$&2.586(6.84)&2.433(9.99)&2.116(17.49)&1.103(1.03)&0.837(0.74)\\
&&$d(\mathcal{\wh S}_{0,t},\mathcal{S}_t)$& 11.720(55.74)& 12.470(84.39)& 3.239(11.88)&2.123(2.81)& 1.425(2.10)\\
&&$d(\mathcal{\wh S}_{t}^{YHKZ},\mathcal{S}_t)$&2.859(0.087)&2.857(0.064)&2.855(0.046)&2.857(0.040)&2.857(0.026)\\
\cline{2-8}
& $(20,20)$&$d(\mathcal{\wh S}_{t},\mathcal{S}_t)$&2.192(11.92)&2.103(30.55)&0.594(7.83)&0.217(0.252)&0.158(0.155)\\
&&$d(\mathcal{\wh S}_{0,t},\mathcal{S}_t)$& 14.573(53.16)& 13.243(122.69)& 1.521(6.51)& 3.187(58.70)& 0.258(0.58)\\
&&$d(\mathcal{\wh S}_{t}^{YHKZ},\mathcal{S}_t)$&2.544(0.075)&2.552(0.057)&2.549(0.041)&2.548(0.033)&2.549(0.022)\\
\cline{2-8}
& $(20,30)$&$d(\mathcal{\wh S}_{t},\mathcal{S}_t)$&6.291(38.00)&3.043(25.88)&0.338(0.78)&0.247(0.37)&0.158(0.17)\\
&&$d(\mathcal{\wh S}_{0,t},\mathcal{S}_t)$& 103.446(243.13)& 47.776(206.85)& 3.099(19.14)& 5.965(103.09)& 0.334(0.61)\\
&&$d(\mathcal{\wh S}_{t}^{YHKZ},\mathcal{S}_t)$&2.846(0.086)&2.847(0.066)&2.849(0.047)&2.849(0.038)&2.849(0.027)\\
\hline
$(0.2,0.4)$& $(7,7)$&$d(\mathcal{\wh S}_{0,t},\mathcal{S}_t)$& 0.815(1.388)& 0.817(6.069)& 0.374(0.413)& 0.366(0.385)& 0.359(0.252)\\
&&$d(\mathcal{\wh S}_{t},\mathcal{S}_t)$&0.652(0.597)&0.382(0.238)&0.340(0.077)&0.338(0.065)&0.348(0.106)\\
&&$d(\mathcal{\wh S}_{t}^{YHKZ},\mathcal{S}_t)$&1.497(0.404)&1.494(0.030)&1.498(0.021)&1.497(0.018)&1.497(0.012)\\
\cline{2-8}
& $(10,15)$&$d(\mathcal{\wh S}_{t},\mathcal{S}_t)$&0.840(7.87)&0.347(1.67)&0.398(3.02)&0.247(0.20)&0.244(0.13)\\
&&$d(\mathcal{\wh S}_{0,t},\mathcal{S}_t)$& 1.727(16.78)& 1.421(10.80)& 1.412(0.88)& 0.335(0.95)& 0.402(2.03)\\
&&$d(\mathcal{\wh S}_{t}^{YHKZ},\mathcal{S}_t)$&1.141(0.03)&1.140(0.02)&1.140(0.02)&1.141(0.01)&1.141(0.01)\\
\cline{2-8}
& $(20,20)$&$d(\mathcal{\wh S}_{t},\mathcal{S}_t)$&0.227(0.33)&0.137(0.14)&0.122(0.02)&0.120(0.03)&0.118(0.02)\\
&&$d(\mathcal{\wh S}_{0,t},\mathcal{S}_t)$& 1.343(18.24)& 2.339(24.49)& 0.443(4.52)& 0.540(7.65)& 0.156(0.34)\\
&&$d(\mathcal{\wh S}_{t}^{YHKZ},\mathcal{S}_t)$&0.848(0.02)&0.850(0.02)&0.849(0.01)&0.849(0.01)&0.849(0.01)\\
\cline{2-8}
& $(20,30)$&$d(\mathcal{\wh S}_{t},\mathcal{S}_t)$&0.756(6.50)&0.111(0.11)&0.100(0.03)&0.099(0.03)&0.097(0.02)\\
&&$d(\mathcal{\wh S}_{0,t},\mathcal{S}_t)$& 4.781(45.62)& 6.976(57.27)& 4.453(48.84)& 5.220(54.17)& 0.595(6.36)\\
&&$d(\mathcal{\wh S}_{t}^{YHKZ},\mathcal{S}_t)$&0.858(0.02)&0.858(0.02)&0.860(0.01)&0.859(0.01)&0.860(0.01)\\
\hline
$(0.6,0.2)$& $(7,7)$&$d(\mathcal{\wh S}_{t},\mathcal{S}_t)$&1.833(2.67)&0.902(1.18)&0.425(0.62)&0.354(0.14)&0.362(0.17)\\
&&$d(\mathcal{\wh S}_{0,t},\mathcal{S}_t)$& 3.379(3.10)& 1.572(1.60)& 0.677(0.70)& 0.471(0.047)& 0.361(0.13)\\
&&$d(\mathcal{\wh S}_{t}^{YHKZ},\mathcal{S}_t)$&2.108(0.06)&2.104(0.05)&2.109(0.03)&2.108(0.03)&2.107(0.02)\\
\cline{2-8}
& $(10,15)$&$d(\mathcal{\wh S}_{t},\mathcal{S}_t)$&2.088(2.75)&1.379(2.20)&0.662(1.11)&0.426(0.38)&0.334(0.27)\\
&&$d(\mathcal{\wh S}_{0,t},\mathcal{S}_t)$& 3.717(3.47)& 2.699(2.84)& 1.453(1.30)& 1.081(0.46)& 0.731(0.41)\\
&&$d(\mathcal{\wh S}_{t}^{YHKZ},\mathcal{S}_t)$&1.775(0.05)&1.773(0.04)&1.772(0.03)&1.773(0.02)&1.770(0.02)\\
\cline{2-8}
&{$(20,20)$}&$d(\mathcal{\wh S}_{t},\mathcal{S}_t)$&19.144(10.97)&14.168(11.54)&4.131(8.14)&0.804(2.67)&0.149(0.38)\\
&&$d(\mathcal{\wh S}_{0,t},\mathcal{S}_t)$& 26.841(9.60)& 22.530(10.69)& 9.525(9.93)& 3.372(5.77)& 0.330(0.65)\\
&&$d(\mathcal{\wh S}_{t}^{YHKZ},\mathcal{S}_t)$&1.433(0.04)&1.437(0.03)&1.436(0.02)&1.435(0.02)&1.435(0.1)\\
\cline{2-8}
& $(20,30)$&$d(\mathcal{\wh S}_{t},\mathcal{S}_t)$&0.851(0.518)&1.070(0.529)&1.007(0.409)&0.882(0.43)&0.663(0.37)\\
&&$d(\mathcal{\wh S}_{0,t},\mathcal{S}_t)$& 30.248(6.90)& 28.640(8.49)& 19.010(11.75)& 10.538(9.80)& 2.266(3.49)\\
&&$d(\mathcal{\wh S}_{t}^{YHKZ},\mathcal{S}_t)$&1.529(0.04)&1.530(0.03)&1.532(0.02)&1.521(0.02)&1.532(0.01)\\
\hline
\end{tabular}
          \end{center}}
\end{table}


\begin{table}[ht]
 \caption{Empirical probabilities (EP) of determining the factor orders by different methods for Example 1 with $(r_1,r_2)=(2,3)$ and $(k_1,k_2)=(1,2)$, where $(p_1,p_2)$ and $T$ are the dimension and the sample size, respectively. $\delta_1$ and $\delta_2$ are the strength parameters of the factors and the errors, respectively. $500$ iterations are used. In the table, $\hat{r}, \hat{r}^0, 
 \hat{r}^{WLC}$ denote the proposed, 
 \cite{gaotsay2021c},  \cite{wang2018} method, respectively.  $k_0$ in (2.15) are set to $1,3,$ and $4$ for $(p_1,p_2)=(20,20)$ and $(\delta_1,\delta_2)=(0,0)$.}
          \label{Table-a8}
\begin{center}
 \setlength{\abovecaptionskip}{0pt}
\setlength{\belowcaptionskip}{3pt}

\begin{tabular}{c|ccc|ccccc}
\hline
 & &&&\multicolumn{5}{c}{$T$}\\
 \cline{5-9}
$k_0$ &$(p_1,p_2)$&$p_1p_2$&EP&$300$&$500$&$1000$&$1500$&$3000$\\
\hline
1 &$(20,20)$&400&$P(\wh r_1=r_1,\wh r_2=r_2)$&0.476&0.776&0.900&0.894&0.948\\
 &&&$P(\wh r_1^0=r_1,\wh r_2^0=r_2)$&0.448&0.736&0.884&0.876&0.948\\
  &&&$P(\wh r_1^{WLC}=r_1,\wh r_2^{WLC}=r_2)$&0.026&0.098&0.282&0.438&0.770\\
\hline
3&$(20,20)$&400&$P(\wh r_1=r_1,\wh r_2=r_2)$&0.434&0.688&0.802&0.846&0.920\\
 &&&$P(\wh r_1^0=r_1,\wh r_2^0=r_2)$&0.380&0.598&0.758&0.832&0.928\\
  &&&$P(\wh r_1^{WLC}=r_1,\wh r_2^{WLC}=r_2)$&0&0&0.002&0.014&0.192\\
 \hline
  4&$(20,20)$&400&$P(\wh r_1=r_1,\wh r_2=r_2)$&0.428&0.692&0.792&0.820&0.912\\
 &&&$P(\wh r_1^0=r_1,\wh r_2^0=r_2)$&0.352&0.596&0.736&0.808&0.916\\
  &&&$P(\wh r_1^{WLC}=r_1,\wh r_2^{WLC}=r_2)$&0&0&0&0&0.05\\
  \hline
\end{tabular}
          \end{center}
\end{table}

\begin{table}[ht]
 \caption{The estimation accuracy of loading matrices defined in (4.1) when $(r_1,r_2)=(2,3)$ and  $(k_1,k_2)=(1,2)$ 
 in Example 1.  
 The sample sizes used are $T=300, 500, 1000, 1500, 3000$. Standard errors are given in the parentheses and 500 iterations are used.  $k_0$ in (2.15) are set to $1,3,$ and $4$ for $(p_1,p_2)=(20,20)$ and $(\delta_1,\delta_2)=(0,0)$.}
          \label{Table-a9}
          \footnotesize{
\begin{center}
 \setlength{\abovecaptionskip}{0pt}
\setlength{\belowcaptionskip}{3pt}

\begin{tabular}{cccccccc}
\hline
&&&\multicolumn{5}{c}{$T$}\\
\cline{4-8}
$k_0$&$(p_1,p_2)$&Method&$300$&$500$&$1000$&$1500$&$3000$\\
\hline
$1$& $(20,20)$&$MD(\wh\bA_1,\bL_1)$&0.071(0.172)&0.042(0.123)&0.033(0.111)&0.032(0.114)&0.016(0.076)\\
&&$MD(\wh\bA_1^o,\bL_1)$& 0.084(0.190)& 0.049(0.136)& 0.035(0.117)& 0.037(0.124)& 0.015(0.072)\\
&&$MD(\wh\bP_1,\bR_1)$&0.265(0.273)&0.110(0.204)&0.042(0.121)&0.041(0.122)&0.023(0.089)\\
&&$MD(\wh\bP_1^o,\bR_1)$&0.269(0.270)&0.125(0.214)&0.048(0.130)&0.047(0.130)&0.025(0.094)\\
\hline
$3$& $(20,20)$&$MD(\wh\bA_1,\bL_1)$&0.074(0.173)&0.069(0.167)&0.058(0.158)&0.045(0.139)&0.026(0.104)\\
&&$MD(\wh\bA_1^o,\bL_1)$& 0.112(0.215)& 0.082(0.184)& 0.067(0.170)& 0.044(0.138)& 0.024(0.098)\\
&&$MD(\wh\bP_1,\bR_1)$&0.288(0.273)&0.135(0.212)&0.069(0.161)&0.055(0.142)&0.031(0.107)\\
&&$MD(\wh\bP_1^o,\bR_1)$&0.282(0.267)&0.168(0.237)&0.082(0.172)&0.064(0.154)&0.028(0.099)\\
\hline
$4$& $(20,20)$&$MD(\wh\bA_1,\bL_1)$&0.083(0.183)&0.069(0.167)&0.060(0.160)&0.055(0.155)&0.029(0.112)\\
&&$MD(\wh\bA_1^o,\bL_1)$& 0.122(0.223)& 0.088(0.190)& 0.069(0.173)& 0.050(0.147)& 0.029(0.112)\\
&&$MD(\wh\bP_1,\bR_1)$&0.283(0.272)&0.132(0.219)&0.074(0.165)&0.061(0.150)&0.031(0.107)\\
&&$MD(\wh\bP_1^o,\bR_1)$&0.288(0.267)&0.163(0.234)&0.094(0.183)&0.070(0.160)&0.029(0.103)\\
\hline
\end{tabular}
          \end{center}}
\end{table}

\begin{table}[ht]
 \caption{The  distance between the extracted common components and the true ones defined in (4.2) when $(r_1,r_2)=(2,3)$ and  $(k_1,k_2)=(1,2)$ 
 in Example 1.  
 The sample sizes used are $T=300, 500, 1000, 1500, 3000$. Standard errors are given in the parentheses and 500 iterations are used.  $\delta_1=\delta_2=0$. $k_0$ in (2.15) are set to $1,3,$ and $4$ for $(p_1,p_2)=(20,20)$ and $(\delta_1,\delta_2)=(0,0)$.}
          \label{Table-a10}
          
\footnotesize{
\begin{center}
 \setlength{\abovecaptionskip}{0pt}
\setlength{\belowcaptionskip}{3pt}

\begin{tabular}{cccccccc}
\hline
&&&\multicolumn{5}{c}{$T$}\\
\cline{4-8}
$k_0$&$(p_1,p_2)$&Method&$300$&$500$&$1000$&$1500$&$3000$\\
\hline
$1$& $(20,20)$&$d(\mathcal{\wh S}_{t},\mathcal{S}_t)$&1.219(1.14)&0.496(0.76)&0.196(0.21)&0.190(0.22)&0.147(0.13)\\
&&$d(\mathcal{\wh S}_{0,t},\mathcal{S}_t)$& 6.849(22.84)& 5.841(78.00)& 1.224(11.34)& 3.806(75.56)& 0.219(0.43)\\
\hline
$3$& $(20,20)$&$d(\mathcal{\wh S}_{t},\mathcal{S}_t)$&2.779(13.76)&0.697(0.915)&0.605(6.28)&0.269(0.371)&0.188(0.247)\\
&&$d(\mathcal{\wh S}_{0,t},\mathcal{S}_t)$& 23.930(83.96)& 16.093(91.32)& 2.720(9.33)& 5.618(79.68)& 0.306(0.83)\\
\hline
$4$& $(20,20)$&$d(\mathcal{\wh S}_{t},\mathcal{S}_t)$&3.633(21.02)&0.709(0.93)&0.674(4.27)&0.312(0.45)&0.200(0.27)\\
&&$d(\mathcal{\wh S}_{0,t},\mathcal{S}_t)$& 28.65(81.47)& 18.70(85.46)& 6.742(55.35)& 5.641(96.80)& 0.402(1.35)\\
\hline
\end{tabular}
          \end{center}}
\end{table}


\begin{table}[ht]
 \caption{Empirical probabilities (EP) of determining the factor orders by different methods for Example 1 with $(r_1,r_2)=(2,3)$ and $(k_1,k_2)=(1,2)$, where $(p_1,p_2)$ and $T$ are the dimension and the sample size, respectively. $\delta_1$ and $\delta_2$ are the strength parameters of the factors and the errors, respectively. $500$ iterations are used. In the table, $\hat{r}, \hat{r}^0, 
 \hat{r}^{WLC}$ denote the proposed, 
 \cite{gaotsay2021c},  \cite{wang2018}  method, respectively.  $k_0$ is set to $2$ and the initial $\bP_{0,1}$ is chosen according to the procedure in Section 2.2.4.}
          \label{Table-a11}
\begin{center}
 \setlength{\abovecaptionskip}{0pt}
\setlength{\belowcaptionskip}{3pt}

\begin{tabular}{c|ccc|ccccc}
\hline
 & &&&\multicolumn{5}{c}{$T$}\\
 \cline{5-9}
$(\delta_1, \delta_2)$ &$(p_1,p_2)$&$p_1p_2$&EP&$300$&$500$&$1000$&$1500$&$3000$\\
\hline
 (0,0) &$(7,7)$&49&$P(\wh r_1=r_1,\wh r_2=r_2)$&0.682&0.932&0.974&0.968&0.964\\
  \cline{2-9}
 &$(10,15)$&150&$P(\wh r_1=r_1,\wh r_2=r_2)$&0.342&0.384&0.418&0.436&0.582\\
    \cline{2-9}

&$(20,20)$&400&$P(\wh r_1=r_1,\wh r_2=r_2)$&0.472&0.750&0.858&0.868&0.934\\
    \cline{2-9}
&$(20,30)$&600&$P(\wh r_1=r_1,\wh r_2=r_2)$&0.374&0.550&0.730&0.768&0.868\\
\hline
\end{tabular}
          \end{center}
\end{table}

\begin{table}[ht]
 \caption{The estimation accuracy of loading matrices defined in (4.1) when $(r_1,r_2)=(2,3)$ and  $(k_1,k_2)=(1,2)$ 
 in Example 1.  
 The sample sizes used are $T=300, 500, 1000, 1500, 3000$. Standard errors are given in the parentheses and 500 iterations are used.  $k_0$ is set to $2$ and the initial $\bP_{0,1}$ is chosen according to the procedure in Section 2.2.4.}
          \label{Table-a12}
          \footnotesize{
\begin{center}
 \setlength{\abovecaptionskip}{0pt}
\setlength{\belowcaptionskip}{3pt}

\begin{tabular}{cccccccc}
\hline
&&&\multicolumn{5}{c}{$T$}\\
\cline{4-8}
$(\delta_1,\delta_2)$&$(p_1,p_2)$&Method&$300$&$500$&$1000$&$1500$&$3000$\\
\hline
$(0,0)$& $(7,7)$&$MD(\wh\bA_1,\bL_1)$&0.070&60.024&0.024&0.023&0.021\\
&&$MD(\wh\bP_1,\bR_1)$&0.149&0.041&0.010&0.010&0.009\\
\cline{2-8}
& $(10,15)$&$MD(\wh\bA_1,\bL_1)$&0.049&0.034&0.023&0.015&0.011\\
&&$MD(\wh\bP_1,\bR_1)$&0.359&0.315&0.296&0.290&0.218\\
\cline{2-8}
& $(20,20)$&$MD(\wh\bA_1,\bL_1)$&0.083&0.057&0.041&0.037&0.018\\
&&$MD(\wh\bP_1,\bR_1)$&0.263&0.113&0.056&0.051&0.029\\
\cline{2-8}
& $(20,30)$&$MD(\wh\bA_1,\bL_1)$&0.086&0.028&0.021&0.013&0.016\\
&&$MD(\wh\bP_1,\bR_1)$&0.317&0.230&0.134&0.118&0.061\\
\hline
\end{tabular}
          \end{center}}
\end{table}

\begin{table}[ht]
 \caption{The  distance between the extracted common components and the true ones defined in (4.2) when $(r_1,r_2)=(2,3)$ and  $(k_1,k_2)=(1,2)$ 
 in Example 1.  
 The sample sizes used are $T=300, 500, 1000, 1500, 3000$. Standard errors are given in the parentheses and 500 iterations are used. $k_0$ is set to $2$ and the initial $\bP_{0,1}$ is chosen according to the procedure in Section 2.2.4.}
          \label{Table-a13}
          
\footnotesize{
\begin{center}
 \setlength{\abovecaptionskip}{0pt}
\setlength{\belowcaptionskip}{3pt}

\begin{tabular}{cccccccc}
\hline
&&&\multicolumn{5}{c}{$T$}\\
\cline{4-8}
$(\delta_1,\delta_2)$&$(p_1,p_2)$&Method&$300$&$500$&$1000$&$1500$&$3000$\\
\hline
$(0,0)$&$(7,7)$& $d(\mathcal{\wh S}_{t},\mathcal{S}_t)$&1.312&0.506&0.382&0.379&0.393\\
\cline{2-8}
& $(10,15)$&$d(\mathcal{\wh S}_{t},\mathcal{S}_t)$&2.256&1.906&1.602&1.107&0.821\\
\cline{2-8}
& $(20,20)$&$d(\mathcal{\wh S}_{t},\mathcal{S}_t)$&2.210&0.548&0.251&0.218&0.163\\
\cline{2-8}
& $(20,30)$&$d(\mathcal{\wh S}_{t},\mathcal{S}_t)$&3.446&1.085&0.547&0.248&0.176\\
\hline
\end{tabular}
          \end{center}}
\end{table}



\section{Appendix: Proofs}

In this section, $C$ denotes a generic constant the value of which may change at different places.  The outline of the proof is as follows. We first state some notation and useful lemmas that will be used to prove the Theorems. The main technique used is the matrix-perturbation theory which is commonly used to show the subspace consistency of PCA. The estimators of loading matrices such as $\bA_1$ and $\bP_1$ are the analytic ones specified in Lemma 1 of \cite{gaotsay2018b} or Lemma 3 of \cite{LamYaoBathia_Biometrika_2011}, which is the reason why we use a direct $\ell_2$-distance between the estimated loading matrix and the true one without using a rotational matrix in the proof below because the rotation has been incorporated into the true loadings implicitly, and the results also hold for the distances between linear spaces spanned by columns of the corresponding matrices as described in the theorems of the main text.  Based on the discussion of some equivalent distances between linear spaces in Section 3 of \cite{gaotsay2021c}, $D(\mathcal{M}(\bH_1),\mathcal{M}(\bH_2))$ and $\|\bH_1\bH_1'-\bH_2\bH_2'\|_F$ are equivalent if $\bH_1$ and $\bH_2$ are two $p\times r$ semi-orthogonal matrices.   In fact, the two measures $D(\mathcal{M}(\bH_1),\mathcal{M}(\bH_2))$ and $\|\bH_1-\bH_2\|_2$ are also equivalent  with the additional constraint $\|\bH_1-\bH_2\|_2=o(1)$, which is the case in the proofs below.    First, it is not hard to establish that 
\[D(\mathcal{M}(\bH_1),\mathcal{M}(\bH_2))^2\leq \|\bH_1'(\bH_1\bH_1'-\bH_2\bH_2')\bH_1\|_2\leq 2\|\bH_1-\bH_2\|_2^2.\]
See the proof of Theorem 1 in \cite{gaotsay2018b}. Furthermore,  note that $\tr(\bH_1'\bH_2)$ and $\tr(\bH_1'\bH_2\bH_2'\bH_1)$ are roughly equal to each other if $\|\bH_1-\bH_2\|_2=o(1)$, then
\[\|\bH_1-\bH_2\|_2^2\leq\|\bH_1-\bH_2\|_F^2=2r-2\tr(\bH_1'\bH_2)\approx 2r-2\tr(\bH_1'\bH_2\bH_2'\bH_1)=\|\bH_1\bH_1'-\bH_2\bH_2'\|_F^2,\]
implying that $D(\mathcal{M}(\bH_1),\mathcal{M}(\bH_2))$ and $\|\bH_1-\bH_2\|_2$  are equivalent in our proofs below.

The key observation in proving the consistency of Algorithm 1 is that the convergence rates of the loading matrices will not be changed after one iteration of the algorithm, and they are already faster than the existing ones, as will be seen in the proof of Theorem 1 below.

Note that $\bZ_{0,t}=\bY_t\bP_{0,1}=[\bz_{1,t}^0,...,\bz_{r_2,t}^0]$, $\bW_{0,t}=\bY_t'\bA_1=[\bw_{1,t}^0,...,\bw_{r_1,t}^0]$ and $\wh\bW_{0,t}=\bY_t'\wh\bA_{0,1}=[\wh\bw_{1,t}^0,...,\wh\bw_{r_1,t}^0]$.  Let $\bSigma_{z_0,ij}(k)=\cov(\bz_{i,t}^0,\bz_{j,t-k}^0)$, $\bSigma_{w_0,ij}(k)=\cov(\bw_{i,t}^0,\bw_{j,t-k}^0)$,  
\[\wh\bSigma_{z_0,ij}(k)=\frac{1}{T}\sum_{t=k+1}^T\bz_{i,t}^0\bz_{j,t-k}^0{'},\quad \wh\bSigma_{w_0,ij}(k)=\frac{1}{T}\sum_{t=k+1}^T\bw_{i,t}^0\bw_{j,t-k}^0{'},\]
and 
\[\wh\bSigma_{\wh w_0,ij}(k)=\frac{1}{T}\sum_{t=k+1}^T\wh\bw_{i,t}^0\wh\bw_{j,t-k}^0{'}.\]

{\bf Proof of Proposition 1.} {The consistency of estimated loading matrices can be 
proven by a similar argument as that in \cite{wang2018} or \cite{gaotsay2021c}.  Though our model is different from the one considered in \cite{gaotsay2021c}, we can drop the cross-terms therein and the resulting model will be similar to ours. Therefore,   the second part of Proposition 1 can be shown by simplifying the one in \cite{gaotsay2021c}.  The convergence rates remain the same, because the cross-term in \cite{gaotsay2021c} is less dominant.  We omit the details here. $\Box$ }

To prove Theorem 1, {we need the following additional lemmas.}
\begin{lemma}
If Assumptions 1--5 hold, then, for $1\leq k\leq k_0$, 
\[\sum_{i=1}^{r_2}\sum_{j=1}^{r_2}\|\bSigma_{z_0,ij}(k)\|_2^2=O_p(p_1^{2-2\delta_1}p_2^{2-2\delta_1})\,\,\text{and}\,\,\sum_{i=1}^{r_1}\sum_{j=1}^{r_1}\|\bSigma_{w_0,ij}(k)\|_2^2=O_p(p_1^{2-2\delta_1}p_2^{2-2\delta_1}).\]
\end{lemma}
{\bf Proof.}  We only prove the first result here as the 
proof of the second one is similar. Note that $\bff_t=\vc(\bF_t)$ and $\bz_{i,t}^0=\bY_t\bp_{01,\sbullet i}=(\bL_1\bF_t\bR_1'+\bE_t)\bp_{01.\sbullet i}$, where $\bp_{01,\sbullet i}$ is the $i$-th column of $\bP_{0,1}$.  As $\bF_t$ and $\bE_t$ are uncorrelated, it follows that
\begin{equation}\label{sig:z00}
\bSigma_{z_0,ij}(k)=\cov(\bz_{i,t}^0,\bz_{j,t-k}^0)=\bL_1(\bp_{01,\sbullet i}'\bR_1\otimes\bI_{r_1})\bSigma_f(k)(\bR_1'\bp_{01,\sbullet i}\otimes\bI_{r_1})\bL_1'.
\end{equation}
Therefore, 
\[\sum_{i,j=1}^{r_2}\|\bSigma_{z_0,ij}(k)\|_2^2\leq C\|\bL_1\|_2^4\|\bR_1\|_2^4\|\bSigma_f(k)\|_2^2=O_p(p_1^{2-2\delta_1}p_2^{2-2\delta_1}).\]
This completes the proof. $\Box$

\begin{lemma}\label{lm2}
If Assumptions 1--5 hold and $p_1\asymp p_1$, then, for $1\leq k\leq k_0$,
\[\sum_{k=1}^{k_0}\sum_{i=1}^{r_2}\sum_{j=1}^{r_2}\|\wh\bSigma_{z_0,ij}(k)-\bSigma_{z_0,ij}(k)\|_2^2=\left\{\begin{array}{cc}
O_p(p_1^{2-2\delta_1}p_2^{2-2\delta_1}T^{-1}),&\text{if}\,\, \delta_1\leq \delta_2,\delta_2\leq1/2,\\
O_p(p_1^{2-2\delta_2}p_2^{2-2\delta_2}T^{-1}),&\text{if}\,\, \delta_1>\delta_2,\delta_2\leq 1/2,\\
O_p(p_1^{2-2\delta_1}p_2^{2-2\delta_1}T^{-1}),&\text{if}\,\, \delta_1\leq 1/2,\delta_2> 1/2,\\
O_p(p_1^2T^{-1}),&\text{if}\,\, \delta_1> 1/2,\delta_2> 1/2.
\end{array}\right..\]
\end{lemma}

{\bf Proof.} Note that
\begin{align}\label{sig:z0}
\wh\bSigma_{z_0,ij}(k)=&\frac{1}{T}\sum_{t=k+1}^T(\bL_1\bF_t\bR_1'\bp_{01\sbullet i}+\bE_t\bp_{01,\sbullet i})(\bL_1\bF_{t-k}\bR_1'\bp_{01\sbullet j}+\bE_{t-k}\bp_{01,\sbullet j})'\notag\\
=&\frac{1}{T}\sum_{t=k+1}^T\bL_1\bF_t\bR_1'\bp_{01,\sbullet i}\bp_{01,\sbullet j}'\bR_1\bF_{t-k}'\bL_1'+\frac{1}{T}\sum_{t=k+1}^T\bL_1\bF_t\bR_1'\bp_{01,\sbullet i}\bp_{01,\sbullet j}'\bE_{t-k}'\notag\\
&+\frac{1}{T}\sum_{t=k+1}^T\bE_t\bp_{01,\sbullet i}\bp_{01,\sbullet j}'\bR_1\bF_{t-k}'\bL_1'+\frac{1}{T}\sum_{t=k+1}^T\bE_t\bp_{01,\sbullet i}\bp_{01,\sbullet j}'\bE_{t-k}\notag\\
=&\Delta_1^{ijk}+\Delta_2^{ijk}+\Delta_3^{ijk}+\Delta_4^{ijk}.
\end{align}
By (\ref{sig:z00}),
\[\Delta_1^{ijk}-\bSigma_{z_0,ij}(k)=\bL_1(\bp_{01,\sbullet i}'\bR_1\otimes\bI_{r_1})\left[\frac{1}{T}\sum_{t=k+1}^T\bff_t\bff_{t-k}'-\bSigma_f(k)\right](\bR_1'\bp_{01,\sbullet i}\otimes\bI_{r_1})\bL_1'.\]
As $r_1$ and $r_2$ are finite,  by Assumptions 1--2 and a similar argument as the proof of Theorem 1 in \cite{gaotsay2018b}, 
\[\|\frac{1}{T}\sum_{t=k+1}^T\bff_t\bff_{t-k}'-\bSigma_f(k)\|_2=O_p(T^{-1/2}).\]
Therefore,
\begin{equation}\label{delta1}
\sum_{i,j=1}^{r_2}\|\Delta_1^{ijk}-\bSigma_{z_0,ij}(k)\|_2^2\leq C\|\bL_1\|_2^4\|\bR_1\|_2^4T^{-1}=O_p(p_1^{2-2\delta_1}p_2^{2-2\delta_1}T^{-1}).
\end{equation}
Furthermore, note that the SVD of $\bL_2$ and $\bR_2$ implies that
\begin{align}\label{ET}
\bE_t=&\bA_2\bD_2^{1/2}\bU_2'\bxi_t\bV_2\bLambda_2^{1/2}\bP_2'+\bfeta_t
=I_{1,t}+I_{2,t}.
\end{align}
Then,
\begin{align}\label{dec:d2}
\Delta_2^{ijk}=&\frac{1}{T}\sum_{t=k+1}^T\bL_1\bF_t\bR_1'\bp_{01,\sbullet i}\bp_{01,\sbullet j}'\bP_2\bLambda_2^{1/2}\bV_2'\bxi_{t-k}'\bU_2\bD_2^{1/2}\bA_2'+\frac{1}{T}\sum_{t=k+1}^T\bL_1\bF_t\bR_1'\bp_{01,\sbullet i}\bp_{01,\sbullet j}'\bfeta_t\notag\\
=&\Delta_{2,1}^{ijk}+\Delta_{2,2}^{ijk}.
\end{align}
By Assumptions 1--4, it is straightforward to show that
\[\sum_{i,j=1}^{r_2}\|\Delta_{2,1}^{ijk}\|_2^2\leq C\|\bL_1\|_2^2\|\bR_1\|_2^2\|\bLambda_2^{1/2}\|_2^2\|\bD_2^{1/2}\|_2^2T^{-1}=O_p(p_1^{2-\delta_1-\delta_2}p_2^{2-\delta_1-\delta_2}T^{-1}),\]
and
\[\sum_{i,j=1}^{r_2}\|\Delta_{2,2}^{ijk}\|_2^2\leq C\|\bL_1\|_2^2\|\bR_1\|_2^2p_1T^{-1}=O_p(p_1^{2-\delta_1}p_2^{1-\delta_1}T^{-1}).\]
Therefore,
\begin{equation}\label{delta2}
\sum_{i,j=1}^{r_2}\|\Delta_2^{ijk}\|_2^2\leq C\max(p_1^{2-\delta_1-\delta_2}p_2^{2-\delta_1-\delta_2}T^{-1},p_1^{2-\delta_1}p_2^{1-\delta_1}T^{-1})=\left\{\begin{array}{ll}
Cp_1^{2-\delta_1-\delta_2}p_2^{2-\delta_1-\delta_2}T^{-1},&\delta_2\leq 1/2,\\
Cp_1^{2-\delta_1}p_2^{1-\delta_1}T^{-1},&\delta_2>1/2.
\end{array}
\right.
\end{equation}
Similarly, we can show that $\sum_{i,j=1}^{r_2}\|\Delta_3^{ijk}\|_2^2=O_p(\max(p_1^{2-\delta_1-\delta_2}p_2^{2-\delta_1-\delta_2}T^{-1},p_1^{2-\delta_1}p_2^{1-\delta_1}T^{-1}))$, which is the same as the above one.  

Turn to $\Delta_4^{ijk}$.  By (\ref{ET}), 
\begin{align}\label{delta4}
\Delta_4^{ijk}=&\frac{1}{T}\sum_{t=k+1}^T(I_{1,t}+I_{2,t})\bp_{01,\sbullet i}\bp_{01,\sbullet j}'(I_{1,t-k}+I_{2,t-k})'\notag\\
=&\frac{1}{T}\sum_{t=k+1}^TI_{1,t}\bp_{01,\sbullet i}\bp_{01,\sbullet j}'I_{1,t-k}'+\frac{1}{T}\sum_{t=k+1}^TI_{1,t}\bp_{01,\sbullet i}\bp_{01,\sbullet j}'I_{2,t-k}'\notag\\
&+\frac{1}{T}\sum_{t=k+1}^TI_{2,t}\bp_{01,\sbullet i}\bp_{01,\sbullet j}'I_{1,t-k}'+\frac{1}{T}\sum_{t=k+1}^TI_{2,t}\bp_{01,\sbullet i}\bp_{01,\sbullet j}'I_{2,t-k}'\notag\\
=&\Delta_{4,1}^{ijk}+\Delta_{4,2}^{ijk}+\Delta_{4,3}^{ijk}+\Delta_{4,4}^{ijk}.
\end{align}
By a similar argument as those in (\ref{dec:d2}), we can show that
\[\sum_{i,j=1}^{r_2}\|\Delta_{4,1}^{ijk}\|_2^2\leq C\|\bD_2^{1/2}\|_2^2\|\bLambda_2^{1/2}\|_2^2\|\bD_2^{1/2}\|_2^2\|\bLambda_2^{1/2}\|_2^2T^{-1}=O_p(p_1^{2-2\delta_2}p_2^{2-2\delta_2}T^{-1}),\]
\[\sum_{i,j=1}^{r_2}\|\Delta_{4,2}^{ijk}\|_2^2\leq C\|\bD_1^{1/2}\|_2^2\|\bLambda_1^{1/2}\|_2^2p_1T^{-1}=O_p(p_1^{2-\delta_2}p_2^{1-\delta_2}T^{-1}),\]
\[\sum_{i,j=1}^{r_2}\|\Delta_{4,3}^{ijk}\|_2^2\leq Cp_1\|\bD_1^{1/2}\|_2^2\|\bLambda_1^{1/2}\|_2^2T^{-1}=O_p(p_1^{2-\delta_2}p_2^{1-\delta_2}T^{-1}),\]
and
\[\sum_{i,j=1}^{r_2}\|\Delta_{4,4}^{ijk}\|_2^2\leq Cp_1p_1T^{-1}=O_p(p_1^{2}T^{-1}).\]
As a result, 
\begin{equation}\label{delta4:s}
\sum_{i,j=1}^{r_2}\|\Delta_4^{ijk}\|_2^2\leq C\max\{p_1^{2-2\delta_2}p_2^{2-2\delta_2}T^{-1},p_1^{2-\delta_2}p_2^{1-\delta_2}T^{-1},p_1^{2}T^{-1}\}=C\left\{\begin{array}{ll}
p_1^{2-2\delta_2}p_2^{2-2\delta_2}T^{-1},&\delta_2\leq 1/2,\\
p_1^2T^{-1},&\delta_2> 1/2.
\end{array}\right.
\end{equation}
It follows from (\ref{sig:z0}) and the above bounds that
\begin{align}\label{sigh:zc}
\sum_{i,j=1}^{r_2}\|\wh\bSigma_{z_0,ij}(k)-\bSigma_{z_0,ij}(k)\|_2^2\leq &\sum_{i,j=1}^{r_2}\|\Delta_{1}^{ijk}-\bSigma_{z_0,ij}(k)\|_2^2+\sum_{i,j=1}^{r_2}\|\Delta_{2}^{ijk})\|_2^2\notag\\
&+\sum_{i,j=1}^{r_2}\|\Delta_{3}^{ijk})\|_2^2+\sum_{i,j=1}^{r_2}\|\Delta_{4}^{ijk})\|_2^2\notag\\
=&\left\{\begin{array}{cc}
O_p(p_1^{2-2\delta_1}p_2^{2-2\delta_1}T^{-1}),&\text{if}\,\, \delta_1\leq \delta_2,\delta_2\leq1/2,\\
O_p(p_1^{2-2\delta_2}p_2^{2-2\delta_2}T^{-1}),&\text{if}\,\, \delta_1>\delta_2,\delta_2\leq 1/2,\\
O_p(p_1^{2-2\delta_1}p_2^{2-2\delta_1}T^{-1}),&\text{if}\,\, \delta_1\leq 1/2,\delta_2> 1/2,\\
O_p(p_1^2T^{-1}),&\text{if}\,\, \delta_1> 1/2,\delta_2> 1/2.
\end{array}\right.
\end{align}
Lemma 2 follows from the assumption that $k_0$ is finite. This completes the proof. $\Box$

\begin{lemma}\label{lm3}
Assume that Assumptions 1-5 hold. If $p_1\asymp p_2$ and $p_1^{\delta_1-\delta_2}p_2^{\delta_1-\delta_2}T^{-1/2}=o(1)$, then
\[\|\wh\bM_{0,1}^*-\bM_{0,1}^*\|_2=\left\{\begin{array}{cc}
O_p(p_1^{2-2\delta_1}p_2^{2-2\delta_1}T^{-1/2}),&\text{if}\,\, \delta_1\leq \delta_2,\delta_2\leq1/2,\\
O_p(p_1^{2-\delta_1-\delta_2}p_2^{2-\delta_1-\delta_2}T^{-1/2}),&\text{if}\,\, \delta_1>\delta_2,\delta_2\leq 1/2,\\
O_p(p_1^{2-2\delta_1}p_2^{2-2\delta_1}T^{-1/2}),&\text{if}\,\, \delta_1\leq 1/2,\delta_2> 1/2,\\
O_p(p_1^{2-\delta_1}p_2^{1-\delta_1}T^{-1/2}),&\text{if}\,\, \delta_1> 1/2,\delta_2> 1/2.
\end{array}\right..\]
\end{lemma}

{\bf Proof.} By Lemmas 1-2, if $p_1\asymp p_2$ and $p_1^{\delta_1-\delta_2} p_2^{\delta_1-\delta_2}T^{-1/2}=o(1)$, 
\begin{align}\label{M:df}
\|\wh\bM_{0,1}^*-\bM_{0,1}^*\|_2\leq & \sum_{k=1}^{k_0}\sum_{i=1}^{r_2}\sum_{j=1}^{r_2}\left\{\|\wh\bSigma_{z_0,ij}(k)-\bSigma_{z_0,ij}(k)\|_2^2+2\|\bSigma_{z_0,ij}(k)\|_2\|\wh\bSigma_{z_0,ij}(k)-\bSigma_{z_0,ij}(k)\|_2\right\}\notag\\
\leq&\sum_{k=1}^{k_0}\sum_{i=1}^{p_2}\sum_{j=1}^{p_2}\|\wh\bSigma_{z_0,ij}(k)-\bSigma_{z_0,ij}(k)\|_2^2+2\sum_{k=1}^{k_0}\left(\sum_{i=1}^{p_2}\sum_{j=1}^{p_2}\|\bSigma_{z_0,ij}(k)\|_2^2\right)^{1/2}\notag\\
&\times\left(\sum_{i=1}^{p_2}\sum_{j=1}^{p_2}\|\wh\bSigma_{z_0,ij}(k)-\bSigma_{z_0,ij}(k)\|_2^2\right)^{1/2}\notag\\
=&\left\{\begin{array}{cc}
O_p(p_1^{2-2\delta_1}p_2^{2-2\delta_1}T^{-1/2}),&\text{if}\,\, \delta_1\leq \delta_2,\delta_2\leq1/2,\\
O_p(p_1^{2-\delta_1-\delta_2}p_2^{2-\delta_1-\delta_2}T^{-1/2}),&\text{if}\,\, \delta_1>\delta_2,\delta_2\leq 1/2,\\
O_p(p_1^{2-2\delta_1}p_2^{2-2\delta_1}T^{-1/2}),&\text{if}\,\, \delta_1\leq 1/2,\delta_2> 1/2,\\
O_p(p_1^{2-\delta_1}p_2^{1-\delta_1}T^{-1/2}),&\text{if}\,\, \delta_1> 1/2,\delta_2> 1/2.
\end{array}\right.
\end{align}
This completes the proof. $\Box$

\begin{lemma}
If Assumptions 1-5 hold, then
\begin{equation}\label{lbdr}
\lambda_{r_1}(\bM_{0,1}^*)\geq Cp_1^{2-2\delta_1}p_2^{2-2\delta_1}.
\end{equation}
\end{lemma}
{\bf Proof.} Note that
\begin{align*}
\bSigma_{z_0,ij}(k)=&\bL_1(\bp_{01,\sbullet i}'\bR_1\otimes\bI_{r_1})\bSigma_f(k)(\bR_1'\bp_{01,\sbullet j}\otimes\bI_{r_1})\bL_1',
\end{align*}
and $\lambda_{r_1}(\bSigma_{f}(k)\bSigma_{f}(k)')\geq C>0$. The result can be established by a similar argument as the Proof of Lemma 5 in \cite{wang2018}. We omit the details. This competes the proof. $\Box$
\begin{lemma}
Suppose Assumptions 1 to 5 hold.  If $p\asymp p_2$ and $p_1^{\delta_1-\delta_2}p_2^{\delta_1-\delta_2}T^{-1/2}=o(1)$, then 
\[\|\wh\bA_{0,1}-\bA_1\|_2=\left\{\begin{array}{ll}
O_p(T^{-1/2}),&\text{if}\,\, \delta_1\leq \delta_2,\delta_2\leq1/2,\\
O_p(p_1^{\delta_1-\delta_2}p_2^{\delta_1-\delta_2}T^{-1/2}),&\text{if}\,\, \delta_1>\delta_2,\delta_2\leq 1/2,\\
O_p(T^{-1/2}),&\text{if}\,\, \delta_1\leq 1/2,\delta_2> 1/2,\\
O_p(p_1^{\delta_1}p_2^{\delta_1-1}T^{-1/2}),&\text{if}\,\, \delta_1> 1/2,\delta_2> 1/2.
\end{array}\right..\]
\end{lemma}
{\bf Proof.} By   the matrix perturbation theorem in Lemma 1 of \cite{gaotsay2018b},
\[\|\wh\bA_{0,1}-\bA_1\|_2\leq \frac{\|\wh\bM_{0,1}^*-\bM_{0,1}^*\|_2}{\lambda_{r_1}(\bM_{0,1}^*)}.\]
Lemma 5 follows from the results in  Lemmas 3 and 4 and the above inequality. This completes the proof. $\Box$

\begin{lemma}
If Assumptions 1--5 hold, then, for $1\leq k\leq k_0$,
\[\sum_{k=1}^{k_0}\sum_{i=1}^{r_1}\sum_{j=1}^{r_1}\|\wh\bSigma_{\wh w_0,ij}(k)-\bSigma_{w_0,ij}(k)\|_2^2=\left\{\begin{array}{cc}
O_p(p_1^{2-2\delta_1}p_2^{2-2\delta_1}T^{-1}),&\text{if}\,\, \delta_1\leq \delta_2,\delta_2\leq1/2,\\
O_p(p_1^{2-2\delta_2}p_2^{2-2\delta_2}T^{-1}),&\text{if}\,\, \delta_1>\delta_2,\delta_2\leq 1/2,\\
O_p(p_1^{2-2\delta_1}p_2^{2-2\delta_1}T^{-1}),&\text{if}\,\, \delta_1\leq 1/2,\delta_2> 1/2,\\
O_p(p_2^2T^{-1}),&\text{if}\,\, \delta_1> 1/2,\delta_2> 1/2.
\end{array}\right.\]
\end{lemma}

{\bf Proof.} Note that 
\[\wh\bW_{a,t}=\bY_t'\bA_1+\bY_t'(\wh\bA_{0.1}-\bA_1)=\bW_{a,t}+(\bR_1\bF_t'\bL_1'+\bE_t')(\wh\bA_{0,1}-\bA_1).\]
Thus,
\[\wh\bw_{i,t}^a=\bw_{i,t}^a+(\bR_1\bF_t'\bL_1'+\bE_t')(\wh\ba_{01,\sbullet i}-\ba_{1,\sbullet i}),\]
and
\begin{align}\label{sigh:w}
\wh\bSigma_{\wh w_a,ij}(k)=&\frac{1}{T}\sum_{t=k+1}^T\wh\bw_{i,t}^a\wh\bw_{j,t-k}^a{'}\notag\\
=&\frac{1}{T}\sum_{t=k+1}^T\bw_{i,t}^a\bw_{j,t-k}^a{'}+\frac{1}{T}\sum_{t=k+1}^T\bw_{i,t}^a(\wh\ba_{01.\sbullet j}-\ba_{1,\sbullet j})'(\bR_1\bF_{t-k}'\bL_1'+\bE_{t-k}')'\notag\\
&+\frac{1}{T}\sum_{t=k+1}^T(\bR_1\bF_t'\bL_1'+\bE_t')(\wh\ba_{01.\sbullet i}-\ba_{1,\sbullet i})\bw_{j,t-k}^a{'}\notag\\
&+\frac{1}{T}\sum_{t=k+1}^T(\bR_1\bF_t'\bL_1'+\bE_t')(\wh\ba_{01.\sbullet i}-\ba_{1,\sbullet i})(\wh\ba_{01.\sbullet j}-\ba_{1,\sbullet j})'(\bR_1\bF_{t-k}'\bL_1'+\bE_{t-k}')'\notag\\
=&J_1^{ijk}+J_2^{ijk}+J_3^{ijk}+J_4^{ijk}.
\end{align}
Therefore,
\begin{equation}\label{sigh:w:d}
\|\wh\bSigma_{\wh w_a,ij}(k)-\bSigma_{w_a,ij}(k)\|_2^2\leq\{\|J_1^{ijk}-\bSigma_{w_a,ij}(k)\|_2^2+\|J_2^{ijk}\|_2^2+\|J_3^{ijk}\|_2^2+\|J_4^{ijk}\|_2^2\}.
\end{equation}
By a similar argument as that of Lemma 3 and switching $p_1$ and $p_2$,  we can show
\begin{equation}\label{j1:d}
\sum_{i=1}^{r_1}\sum_{j=1}^{r_1}\|J_1^{ijk}-\bSigma_{w_a,ij}(k)\|_2^2=\left\{\begin{array}{cc}
O_p(p_1^{2-2\delta_1}p_2^{2-2\delta_1}T^{-1}),&\text{if}\,\, \delta_1\leq \delta_2,\delta_2\leq1/2,\\
O_p(p_1^{2-2\delta_2}p_2^{2-2\delta_2}T^{-1}),&\text{if}\,\, \delta_1>\delta_2,\delta_2\leq 1/2,\\
O_p(p_1^{2-2\delta_1}p_2^{2-2\delta_1}T^{-1}),&\text{if}\,\, \delta_1\leq 1/2,\delta_2> 1/2,\\
O_p(p_2^2T^{-1}),&\text{if}\,\, \delta_1> 1/2,\delta_2> 1/2.
\end{array}\right.
\end{equation}

Next, we only consider $J_2^{ijk}$ as the remaining terms are of similar or smaller orders.  
Note that
\begin{align}\label{j2}
\|J_2^{ijk}\|_2^2\leq &C\|\frac{1}{T}\sum_{t=k+1}^T\bR_1\bF_t'\bL_1'\ba_{1,\sbullet i}(\wh\ba_{01.\sbullet j}-\ba_{1,\sbullet j})'\bL_1\bF_{t-k}\bR_1'\|_2^2\notag\\
\leq&C\|\bR_1\|_2^4\|\bL_1\|_2^4\|\wh\bA_{0,1}-\bA_1\|_2^2\notag\\
\leq &\left\{\begin{array}{cc}
O_p(p_1^{2-2\delta_1}p_2^{2-2\delta_1}T^{-1}),&\text{if}\,\, \delta_1\leq \delta_2,\delta_2\leq1/2,\\
O_p(p_1^{2-2\delta_2}p_2^{2-2\delta_2}T^{-1}),&\text{if}\,\, \delta_1>\delta_2,\delta_2\leq 1/2,\\
O_p(p_1^{2-2\delta_1}p_2^{2-2\delta_1}T^{-1}),&\text{if}\,\, \delta_1\leq 1/2,\delta_2> 1/2,\\
O_p(p_1^2T^{-1}),&\text{if}\,\, \delta_1> 1/2,\delta_2> 1/2,
\end{array}\right.
\end{align}
which is the same as (\ref{j1:d}) if $p_1\asymp p_2$. Furthermore, 
if $p_1\asymp p_2$, the stochastic upper bounds in (\ref{j1:d}) are dominant in all terms of (\ref{sigh:w:d}). Lemma 6 follows from the assumptions that $k_0$, $r_1$ and $r_2$ are finite. This completes the proof. $\Box$

\begin{lemma}
If Assumptions 1--5 hold, then 
\[\lambda_{r_2}(\bM_{0,2}^*)\geq Cp_1^{2-2\delta_1}p_2^{2-2\delta_1}.\]
Furthermore, if $p_1^{\delta_1-\delta_2}p_2^{\delta_1}T^{-1/2}=o(1)$, then 
\[\|\wh\bM_{0,2}^*-\bM_{0,2}^*\|_2=\left\{\begin{array}{cc}
O_p(p_1^{2-2\delta_1}p_2^{2-2\delta_1}T^{-1/2}),&\text{if}\,\, \delta_1\leq \delta_2,\delta_2\leq1/2,\\
O_p(p_1^{2-\delta_1-\delta_2}p_2^{2-\delta_1-\delta_2}T^{-1/2}),&\text{if}\,\, \delta_1>\delta_2,\delta_2\leq 1/2,\\
O_p(p_1^{2-2\delta_1}p_2^{2-2\delta_1}T^{-1/2}),&\text{if}\,\, \delta_1\leq 1/2,\delta_2> 1/2,\\
O_p(p_2^{2-\delta_1}p_1^{1-\delta_1}T^{-1/2}),&\text{if}\,\, \delta_1> 1/2,\delta_2> 1/2.
\end{array}\right..\]
\end{lemma}
{\bf Proof.} The Proof is similar to those in Lemmas 3 and 4.  In fact, we only need to switch the positions between $p_1$ and $p_2$ in Lemma~\ref{lm3}. We omit the details. $\Box$

\begin{lemma}
Suppose Assumptions 1--5 hold. If $p_1\asymp p_2$ and $p_1^{\delta_1-\delta_2}p_2^{\delta_1-\delta_2}T^{-1/2}=o(1)$, then 
\[\|\wh\bP_{1,1}-\bP_1\|_2=\left\{\begin{array}{ll}
O_p(T^{-1/2}),&\text{if}\,\, \delta_1\leq \delta_2,\delta_2\leq1/2,\\
O_p(p_1^{\delta_1-\delta_2}p_2^{\delta_1-\delta_2}T^{-1/2}),&\text{if}\,\, \delta_1>\delta_2,\delta_2\leq 1/2,\\
O_p(T^{-1/2}),&\text{if}\,\, \delta_1\leq 1/2,\delta_2> 1/2,\\
O_p(p_2^{\delta_1}p_1^{\delta_1-1}T^{-1/2}),&\text{if}\,\, \delta_1> 1/2,\delta_2> 1/2.
\end{array}\right.\]

\end{lemma}

{\bf Proof.} By a similar argument as that of Lemma 5, we have 
\[\|\wh\bP_{1,1}-\bP_1\|_2\leq \frac{\|\wh\bM_{0,2}^*-\bM_{0,2}^*\|_2}{\lambda_{r_2}(\bM_{0,2})}.\]
Lemma 8 then follows from the results in  Lemma 7 and the above inequality.  This completes the proof. $\Box$

\begin{lemma}
Assume that Assumptions 1-5 hold.  We have
\[\sum_{i=1}^{p_2}\|\bOmega_{y_i}(\bQ_1\otimes\bB_1)\|_2^2=O_p(p_1^{2-2\delta_2}p_2^{2-2\delta_2}+p_2),\]
and
\begin{align}
\sum_{i=1}^{p_2}\|\wh\bOmega_{y_i}(\wh\bQ_1\otimes\wh\bB_1)-\bOmega_{y_i}(\bQ_1\otimes\bB_1)\|_2^2=& O_p\left((p_1^{2-2\delta_2}p_2^{2-2\delta_2}+p_2)(\|\wh\bQ_1-\bQ_1\|_2^2+\|\wh\bB_1-\bB_1\|_2^2)\right.\notag\\
&\left.+p_1^2p_2^2T^{-1}\right).
\end{align}

Consequently, 
\begin{align}
\|\wh\bS_1-\bS_1\|_2=&O_p\left(p_1^2p_2^2T^{-1}+p_1^{2-\delta_2}p_2^{2-\delta_2}T^{-1/2}+p_1p_2^{3/2}T^{-1/2}\right.\notag\\
&\left.+(p_1^{2-2\delta_2}p_2^{2-2\delta_2}+p_1^{1-\delta_2}p_2^{3/2-\delta_2}+p_2)(\|\wh\bQ_1-\bQ_1\|_2+\|\wh\bB_1-\bB_1\|_2)\right).
\end{align}
\end{lemma}
{\bf Proof.} The proof can be carried out by a similar argument  as those of Lemma 2 and Lemma 4 in the supplement of \cite{gaotsay2021c}. We omit the details. $\Box$
\\

{\bf Proof of Theorem 1.}  By Lemma 5,  we start with a non-random matrix $\bP_{0,1}$ and construct $\wh\bM_{0,1}^*$ by the projected data $\bY_t'\bP_{0,1}$.  The obtained $\wh\bA_{0,1}$ satisfies 
\begin{equation}\label{a01h}
\|\wh\bA_{0,1}-\bA_1\|_2=\left\{\begin{array}{ll}
O_p(T^{-1/2}),&\text{if}\,\, \delta_1\leq \delta_2,\delta_2\leq1/2,\\
O_p(p_1^{\delta_1-\delta_2}p_2^{\delta_1-\delta_2}T^{-1/2}),&\text{if}\,\, \delta_1>\delta_2,\delta_2\leq 1/2,\\
O_p(T^{-1/2}),&\text{if}\,\, \delta_1\leq 1/2,\delta_2> 1/2,\\
O_p(p_1^{\delta_1}p_2^{\delta_1-1}T^{-1/2}),&\text{if}\,\, \delta_1> 1/2,\delta_2> 1/2.
\end{array}\right.
\end{equation}
With $\wh\bA_{0,1}$ being random and by Lemma 8, we project the transpose of the data on to $\wh\bA_{0,1}$ as $\bY_t'\wh\bA_{0,1}$, and the estimator $\wh\bP_{1,1}$ satisfies
\begin{equation}\label{p11h}
\|\wh\bP_{1,1}-\bP_1\|_2=\left\{\begin{array}{ll}
O_p(T^{-1/2}),&\text{if}\,\, \delta_1\leq \delta_2,\delta_2\leq1/2,\\
O_p(p_1^{\delta_1-\delta_2}p_2^{\delta_1-\delta_2}T^{-1/2}),&\text{if}\,\, \delta_1>\delta_2,\delta_2\leq 1/2,\\
O_p(T^{-1/2}),&\text{if}\,\, \delta_1\leq 1/2,\delta_2> 1/2,\\
O_p(p_2^{\delta_1}p_1^{\delta_1-1}T^{-1/2}),&\text{if}\,\, \delta_1> 1/2,\delta_2> 1/2.
\end{array}\right.
\end{equation}
If we compare the stochastic bounds of the above two distances, the one for $\wh\bP_{1,1}$ simply switches the roles of $p_1$ and $p_2$ in that for $\wh\bA_{0,1}$. Therefore, if we continue the iteration and construct $\wh\bM_{1,1}^*$ based on $\bY_t\wh\bP_{1,1}$ with a given $\wh\bP_{1,1}$, by a similar argument as the proof for $\wh\bP_{1,1}$ with a given $\wh\bA_{0,1}$,  we can show that
\[\|\wh\bA_{1,1}-\bA_1\|_2=\left\{\begin{array}{ll}
O_p(T^{-1/2}),&\text{if}\,\, \delta_1\leq \delta_2,\delta_2\leq1/2,\\
O_p(p_1^{\delta_1-\delta_2}p_2^{\delta_1-\delta_2}T^{-1/2}),&\text{if}\,\, \delta_1>\delta_2,\delta_2\leq 1/2,\\
O_p(T^{-1/2}),&\text{if}\,\, \delta_1\leq 1/2,\delta_2> 1/2,\\
O_p(p_1^{\delta_1}p_2^{\delta_1-1}T^{-1/2}),&\text{if}\,\, \delta_1> 1/2,\delta_2> 1/2.
\end{array}\right.\]
which is the same as that in (\ref{a01h}) for $\wh\bA_{0,1}$.  Thus,  the obtained $\wh\bP_{2,1}$ with a given $\wh\bA_{1,1}$ also satisfies the stochastic bounds in (\ref{p11h}).  Consequently, the final estimators $\wh\bA_1$ and $\wh\bP_1$ satisfy the results in (\ref{a01h}) and (\ref{p11h}), respectively. This proved the first part of Theorem 1.\\

For the second part, by a similar argument as Lemma 4, we have 
\[\lambda_{k_1}(\bS_1)\geq Cp_1^{2-2\delta_2}p_2^{2-2\delta_2}.\]
Therefore, we apply the matrix perturbation theory again as that in Lemma 5,
\begin{align*}
\|\wh\bA_2-\bA_2\|_2\leq\frac{\|\wh\bS_1-\bS_1\|_2}{\lambda_{k_1}(\bS_1)}=&O_p\left(p_1^{\delta_2}p_2^{\delta_2}T^{-1/2}+p_1^{-1+2\delta_2}p_2^{-1/2+2\delta_2}T^{-1/2}\right.\\
&\left.+(1+p_1^{-1+\delta_2}p_2^{-1/2+\delta_2}+p_1^{-2+2\delta_2}p_2^{-1+2\delta_2})(\|\wh\bQ_1-\bQ_1\|_2+\|\wh\bB_1-\bB_1\|_2)\right).
\end{align*}
Thus,  the orthogonal matrix $\wh\bB_2$ also satisfies 
\begin{align}
\|\wh\bB_2-\bB_2\|_2\leq\frac{\|\wh\bS_1-\bS_1\|_2}{\lambda_{k_1}(\bS_1)}=&O_p\left(p_1^{\delta_2}p_2^{\delta_2}T^{-1/2}+p_1^{-1+2\delta_2}p_2^{-1/2+2\delta_2}T^{-1/2}\right.\\
&\left.+(1+p_1^{-1+\delta_2}p_2^{-1/2+\delta_2}+p_1^{-2+2\delta_2}p_2^{-1+2\delta_2})(\|\wh\bQ_1-\bQ_1\|_2+\|\wh\bB_1-\bB_1\|_2)\right).\notag
\end{align}
By a similar argument as above and switching the roles of $p_1$ and $p_2$, we can show that
\begin{align}
\|\wh\bQ_2-\bQ_2\|_2=&O_p\left(p_2^{\delta_2}p_1^{\delta_2}T^{-1/2}+p_2^{-1+2\delta_2}p_1^{-1/2+2\delta_2}T^{-1/2}\right.\\
&\left.+(1+p_2^{-1+\delta_2}p_1^{-1/2+\delta_2}+p_2^{-2+2\delta_2}p_1^{-1+2\delta_2})(\|\wh\bQ_1-\bQ_1\|_2+\|\wh\bB_1-\bB_1\|_2)\right).\notag
\end{align}
Combining with the results in (\ref{a01h}) and (\ref{p11h}), we can obtain that 
\begin{equation}\label{b2h}
\|\wh\bB_{2}-\bB_2\|_2=\left\{\begin{array}{ll}
O_p(p_1^{\delta_2}p_2^{\delta_2}T^{-1/2}),&\text{if}\,\, \delta_1\leq \delta_2,\delta_2\leq1/2,\\
O_p(p_1^{\delta_2}p_2^{\delta_2}T^{-1/2}+p_1^{\delta_1-\delta_2}p_2^{\delta_1-\delta_2}T^{-1/2}),&\text{if}\,\, \delta_1>\delta_2,\delta_2\leq 1/2,\\
O_p(p_1^{\delta_2}p_2^{\delta_2}T^{-1/2}+p_1^{-1+2\delta_2}p_2^{-1/2+2\delta_2}T^{-1/2}),&\text{if}\,\, \delta_1\leq 1/2,\delta_2> 1/2,\\
O_p(p_1^{\delta_1}p_2^{\delta_1-1}T^{-1/2}+p_1^{\delta_2}p_2^{\delta_2}T^{-1/2}+p_1^{-1+2\delta_2}p_2^{-1/2+2\delta_2}T^{-1/2}),&\text{if}\,\, \delta_1> 1/2,\delta_2> 1/2,
\end{array}\right.
\end{equation}
and
\begin{equation}\label{q2h}
\|\wh\bQ_{2}-\bQ_2\|_2=\left\{\begin{array}{ll}
O_p(p_1^{\delta_2}p_2^{\delta_2}T^{-1/2}),&\text{if}\,\, \delta_1\leq \delta_2,\delta_2\leq1/2,\\
O_p(p_1^{\delta_2}p_2^{\delta_2}T^{-1/2}+p_1^{\delta_1-\delta_2}p_2^{\delta_1-\delta_2}T^{-1/2}),&\text{if}\,\, \delta_1>\delta_2,\delta_2\leq 1/2,\\
O_p(p_1^{\delta_2}p_2^{\delta_2}T^{-1/2}+p_2^{-1+2\delta_2}p_1^{-1/2+2\delta_2}T^{-1/2}),&\text{if}\,\, \delta_1\leq 1/2,\delta_2> 1/2,\\
O_p(p_2^{\delta_1}p_1^{\delta_1-1}T^{-1/2}+p_1^{\delta_2}p_2^{\delta_2}T^{-1/2}+p_2^{-1+2\delta_2}p_1^{-1/2+2\delta_2}T^{-1/2}),&\text{if}\,\, \delta_1> 1/2,\delta_2> 1/2.
\end{array}\right.
\end{equation}
This completes the proof. $\Box$

{\bf Proof of Theorem 2.} By the definition of $\wh\bX_t$ and the Model (2.1),   we can verify that 
\[\wh\bX_t=\bH_L\bX_t\bH_R'+(\wh\bB_2^*{'}\wh\bA_1)^{-1}\wh\bB_2^*{'}\bE_t\wh\bQ_2^*(\wh\bP_2'\wh\bQ_2^*)^{-1},\]
where $\bH_L=(\wh\bB_2^*{'}\wh\bA_1)^{-1}\wh\bB_2^*{'}\bA_1$ and $\bH_R=(\wh\bQ_2^*{'}\wh\bP_1)^{-1}\wh\bQ_2^*{'}\bP_1$. Note that $\bE_t$ can be expressed as $\bE_t=\bA_2\bD_2^{1/2}\bU_2'\bxi_t\bV_2\bLambda_2^{1/2}\bP_2'+\bfeta_t$ where $\bA_2,\bD_2,\bU_2,\bV_2,\bLambda_2$, and $\bP_2$ are  defined in Section 2.1 of the main text,  then
\begin{align}\label{pi4:dec}
\wh\bX_t-\bH_L\bX_t\bH_R'=&(\bC_1'\wh\bB_2'\wh\bA_1)^{-1}\bC_1'\wh\bB_2'\bA_{2}\bD_{2}^{1/2}\bU_{2}'\bxi_t\bV_{2}\bLambda_{2}^{1/2}\bP_{2}'\wh\bQ_2\bC_2(\wh\bP_1'\wh\bQ_2\bC_2)^{-1}\notag\\
&+(\bC_1'\wh\bB_2'\wh\bA_1)^{-1}\bC_1'\wh\bB_2'\bfeta_t\wh\bQ_2\bC_2(\wh\bP_1'\wh\bQ_2\bC_2)^{-1}\notag\\
=:&I+II,
\end{align}
 where $\bC_1$ and $\bC_2$ are defined in Section 2.2.3 such that $\wh\bB_2^*=\wh\bB_2\bC_1$ and $\wh\bQ_2^*=\wh\bQ_2\bC_2$ are the estimated subspace matrices. By a similar argument as the proof of Theorem 3 in \cite{gaotsay2021c}, we can show that
 \begin{align}\label{pi41:nm}
 p_1^{-1/2}p_2^{-1/2}\|I\|_2\leq& Cp_1^{-1/2}p_2^{-1/2}\|\wh\bB_2'\bA_{2}\bD_{1}^{1/2}\|_2\|\bU_{2}'\bxi_t\bV_{2}\|_2\|\bLambda_{1}^{1/2}\bP_{2}'\wh\bQ_2\|_2\notag\\
 \leq & Cp_1^{-\delta_2/2}p_2^{-\delta_2/2}\|\wh\bB_2-\bB_2\|_2 \|\wh\bQ_2-\bQ_2\|_2,
 \end{align}
 where we use the fact that $\bU_2'\bxi_t\bV_2$ is a $k_1\times k_2$ random matrix of finite norms. 
 
 For $II$, note that it can be bounded by
 \begin{align}\label{II:xt}
  II\leq& \|\bC_2'(\wh\bB_2-\bB_2)'\bfeta_t(\wh\bQ_2-\bQ_2)\bC_2\|_2+\|\bC_2'(\wh\bB_2-\bB_2)'\bfeta_t\bQ_2\bC_2\|_2\notag\\
  &+\|\bC_1'\bB_2'\bfeta_t(\wh\bQ_2-\bQ_2)\bC_2\|_2+\|\bC_1'\bB_2'\bfeta_t\bQ_2\bC_2\|_2\notag\\
  =&II_1+II_2+II_3+II_4.
 \end{align}
 According to Assumption 5, we can treat $\bC_1$ as  non-random orthonormal matrices.  It is not hard to show that
 \[II_1\leq \|\wh\bB_2-\bB_2\|_2\|\bfeta_t\|_2\|\wh\bQ_2-\bQ\|_2=O_p(p_1^{1/2}p_2^{1/2} \|\wh\bB_2-\bB_2\|_2\|\wh\bQ_2-\bQ\|_2),\]
  \[II_2\leq \|\wh\bB_2-\bB_2\|_2\|\Var(\vc(\bfeta_t\bQ_2\bC_2))\|_2=O_p(p_1^{1/2} \|\wh\bB_2-\bB_2\|_2),\]
 \[II_3\leq \|\Var(\vc(\bC_1'\bB_2'\bfeta_t))\|_2\|\wh\bQ_2-\bQ\|_2=O_p(p_2^{1/2}\|\wh\bQ_2-\bQ\|_2),\]
and
 \[II_4\leq \|\Var(\vc(\bC_1'\bB_2'\bfeta_t\bQ_2\bC_2))\|_2=O_p(1).\]
 Therefore, 
 \[p_1^{-1/2}p_2^{-1/2}\|II\|_2=O_p(\|\wh\bB_2-\bB_2\|_2\|\wh\bQ_2-\bQ\|_2+p_2^{-1/2} \|\wh\bB_2-\bB_2\|_2+p_1^{-1/2}\|\wh\bQ_2-\bQ\|_2+(p_1p_2)^{-1/2}).\]
 Combining with (\ref{pi41:nm}), we obtain that
\begin{equation}\label{pi4:st}
\frac{1}{\sqrt{p_1p_2}}\|\wh\bX_t-\bH_L\bX_t\bH_R'\|_2\leq O_p(\|\wh\bB_2-\bB_2\|_2\|\wh\bQ_2-\bQ\|_2+p_2^{-1/2} \|\wh\bB_2-\bB_2\|_2+p_1^{-1/2}\|\wh\bQ_2-\bQ\|_2+(p_1p_2)^{-1/2}).
\end{equation}
This completes the proof. $\Box$

{\bf Proof of Theorem 3.} (i) The proof of Theorem 3(i) concerning the overall consistency of the extracted common component is similar to that of Theorem 3 in the supplement of \cite{gaotsay2021c}.  In fact, it is the same as (\ref{x:dif}) below.  We omit the details here and the proof can be built on  (\ref{x:dif}) below. \\
(ii) Similar to the technique used in the proofs above,  we assume $\wh\bA_1$ and $\wh\bP_1$ are the estimators of $\bA_1$ and $\bP_1$ ignoring the rotation matrices because they are already incorporated into the true ones by the construction method in Lemma 3 of \cite{LamYaoBathia_Biometrika_2011}.   By the expressions of $\bH_L$ and $\bH_R$ in Theorem~2, it can be shown that
\[\|\bH_L-\bI_{r_1}\|_2\leq C\|\wh\bA_1-\bA_1\|_2\,\,\text{and}\,\, \|\bH_R-\bI_{r_2}\|_2\leq C\|\wh\bP_1-\bP_1\|_2.\]
Consequently, by an elementary argument, we have
\begin{align*}
\wh\bX_t-\bX_t=&\wh\bX_t-\bH_L\bX_t\bH_R'+\bH_L\bX_t\bH_R'-\bH_L\bX_t+\bH_L\bX_t-\bX_t\\
=&\wh\bX_t-\bH_L\bX_t\bH_R'+\bH_L\bX_t(\bH_R-\bI_{r_2})'+(\bH_L-\bI_{r_1})\bX_t.
\end{align*}
Then, it follows that
\begin{align}\label{x:dif}
\frac{1}{\sqrt{p_1p_2}}\|\wh\bX_t-\bX_t\|_2\leq &\frac{1}{\sqrt{p_1p_2}}\|\wh\bX_t-\bH_L\bX_t\bH_R'\|_2+\frac{1}{\sqrt{p_1p_2}}\|\bH_L\bX_t\|_2\|\bH_R-\bI_{r_2}\|_2\notag\\
&+\frac{1}{\sqrt{p_1p_2}}\|\bH_L-\bI_{r_1}\|_2\|\bX_t\|_2\notag\\
\leq &C\left(\|\wh\bB_2-\bB_2\|_2\|\wh\bQ_2-\bQ\|_2+p_1^{-1/2}\|\wh\bB_2-\bB_2\|_2+p_2^{-1/2}\|\wh\bQ_2-\bQ_2\|_2+(p_1p_2)^{-1/2}\right.\notag\\
&\left.+(p_1p_2)^{-\delta_1/2}\|\wh\bA_1-\bA_1\|_2+(p_1p_2)^{-\delta_1/2}\|\wh\bP_1-\bP_1\|_2\right).
\end{align}
For simplicity, we denote the upper bound of $\|\wh\bX_t-\bX_t\|_2$ as $\omega$, which is equal to the rate on the right-hand side of (\ref{x:dif}) multiplied by $\sqrt{p_1p_2}$.  We consider the difference between the $(i,j)$-th elements of $\wh\bA_1\wh\bX_t\wh\bP_1$ and $\bA_1\bX_t\bP_1$,
\begin{align}\label{axp:ij}
\wh\ba_{1,i}'\wh\bX_t\wh\bp_{1,j}-\ba_{1,i}'\bX_t\bp_{1,j}=&(\wh\ba_{1,i}-\ba_{1,i})'(\wh\bX_t-\bX_t)(\wh\bp_{1,j}-\bp_{1,j})+(\wh\ba_{1,i}-\ba_{1,i})'(\wh\bX_t-\bX_t)\bp_{1,j}\notag\\
&+(\wh\ba_{1,i}-\ba_{1,i})'\bX_t(\wh\bp_{1,j}-\bp_{1,j})+(\wh\ba_{1,i}-\ba_{1,i})'\bX_t\bp_{1,j}\\
&+\ba_{1,i}'(\wh\bX_t-\bX_t)(\wh\bp_{1,j}-\bp_{1,j})+\ba_{1,i}'(\wh\bX_t-\bX_t)\bp_{1,j}+\ba_{1,i}'\bX_t(\wh\bp_{1,j}-\bp_{1,j}).\notag
\end{align}
Noting that $\max_{1\leq i\leq p_1}\|\wh\ba_{1,i}-\ba_{1,i}\|_2\leq \|\wh\bA_1-\bA_1\|_2$, $\max_{1\leq j\leq p_2}\|\wh\bp_{1,j}-\bp_{1,j}\|_2\leq \|\wh\bP_1-\bP_1\|_2$,  $\|\ba_{1,i}\|_2\asymp p_1^{-1/2}$, and $\|\bp_{1,j}\|_2\asymp p_2^{-1/2}$,  we obtain that
\begin{align}\label{axp:ij:rate}
\left|\wh\ba_{1,i}'\wh\bX_t\wh\bp_{1,j}-\ba_{1,i}'\bX_t\bp_{1,j}\right|\leq & C\omega\|\wh\bA_1-\bA_1\|_2\|\wh\bP_1-\bP_1\|_2+Cp_2^{-1/2}\omega\|\wh\bA_1-\bA_1\|_2\\
&+C(p_1p_2)^{(1-\delta_1)/2}\|\wh\bA_1-\bA_1\|_2\|\wh\bP_1-\bP_1\|_2+Cp_1^{(1-\delta_1)/2}p_2^{-\delta_1/2}\|\wh\bA_1-\bA_1\|_2\notag\\
&+Cp_1^{-1/2}\omega\|\wh\bP_1-\bP_1\|_2+p_1^{-1/2}p_2^{-1/2}\omega+Cp_1^{-\delta_1/2}p_2^{(1-\delta_1)/2}\|\wh\bP_1-\bP_1\|_2.\notag
\end{align}
In particular, if $\delta_1=\delta_2=0$,  we can obtain from (\ref{axp:ij:rate}) that
\begin{equation}\label{axp:dlt0}
\left|\wh\ba_{1,i}'\wh\bX_t\wh\bp_{1,j}-\ba_{1,i}'\bX_t\bp_{1,j}\right|\leq C\sqrt{\frac{p_1}{T}}+C\sqrt{\frac{p_2}{T}}+C\frac{\sqrt{p_1p_2}}{T}+C\frac{1}{\sqrt{p_1p_2}}.
\end{equation}
This completes the proof. $\Box$

\begin{remark}
(i) Under the additional conditions in Assumption 7 of the main text,  we can obtain that $\max_{1\leq i\leq p_1}\|\wh\ba_{1,i}-\ba_{1,i}\|_2\leq C \sqrt{\frac{1}{Tp_1}}$ and $\max_{1\leq j\leq p_2}\|\wh\bp_{1,j}-\bp_{1,j}\|_2\leq C\sqrt{\frac{1}{Tp_2}}$ when $\delta_1=\delta_2=0$,  as shown in Theorem 5 below. Consequently, we can improve the rates in (\ref{axp:dlt0}) as
\begin{equation}\label{axp:dlt0:ad}
\left|\wh\ba_{1,i}'\wh\bX_t\wh\bp_{1,j}-\ba_{1,i}'\bX_t\bp_{1,j}\right|=O_p(\sqrt{\frac{1}{T}}+\sqrt{\frac{1}{p_1p_2}}),
\end{equation}
which is in line with the results of Theorem 3 in \cite{Bai_Econometrica_2003} if we treat the total dimension as $N=p_1p_2$ therein. In particular, if $p_1=1$ or $p_2=1$, the result in \ref{axp:dlt0:ad} is exactly the same as that in Theorem 3 of \cite{Bai_Econometrica_2003}. We also note that the non-asymptotic rate in (\ref{axp:dlt0:ad}) is slightly different from that in Theorem 3.5(2) of \cite{yu2022}.  The is due to that the essential sample sizes used in our paper and that in \cite{yu2022} are different.  In fact, the rate in (\ref{axp:dlt0:ad}) is equivalent to the one in Theorem 3.5(2) of \cite{yu2022} if we replace $T^{-1/2}$ by $(Tp_1)^{-1/2}+(Tp_2)^{-1/2}$ since the estimation of the common components involves the {estimated front} and back loading matrices.\\
(ii) When $\delta_1=0$ and $\delta_2=1$, the factors are {strong} and the noises are not prominent, it is not hard to obtain that the convergence rate is the same as that in (\ref{axp:dlt0:ad}),  and it is equivalent to the counterpart in \cite{yu2022} according to the discussion above, but we require {a stronger} condition $(p_1p_2)^3=o(T)$ because the estimation of $\bB_2$ and $\bQ_2$ is more difficult when there is no diverging components in the covariance of the noise terms.
\end{remark}

\vskip 0.2cm
\noindent
{\bf Proof of Theorem 4.} (i) The proof follows closely the arguments in the proof of Theorem 6 in \cite{gaotsay2018b}. Let $ u_{ij,t}=\bb_{1,i}'\bY_t\bq_{1,j}$ be the $(i,j)$-th element of $\bB_1'\bY_t\bQ_1$, where $\bb_{1,i}$ and $\bq_{1,j}$ are the $i$-th and $j$-th columns of $\bB_1$ and $\bQ_1$, respectively. By the proof of Theorem 3 in \cite{changyaozhou2017}, we only need to show  that
\begin{equation}\label{dif:uhat}
\frac{1}{T}\sum_{t=1}^T (\wh u_{ij,t}-u_{ij,t})^2=o_p(1),\quad 1\leq i\leq p_1-r_1,\,\,1\leq j\leq p_2-r_2,
\end{equation}
where $\wh u_{ij,t}=\wh\bb_{1,i}'\bY_t\wh\bq_{1,j}$ and $\wh\bb_{1,i}$ and $\wh\bq_{1,j}$ are the $i$-th and $j$-th column of $\wh\bB_1$ and $\wh\bQ_1$ if we ignore some orthogonal rotations. Note that
\[\wh\bb_{1,i}'\bY_t\wh\bq_{1,j}=\wh\bb_{1,i}'\bA_1\bX_t\bP_1'\wh\bq_{1,j}+\wh\bb_{1,i}'\bE_t\wh\bq_{1,j},\,\,\text{and}\,\,\bb_{1,i}'\bY_t\bq_{1,j}=\bb_{1,i}'\bE_t\bq_{1,j}.\]
Then it suffices to show
\begin{equation}\label{gt1}
\frac{1}{T}\sum_{t=1}^T(\wh\bb_{1,i}'\bA_1\bX_t\bP_1'\wh\bq_{1,j})^2=o_p(1),
\end{equation}
and
\begin{equation}\label{gt2}
\frac{1}{T}\sum_{t=1}^T(\wh\bb_{1,i}'\bE_t\wh\bq_{1,j}-\bb_{1,i}'\bE_t\bq_{1,j})^2=o_p(1).
\end{equation}
Recall that $\bL_1=\bA_1\bW_1$ and $\bR_1=\bP_1\bG_1$ with $\|\bW_1\|_2\asymp p_1^{(1-\delta_1)/2}$ and $\|\bG_1\|_2\asymp p_2^{(1-\delta_1)/2}$, and since
\begin{equation*}
(\wh\bb_{1,i}'\bA_1\bX_t\bP_1'\wh\bq_{1,j})^2=(\wh\bq_{1,j}'\bP_1\bG_1\otimes\wh\bb_{1,i}'\bA_1\bW_1)\bff_t\bff_t'(\bG_1'\bP_1'\wh\bq_{1,j}\otimes\bW_1'\bA_1'\wh\bb_{1,i}),
\end{equation*}
then
\begin{align}\label{byp:hat}
\left|\frac{1}{T}\sum_{t=1}^n(\wh\bb_{1,i}'\bA_1\bX_t\bP_1'\wh\bq_{1,j})^2\right|\leq &C\|\wh\bQ_1-\bQ_1\|_2^2\|\wh\bB_1-\bB_1\|_2^2\|\bW_1\|_2^2\|\bG_1\|_2^2\notag\\
=&\left\{\begin{array}{ll}
O_p(p_1^{1-\delta_1}p_2^{1-\delta_1}T^{-2}),&\text{if}\,\, \delta_1\leq \delta_2,\delta_2\leq1/2,\\
O_p(p_1^{1+3\delta_1-4\delta_2}p_2^{1+3\delta_1-4\delta_2}T^{-2}),&\text{if}\,\, \delta_1>\delta_2,\delta_2\leq 1/2,\\
O_p(p_1^{1-\delta_1}p_2^{1-\delta_1}T^{-2}),&\text{if}\,\, \delta_1\leq 1/2,\delta_2> 1/2,\\
O_p(p_1^{3\delta_1-1}p_2^{3\delta_1-1}T^{-2}),&\text{if}\,\, \delta_1> 1/2,\delta_2> 1/2.
\end{array}\right.
\end{align}
On the other hand,
\[\wh\bb_{1,i}'\bE_t\wh\bq_{1,j}-\bb_{1,i}'\bE_t\bq_{1,j}=(\wh\bq_{1,j}'\otimes\wh\bb_{1,i}'-\bq_{1,j}'\otimes\bb_{1,i}')([\bR_2\otimes\bL_2]\vc(\bxi_t)+\vc(\bfeta_t)),\]
and then
\begin{align}\label{bypd:hat}
&\left|\frac{1}{T}\sum_{t=1}^T(\wh\bb_{1,i}'\bE_t\wh\bq_{1,j}-\bb_{1,i}'\bE_t\bq_{1,j})^2\right|\notag\\
\leq& (\|\wh\bb_{1,i}-\bb_{1,i}\|_2+\|\wh\bq_{1,j}-\bq_{1,j}\|_2)^2\|\bL_2\|_2^2\|\bR_2\|_2^2(\|\frac{1}{T}\sum_{t=1}^T\vc(\bxi_t)\vc(\bxi_t)'\|_2+\|\frac{1}{T}\sum_{t=1}^T\vc(\bfeta_t)\vc(\bfeta_t)'\|_2)\notag\\
\leq & C(\|\wh\bB_1-\bB_1\|_2^2+\|\wh\bQ_1-\bQ_1\|_2^2)\|\bL_2\|_2^2\|\bR_2\|_2^2\notag\\
=& \left\{\begin{array}{ll}
O_p(p_1^{1-\delta_2}p_2^{1-\delta_2}T^{-1}),&\text{if}\,\, \delta_1\leq \delta_2,\delta_2\leq1/2,\\
O_p(p_1^{1+2\delta_1-3\delta_2}p_2^{1+2\delta_1-3\delta_2}T^{-1}),&\text{if}\,\, \delta_1>\delta_2,\delta_2\leq 1/2,\\
O_p(p_1^{1-\delta_2}p_2^{1-\delta_2}T^{-1}),&\text{if}\,\, \delta_1\leq 1/2,\delta_2> 1/2,\\
O_p(p_1^{1+2\delta_1-\delta_2}p_2^{2\delta_1-\delta_2-1}T^{-1}+p_1^{2\delta_1-\delta_2-1}p_2^{1+2\delta_2-\delta_2}T^{-1}),&\text{if}\,\, \delta_1> 1/2,\delta_2> 1/2.
\end{array}\right.
\end{align}
where we use the property that $\|\frac{1}{T}\sum_{t=1}^T\vc(\bxi_t)\vc(\bxi_t)'\|_2$ and   $\|\frac{1}{T}\sum_{t=1}^T\vc(\bfeta_t)\vc(\bfeta_t)'\|_2$ are bounded since $\vc(\bxi_t)$ is sub-Gaussian; see the proof of Lemma 4 in \cite{gaotsay2018b} or Theorem 4.3.5 of \cite{vershynin2018}. Therefore, we require the rates in (\ref{byp:hat}) and (\ref{bypd:hat}) are of order $o(1)$ under each scenario of $(\delta_1,\delta_2)$.\\

(ii) To show the consistency of the test statistic in \cite{Tsay_2018}, let $\wh\bb_{1,i}$ and $\wh\bq_{1,j}$ be the $i$-th and $j$-th column of $\wh\bB_1$ and $\wh\bQ_1$ if we ignore some orthogonal rotations.   Then
\begin{align}\label{byq-d}
\wh\bb_{1,i}'\bY_{t}\wh\bq_{1,j}=&\wh\bb_{1,i}'\bL_1\bF_t\bR_1'\wh\bq_{1,j}+\wh \bb_{1,i}'\bE_t \wh \bq_{1,j}\notag\\
&=\wh\bb_{1,i}'\bL_1\bF_t\bR_1'\wh\bq_{1,j}+(\wh\bb_{1,i}-\bb_{1,i})'\bE_t\wh\bq_{1,j}+\bb_{1,i}'\bE_t(\wh\bq_{1,j}-\bq_{1,j})+\bb_{1,i}'\bE_t\bq_{1,j}'\notag\\
=&:\alpha_1+\alpha_2+\alpha_3+\alpha_4.
\end{align}
We require that the  effect of $\alpha_1,\alpha_2$, and $\alpha_3$ on the white noise term $\alpha_4$ is asymptotically negligible.
By Assumptions 3, 4, and 6, we can show that for any unit vector $\bv_1\in\mathbb{R}^{r_1r_2}$,  $\bv_2\in\mathbb{R}^{k_1k_2}$, and $\bv_3in R^{p_1p_2}$ such that 
\[P(\max_{1\leq t\leq T}|\bv_1'\vc(\bF_t)|>x)\leq CT\exp(-Cx^2),\]
\[P(\max_{1\leq t\leq T}|\bv_2'\vc(\bxi_t)|>x)\leq CT\exp(-Cx^2),\]
and
\[P(\max_{1\leq t\leq T}|\bv_3'\vc(\bfeta_t)|>x)\leq CT\exp(-Cx^2),\]
which can be used to obtain the maximal magnitude of each random variables.

First, it is not hard to see that
\begin{align}\label{alp:1}
\max_{1\leq i\leq v_1,1\leq j\leq v_2}\max_{1\leq t\leq T}|\alpha_1|\leq& C(\|\wh\bB_1-\bB_1\|_2\|\wh\bQ_1-\bQ_1\|_2)\max_{1\leq t\leq T}\|\bL_1\bF_t\bR_1\|_2\notag\\
\leq &\left\{\begin{array}{ll}
O_p(p_1^{(1-\delta_1)/2}p_2^{(1-\delta_1)/2}T^{-1}\sqrt{\log(T)}),&\text{if}\,\, \delta_1\leq \delta_2,\delta_2\leq1/2,\\
O_p(p_1^{1/2+3\delta_1/2-2\delta_2}p_2^{1/2+3\delta_1/2-2\delta_2}T^{-1}\sqrt{\log(T)}),&\text{if}\,\, \delta_1>\delta_2,\delta_2\leq 1/2,\\
O_p(p_1^{1/2-\delta_1/2}p_2^{1/2-\delta_1/2}T^{-1}\sqrt{\log(T)}),&\text{if}\,\, \delta_1\leq 1/2,\delta_2> 1/2,\\
O_p(p_1^{3\delta_1/2-1/2}p_2^{3\delta_1/2-1/2}T^{-1}\sqrt{\log(T)}),&\text{if}\,\, \delta_1> 1/2,\delta_2> 1/2.
\end{array}\right.
\end{align}

Similarly, we can show that
\begin{align*}
\max_{1\leq i\leq v_1,1\leq j\leq v_2}\max_{1\leq t\leq T}\|\alpha_2\|_2\leq& C(\|\wh\bB_1-\bB_1\|_2)(\|\bL_2\|_2\|\bR_2\|_2+\sqrt{p_1p_2\log(Tp_1p_2)})\notag\\
\leq& \left\{\begin{array}{ll}
O_p(p_1^{1/2}p_2^{1/2}T^{-1/2}\sqrt{\log(Tp_1p_2)}),&\text{if}\,\, \delta_1\leq \delta_2,\delta_2\leq1/2,\\
O_p(p_1^{1/2+\delta_1-\delta_2}p_2^{1/2+\delta_1-\delta_2}T^{-1/2}\sqrt{\log(Tp_1p_2)}),&\text{if}\,\, \delta_1>\delta_2,\delta_2\leq 1/2,\\
O_p(p_1^{1/2}p_2^{1/2}T^{-1/2}\sqrt{\log(Tp_1p_2)}),&\text{if}\,\, \delta_1\leq 1/2,\delta_2> 1/2,\\
O_p(p_1^{\delta_1+1/2}p_2^{\delta_1-1/2}T^{-1/2}\sqrt{\log(Tp_1p_2)}),&\text{if}\,\, \delta_1> 1/2,\delta_2> 1/2,
\end{array}\right.
\end{align*}
and 
\[\max_{1\leq i\leq v_1,1\leq j\leq v_2}\max_{1\leq t\leq T}\|\alpha_3\|_2\leq  \left\{\begin{array}{ll}
O_p(p_1^{1/2}p_2^{1/2}T^{-1/2}\sqrt{\log(Tp_1p_2)}),&\text{if}\,\, \delta_1\leq \delta_2,\delta_2\leq1/2,\\
O_p(p_1^{1/2+\delta_1-\delta_2}p_2^{1/2+\delta_1-\delta_2}T^{-1/2}\sqrt{\log(Tp_1p_2)}),&\text{if}\,\, \delta_1>\delta_2,\delta_2\leq 1/2,\\
O_p(p_1^{1/2}p_2^{1/2}T^{-1/2}\sqrt{\log(Tp_1p_2)}),&\text{if}\,\, \delta_1\leq 1/2,\delta_2> 1/2,\\
O_p(p_2^{\delta_1+1/2}p_1^{\delta_1-1/2}T^{-1/2}\sqrt{\log(Tp_1p_2)}),&\text{if}\,\, \delta_1> 1/2,\delta_2> 1/2.
\end{array}\right.\]
If $p_1\asymp p_2$, the asymptotic rate of $\alpha_2$ is the same as that of $\alpha_3$. Therefore, we only need to require the rates in $\alpha_1$ and $\alpha_2$ are of order $o(1)$ under each scenario of $(\delta_1,\delta_2)$.
This completes the proof. $\Box$

 {\bf Proof of Theorem 5.} Suppose we have the estimators $\wh\bA_1$ and $\wh\bP_1$ obtained by our proposed method, and they satisfy the properties in Theorem 1 of the main text.  We only show the limiting distributions of the estimated row loadings of $\wh\bA_1$ since it is similar for those of $\wh\bP_1$.  Let 
 \[\wh\bZ_t=\bY_t\wh\bP_1=\bA_1\bX_t\bP_1'\wh\bP_1+\bE_t\wh\bP_1.\]
 According to Theorem 1, each column of $\wh\bP_1$ is convergent, and we denote the limit of $\wh\bp_{1,\sbullet i}$ as $\bbeta_{1,\sbullet i}$ for $1\leq i\leq r_2$.  Then, the $i$-th column of $\wh\bZ_t$ is
 \[\wh\bz_{i,t}=\bY_t\wh\bp_{1\sbullet i}=\bA_1\bX_t\bP_1'\wh\bp_{1,\sbullet i}+\bE_t\wh\bp_{1,\sbullet i}.\]
 By definition, 
 \[\wh\bSigma_{z,ij}(k)=\frac{1}{T}\sum_{t=k+1}^T \wh\bz_{i,t}\wh\bz_{j,t-k}'=\frac{1}{T}\sum_{t=k+1}^T\bY_t\wh\bp_{1,\sbullet i}\wh\bp_{1,\sbullet j}'\bY_{t-k}',\]
 then,
 \begin{align}\label{m1:eigen}
  \wh\bM_1:=&\sum_{k=1}^{k_0}\sum_{i=1}^{r_2}\sum_{j=1}^{r_2}\wh\bSigma_{z,ij}(k)\wh\bSigma_{z,ij}(k)'\notag\\
  =&\sum_{k=1}^{k_0}\sum_{i=1}^{r_2}\sum_{j=1}^{r_2}\frac{1}{T}\sum_{t=k+1}^T(\bA_1\bX_t\bP_1'\wh\bp_{1,\sbullet i}+\bE_t\wh\bp_{1,\sbullet i})\wh\bp_{1,\sbullet j}'\bY_{t-k}'\frac{1}{T}\sum_{t=k+1}^T\bY_{t-k}\wh\bp_{1,\sbullet j}\wh\bp_{1,\sbullet i}'\bY_{t}'\notag\\
  =&\sum_{k=1}^{k_0}\sum_{i=1}^{r_2}\sum_{j=1}^{r_2}\frac{1}{T}\sum_{t=k+1}^T\bA_1\bX_t\bP_1'\wh\bp_{1,\sbullet i}\wh\bp_{1,\sbullet j}'\bY_{t-k}'\frac{1}{T}\sum_{t=k+1}^T\bY_{t-k}\wh\bp_{1,\sbullet j}\wh\bp_{1,\sbullet i}'\bY_{t}'\notag\\
  &+\sum_{k=1}^{k_0}\sum_{i=1}^{r_2}\sum_{j=1}^{r_2}\frac{1}{T}\sum_{t=k+1}^T\bE_t\wh\bp_{1,\sbullet i}\wh\bp_{1,\sbullet j}'\bY_{t-k}'\frac{1}{T}\sum_{t=k+1}^T\bY_{t-k}\wh\bp_{1,\sbullet j}\wh\bp_{1,\sbullet i}'\bY_{t}'.
 \end{align}
 Since $\wh\bA_1$ consists of the eigenvectors associated with the top $r$ eigenvalues of $\wh\bM_1$, then
 \begin{equation}\label{m1:eigen:1}
 \wh\bM_1\wh\bA_1=\wh\bA_1\wh\bV_1,
 \end{equation}
 where $\wh\bV_1$ is a diagonal matrix consisting of the top $r$ eigenvalues of $\wh\bM_1$ as its diagonal elements.  Moreover, it is not hard to see that $\wh\bV_1\asymp (p_1p_2)^2$ by Assumptions 1--4.  By a similar argument as that in \cite{lamyao2012}, there exists a diagonal matrix $\bV_1$ such that $\wh\bV_1/(p_1p_2)^2\rightarrow_p \bV_1$ where $\bV_1\asymp O(1)$. It follows from (\ref{m1:eigen}) and (\ref{m1:eigen:1}) that
 \begin{align}\label{a1hat:exp}
 \wh\bA_1=\wh\bM_1\wh\bA_1\wh\bV_1^{-1}=&\sum_{k=1}^{k_0}\sum_{i=1}^{r_2}\sum_{j=1}^{r_2}\frac{1}{T}\sum_{t=k+1}^T\bA_1\bX_t\bP_1'\wh\bp_{1,\sbullet i}\wh\bp_{1,\sbullet j}'\bY_{t-k}'\frac{1}{T}\sum_{t=k+1}^T\bY_{t-k}\wh\bp_{1,\sbullet j}\wh\bp_{1,\sbullet i}'\bY_{t}'\wh\bA_1\wh\bV_1^{-1}\notag\\
  &+\sum_{k=1}^{k_0}\sum_{i=1}^{r_2}\sum_{j=1}^{r_2}\frac{1}{T}\sum_{t=k+1}^T\bE_t\wh\bp_{1,\sbullet i}\wh\bp_{1,\sbullet j}'\bY_{t-k}'\frac{1}{T}\sum_{t=k+1}^T\bY_{t-k}\wh\bp_{1,\sbullet j}\wh\bp_{1,\sbullet i}'\bY_{t}'\wh\bA_1\wh\bV_1^{-1}.
 \end{align}
 Let
 \begin{equation}\label{H1:rot}
 \bH_{1,T}'=\sum_{k=1}^{k_0}\sum_{i=1}^{r_2}\sum_{j=1}^{r_2}\frac{1}{T}\sum_{t=k+1}^T\bX_t\bP_1'\wh\bp_{1,\sbullet i}\wh\bp_{1,\sbullet j}'\bY_{t-k}'\frac{1}{T}\sum_{t=k+1}^T\bY_{t-k}\wh\bp_{1,\sbullet j}\wh\bp_{1,\sbullet i}'\bY_{t}'\wh\bA_1\wh\bV_1^{-1},
 \end{equation}
 be a rotation matrix,  it is not hard to see that $\bH_{1,T}=O_p(1)$ and $\bH_{1,T}^{-1}=O_p(1)$ according to the proofs of Lemma \ref{lm2} and Theorem 1 above. Then,
 \begin{equation}\label{ahat:minus}
 \wh\bA_1-\bA_1\bH_{1,T}'=\sum_{k=1}^{k_0}\sum_{i=1}^{r_2}\sum_{j=1}^{r_2}\frac{1}{T}\sum_{t=k+1}^T\bE_t\wh\bp_{1,\sbullet i}\wh\bp_{1,\sbullet j}'\bY_{t-k}'\frac{1}{T}\sum_{t=k+1}^T\bY_{t-k}\wh\bp_{1,\sbullet j}\wh\bp_{1,\sbullet i}'\bY_{t}'\wh\bA_1\wh\bV_1^{-1}.
 \end{equation}
 By the proof of Theorem 1,  we have two important results. First, under the assumption that $\bE_t$ is a white noise, we can show that
 \begin{equation}\label{a1hat:rate:w}
 \|\wh\bA_1-\bA_1\bH_{1,T}'\|_2=O_p(T^{-1/2}),\,\,\text{if}\,\, \delta_1=\delta_2=0,
 \end{equation}
 which is the same as the result in Theorem 1. In addition, if $\bfeta_t$ is serially correlated and so is the idiosyncratic term $\bE_t$, we can show that there is an additional term in the error rate as follows:
  \begin{equation}\label{a1hat:rate:s}
 \|\wh\bA_1-\bA_1\bH_{1,T}'\|_2=O_p((p_1p_2)^{-1}+T^{-1/2}),\,\,\text{if}\,\, \delta_1=\delta_2=0.
 \end{equation}
 This issue is discussed in Section~\ref{sec3:s} below on the factor model with serially correlated idiosyncratic noises. We omit the details here. 
 
 For each $1\leq l\leq p_1$, (\ref{ahat:minus}) implies that each row vector can be expresses as
 \begin{equation}\label{al:rate}
 \wh\ba_{1,l\sbullet}-\bH_{1,T}\ba_{1,l\sbullet}=\wh\bV_1^{-1}\wh\bA_1'\sum_{k=1}^{k_0}\sum_{i=1}^{r_2}\sum_{j=1}^{r_2}\frac{1}{T}\sum_{t=k+1}^T\bY_t\wh\bp_{1,\sbullet i}\wh\bp_{1,\sbullet j}'\bY_{t-k}'\frac{1}{T}\sum_{t=k+1}^T\bY_{t-k}\wh\bp_{1,\sbullet j}\wh\bp_{1,\sbullet i}'\be_{l\sbullet,t}.
 \end{equation}
 One immediate result from (\ref{al:rate}) is that
 \begin{equation}\label{a1:row:rate}
 \|\wh\ba_{1,l\sbullet}-\bH_{1,T}\ba_{1,l\sbullet}\|_2=O_p(\frac{1}{\sqrt{Tp_1}}),
 \end{equation}
 which can be shown by a similar argument as that in the proof of Lemma~\ref{lm2} above.
 
 By Theorem 1, there exists $\bbeta_{1,\sbullet i}$ such that $\wh\bp_{1,\sbullet i}\rightarrow_p \bbeta_{1,\sbullet i}$. By (\ref{a1hat:rate:w}),  (\ref{al:rate}) can be expressed as
   \begin{align}\label{al:rate:app}
 \sqrt{p_1T}(\wh\ba_{1,l\sbullet}-&\bH_{1,T}\ba_{1,l\sbullet})\notag\\
 =&\frac{\sqrt{p_1T}}{(p_1p_2)^2}\bV_1^{-1}\bH_1\bA_1'\sum_{k=1}^{k_0}\sum_{i=1}^{r_2}\sum_{j=1}^{r_2}\frac{1}{T}\sum_{t=k+1}^T\bY_t\bbeta_{1,\sbullet i}\bbeta_{1,\sbullet j}'\bY_{t-k}'\frac{1}{T}\sum_{t=k+1}^T\bY_{t-k}\bbeta_{1,\sbullet j}\bbeta_{1,\sbullet i}'\be_{l\sbullet,t}+o_p(1).
 \end{align}
 By Assumption 7, we conclude that
 \[ \sqrt{p_1T}(\wh\ba_{1,l\sbullet}-\bH_{1,T}\ba_{1,l\sbullet})\rightarrow_d N({\bf 0}, \bV_1^{-1}\bH_1\bSigma_1\bH_1'\bV_1^{-1}).\]
The proof for $\sqrt{p_2T}(\wh\bp_{1,l\sbullet}-\bH_{2,T}\bp_{1,l\sbullet})$ is similar, and we omit the details.
This complete the proof. $\Box$

 \section{{Serially-Correlated Idiosyncratics}}\label{sec3:s}
 \subsection{Convergence in Vector Factor Models}
 In this section, we briefly {discuss the extension of the proposed method to extract dynamically dependent common factors} when the idiosyncratic noises are serially correlated.  Since the autocovariance-based eigenanalysis method is built on the factor modeling framework in \cite{LamYaoBathia_Biometrika_2011} and \cite{lamyao2012}, we start with the  factor models as outlined in \cite{LamYaoBathia_Biometrika_2011}, and the results for matrix-variate factor models can be similarly established. Our objective is to show the continued efficacy of the proposed method in the presence of serially correlated noises by employing an alternative proof technique.  
 
 By an abuse of notation, we consider the following factor model in \cite{lamyao2012} when the $p$-dimensional data $\by_t$ are centered:
 \begin{equation}\label{ft:ly}
 \by_t=\bQ\bff_t+\bve_t=\bA\bx_t+\bve_t,t=1,...,T,
 \end{equation}
 where $\bQ\in R^{p\times r}$ is the loading matrix, $\bff_t$ is an $r$-dimensional common factor process, and $\bve_t$ is a $p$-dimensional idiosyncratic term, which can be serially correlated.  To make Model (\ref{ft:ly}) identifiable, we rewrite $\bQ\bff_t$ as $\bA\bx_t$ with $\bA'\bA=\bI_r$,  and assume $\{\bx_t\}$ and $\{\bve_t\}$ are uncorrelated with each other.  We borrow the settings of the parameters in \cite{LamYaoBathia_Biometrika_2011} and \cite{lamyao2012} in this section and show their method {continues to work}.  Under the Assumption that $\bA'\bA=\bI_r$ and the conditions on the factor strengths in \cite{LamYaoBathia_Biometrika_2011}, we conclude that the strength of the factors $\|\bx_t\|_2\asymp p^{(1-\delta)/2}$, where $\delta\in [0,1)$ characterizes the factor strength.  We exclude the case when $\delta=1$ to distinguish the factors from the noises, and $\delta=0$ corresponds to the case when all factors are strong.
 
 {Letting} $\bSigma_y(k)=\Cov(\by_t,\by_{t-k})$,  $\bSigma_x(k)=\Cov(\bx_t,\bx_{t-k})$, and $\bSigma_{x\ve}(k)=\Cov(\bx_t,\bve_{t-k})={\bf 0}$ be the true covariance matrices, and
 \[\wh\bSigma_y(k)=\frac{1}{T}\sum_{t=k+1}^T\by_t\by_{t-k}, \wh\bSigma_x(k)=\frac{1}{T}\sum_{t=k+1}^T\bx_t\bx_{t-k}',\text{and}\,\,\wh\bSigma_{x\ve}(k)=\frac{1}{T}\sum_{t=k+1}^T\bx_t\bve_{t-k}',\]
 be {their sample counterparts},  we construct the following matrix
 \begin{equation}
 \wh\bM=\sum_{k=1}^{k_0}\wh\bSigma_y(k)\wh\bSigma_y(k)',
 \end{equation}
 which is the same as the one in (2.5) of the main article if $p_2=1$.  According to \cite{LamYaoBathia_Biometrika_2011},  the estimated loading matrix $\wh\bA$ of $\bA$ consists of the eigenvectors of $\wh\bM$ associated with the {leading} $r$ eigenvalues as its columns.
 {For simplicity,} we assume the number of factors $r$ is known.   We first provide some conditions on the factors and idiosyncratic noises.
 
\begin{condition}\label{c1}
The process $\{\bff_t\}$ is $\alpha$-mixing with the mixing coefficient satisfying the condition $\sum_{k=1}^\infty\alpha_p(k)^{1-2/\gamma}<\infty$ for some $\gamma>2$, where
\[\alpha_p(k)=\sup_{i}\sup_{A\in\mathcal{F}_{-\infty}^i,B\in \mathcal{F}_{i+k}^\infty}|P(A\cap B)-P(A)P(B)|,
\]
and $\mathcal{F}_i^j$ is the $\sigma$-field generated by $\{\bff_t:i\leq t\leq j\}$.
\end{condition}

\begin{condition}\label{c2}
The idiosyncratic term $\bve_t$ is stationary with $\|\Cov(\bve_t,\bve_{t-k})\|_2\leq C$  for $0\leq k\leq k_0$, and satisfies Conditions (A1)--(A3) of \cite{han2020}.
\end{condition}
 \begin{condition}\label{c3}
 For any $i=1,...,r$ and $1\leq j\leq p$, $E|f_{i,t}|^{2\gamma}\leq C_1$ and $E|\ve_{j,t}|^{2\gamma}<C_1$ for some positive constant $C_1>0$, where $f_{i,t}$ and $\ve_{j,t}$ are the $i$-th and the $j$-th components of $\bff_t$ and $\bve_t$, respectively, and $\gamma$ is given in Condition~\ref{c1}.
 \end{condition}
 \begin{condition}\label{c4}
 There exists positive constants $C_2,C_3>0$ such that $C_2p^{1-\delta}\leq \|\bSigma_x(k)\|_{\min}\leq \|\bSigma_x(k)\|_2\leq C_3p^{1-\delta}$ for all $k=1,...,k_0$.
 \end{condition}
 Condition~\ref{c1} controls the dependence of the data. Condition~\ref{c2} imposes some moment and dependence assumptions on the noise term $\bve_t$ in order to establish the moment bounds of the autocovariance of $\bve_t$ as that in \cite{han2020}.  Conditons~\ref{c3} and \ref{c4} are commonly used; see \cite{lamyao2012} and \cite{changguoyao2015}.
 
  Under the above mild conditions on the factors and idiosyncratic terms, we have the following theorem.
 \begin{theorem}\label{thm:a1}
 Assume that Conditions~\ref{c1}--\ref{c4} hold.  If $p^{\delta/2}T^{-1/2}=o(1)$, there exists a rotational matrix $\bH$ such that
 \[\|\wh\bA_1-\bA\bH'\|_2=O_p(p^{\delta/2}T^{-1/2}+p^{-(1-\delta)})\rightarrow_p 0, \]
 as $p,T\rightarrow \infty$.  In particular, if $\{\bve_t\}$ is a white noise sequence, then
  \[\|\wh\bA_1-\bA\bH'\|_2=O_p(p^{\delta/2}T^{-1/2})\rightarrow_p 0, \]
  which is the same as the one in Theorem 2 of \cite{LamYaoBathia_Biometrika_2011}.
 \end{theorem}
 \begin{remark}
 (i) Theorem~\ref{thm:a1}
implies that the autocovariance-based eigenanalysis method still works if the idiosyncratic terms are serially correlated. When the idiosyncratic term is a white noise, the convergence rate of the estimated loading matrix is the same as the one in \cite{LamYaoBathia_Biometrika_2011}. \\
(ii) When $\delta=0$ and all factors are strong, we can obtain that 
\begin{equation}\label{ahat:dif}
 \|\wh\bA_1-\bA\bH'\|_2=O_p(T^{-1/2}+p^{-1})\rightarrow_p 0,
\end{equation}
if the noises are serially correlated. 
 Note that we have an additional term $1/p$ in the upper bound compared with the one 
 {in the case of} white idiosyncratic noises, which is understandable since the serial correlations may {introduce additional} errors.  Moreover,  for strong factors, the  rate is smaller than $T^{-1/2}+p^{-1/2}$, which is a traditional one obtained by \cite{BaiNg2023}. The reason is that we adopted the upper bounds of the sample autocovaraince matrices obtained in \cite{han2020}, which are helpful to reduce the  rate  in (\ref{ahat:dif}).
 \end{remark}
 
 \vskip 0.5cm
 {\bf Proof of Theorem A.1. } For simplicity, we consider the case when $k_0=1$ in $\wh\bM$ since the results still hold for a general finite $k_0$.
 Note that
 \[\wh\bSigma_y(k)=\bA\wh\bSigma_x(k)\bA'+\bA\wh\bSigma_{x\ve}(k)+\wh\bSigma_{x\ve}(k)'\bA'+\wh\bSigma_\ve(k).\]
 It follows that

 \begin{align}
 \wh\bM=\wh\bSigma_y(k)\wh\bSigma_y(k)'=&\bA(\wh\bSigma_x(k)\bA'+\bA\wh\bSigma_{x\ve}(k))(\bA\wh\bSigma_x(k)\bA'+\bA\wh\bSigma_{x\ve}(k)+\wh\bSigma_{x\ve}(k)'\bA'+\wh\bSigma_\ve(k))'\notag\\
 &+(\wh\bSigma_{x\ve}(k)'\bA'+\wh\bSigma_\ve(k))(\bA\wh\bSigma_x(k)\bA'+\bA\wh\bSigma_{x\ve}(k)+\wh\bSigma_{x\ve}(k)'\bA'+\wh\bSigma_\ve(k))'.
 \end{align}
 Let $\wh\bV\in R^{r\times r}$ be a diagonal matrix consisting of the top $r$ eigenvalues of $\wh\bM$ as its diagonal elements, it is not hard to show that $\wh\bV\asymp p^{2(1-\delta)}$.  By the relationship
 \[\wh\bM\wh\bA=\wh\bA\wh\bV,\]
 we obtain that
 \[\wh\bA-\bA\bH'=[\wh\bSigma_{x\ve}(k)'\bA'+\wh\bSigma_\ve(k)][\bA\wh\bSigma_x(k)\bA'+\bA\wh\bSigma_{x\ve}(k)+\wh\bSigma_{x\ve}(k)'\bA'+\wh\bSigma_\ve(k)]'\wh\bA\wh\bV^{-1}\]
 where $\bH'=(\wh\bSigma_x(k)\bA'+\bA\wh\bSigma_{x\ve}(k))(\bA\wh\bSigma_x(k)\bA'+\bA\wh\bSigma_{x\ve}(k)+\wh\bSigma_{x\ve}(k)'\bA'+\wh\bSigma_\ve(k))'\wh\bA\wh\bV^{-1}$.
 We will focus on the $\ell_2$-norm of the matrices of interest in this proof, and it is equivalent to the $F$-norm since $r$ is finite.
 Under the conditions in Theorem~\ref{thm:a1}, we observe that
\[\|\wh\bSigma_{x\ve}(k)\|_2\leq \|\wh\bSigma_{x\ve}(k)\|_F\leq Cp^{1-\delta/2}T^{-1/2}.\]
 By Theorem~2.1 of \cite{han2020}, we have
 \begin{equation}\label{sig:e}
  \|\wh\bSigma_\ve(k)\|_2\leq \|\wh\bSigma_\ve(k)-E\wh\bSigma_\ve(k)\|_2+\|E\wh\bSigma_\ve(k)\|_2\leq O_p(1+\sqrt{\frac{p\log(p)}{T}}+\frac{p\log(p)}{T}), 
 \end{equation}
 where the term $\log(p)$  can be removed if $\bve_t$ is a Gaussian sequence according to Theorem 2.2 of \cite{han2020}.
 By an elementary argument, we can obtain that
 \begin{align}\label{rate:A1}
 \|\wh\bA-\bA\bH'\|_2\leq& Cp^{-2(1-\delta)}\left(p^{1-\delta/2}T^{-1/2}+1+\sqrt{\frac{p\log(p)}{T}}+\frac{p\log(p)}{T}\right)\notag\\
 &\times \left(p^{1-\delta}+p^{1-\delta/2}T^{-1/2}+1+\sqrt{\frac{p\log(p)}{T}}+\frac{p\log(p)}{T}\right)\notag\\
 \leq & Cp^{-2(1-\delta)}\left(p^{1-\delta/2}T^{-1/2}+1+\sqrt{\frac{p\log(p)}{T}}+\frac{p\log(p)}{T}\right)\notag\\
 &\times \left(p^{1-\delta}+p^{1-\delta/2}T^{-1/2}+\sqrt{\frac{p\log(p)}{T}}+\frac{p\log(p)}{T}\right).
 \end{align}
 The upper bound in (\ref{rate:A1}) looks complicated, but we can further simplify the expression.  It is known that the consistency of the estimated loading matrix requires $p^{\delta/2}T^{-1/2}=o(1)$. See, for example,  Theorem 2 of \cite{LamYaoBathia_Biometrika_2011}.  Therefore, we consider the case when $p^{\delta/2}T^{-1/2}=o(1)$ and try to simplify the rate in (\ref{rate:A1}).  First, we divide the expression in each of the big parentheses by a factor of $p^{1-\delta}$, and obtain that
 \begin{align}\label{A:m}
  \|\wh\bA-\bA\bH'\|_2\leq& C\left(p^{\delta/2}T^{-1/2}+p^{-(1-\delta)}+(\sqrt{\frac{p\log(p)}{T}}+\frac{p\log(p)}{T})/p^{1-\delta}\right)\notag\\
  &\times \left(1+p^{\delta/2}T^{-1/2}+(\sqrt{\frac{p\log(p)}{T}}+\frac{p\log(p)}{T})/p^{1-\delta}\right)\notag\\
  \leq&C(p^{\delta/2}T^{-1/2}+p^{-(1-\delta)})(1+p^{\delta/2}T^{-1/2})\notag\\
  \leq &C(p^{\delta/2}T^{-1/2}+p^{-(1-\delta)}),
\end{align}  
 where we used the condition $p^{\delta/2}T^{-1/2}=o(1)$. 
 
 We consider the special case when $\bve_t$ is a white noise sequence.   The upper bound in (\ref{sig:e}) reduces to
  \begin{equation}\label{sig:ee}
  \|\wh\bSigma_\ve(k)\|_2\leq \|\wh\bSigma_\ve(k)-E\wh\bSigma_\ve(k)\|_2+\|E\wh\bSigma_\ve(k)\|_2\leq O_p(\sqrt{\frac{p\log(p)}{T}}+\frac{p\log(p)}{T}), 
 \end{equation}
 where $E\wh\bSigma_\ve(k)={\bf 0}$. By a similar argument as that in (\ref{A:m}), we can obtain that
  \begin{align}\label{A:W}
 \|\wh\bA-\bA\bH'\|_2\leq& C\left(p^{\delta/2}T^{-1/2}+(\sqrt{\frac{p\log(p)}{T}}+\frac{p\log(p)}{T})/p^{1-\delta}\right)\notag\\
  &\times \left(1+p^{\delta/2}T^{-1/2}+(\sqrt{\frac{p\log(p)}{T}}+\frac{p\log(p)}{T})/p^{1-\delta}\right)\notag\\
  \leq&C(p^{\delta/2}T^{-1/2})(1+p^{\delta/2}T^{-1/2})\notag\\
  \leq &C(p^{\delta/2}T^{-1/2}),
 \end{align}
 which is the same as the one in Theorem 2 of \cite{LamYaoBathia_Biometrika_2011} when the idiosyncratic term is a white noise sequence.
 This completes the proof. $\Box$
 
  \subsection{Discussion on the Convergence in Matrix Factor Models}
 
In this section,  we briefly discuss the extension of {the above proof techniques}  to the matrix-variate factor models.  Note that the matrix factor models considered can be written as
\begin{equation}\label{tr:ft}
\bY_t=\bL\bF_t\bR'+\bE_t=\bA_1\bX_t\bP_1+\bE_t,t=1,...,T,
\end{equation}
 which is the same as the ones in (1.1) and (2.1) of the main text.  We only take the one in \cite{wang2018} as an example and explain why their method {continues to} work when $\bE_t$ is serially correlated.  The steps to show the consistency of the estimated loadings using the method in \cite{wang2018} are as follows, which is similar to the approach in the proof of Theorem 5 above.
 
 {\bf Step 1.}  Construct a positive  semi-definite matrix 
 \begin{equation}\label{m1hat:1}
\wh\bM_1=\sum_{k=1}^{k_0}\sum_{i=1}^{p_2}\sum_{j=1}^{p_2}\wh\bSigma_{y,ij}(k)\wh\bSigma_{y,ij}(k)'.
\end{equation}

 {\bf Step 2.}  Note that $\wh\bM_1\wh\bA_1=\wh\bA_1\wh\bV_1$, where $\wh\bV_1\in R^{r_1\times r_1}$ is a diagonal matrix consisting of the top $r_1$ eigenvalues of $\wh\bM_1$ as its diagonal elements.  Then, $\wh\bA_1=\wh\bM\wh\bA_1\wh\bV_1^{-1}$. By a similar argument as the proof of  Theorem~\ref{thm:a1} above, 
 there exists a rotational matrix $\bH_1\in R^{r_1\times r_1}$ such that
 \[\wh\bA_1-\bA_1\bH_1'=\bPi,\]
 where $\bA_1\bH_1$ is the leading term of $\wh\bM\wh\bA_1\wh\bV_1^{-1}$.
 
 {\bf Step 3. } Under some regularity conditions, we can show that $\|\bPi\|_2\rightarrow_p 0$ as $p_1,p_2,T\rightarrow \infty$.  In fact, the convergence rate of $\|\wh\bA_1-\bA_1\bH_1'\|_2$ can be established following a similar approach as the proofs of Theorem 5 of the main text and Theorem~\ref{thm:a1} above.  In particular,  the convergence error bound is $\|\wh\bA_1-\bA_1\bH_1'\|_2=O_p((p_1p_2)^{-1}+T^{-1/2})$ if the idiosyncratic term is serially correlated.
 
The consistency for $\wh\bP_1$ can be established in a similar way. We omit the details to save space. Therefore, the autocovariance-based method still works for matrix-variate factor models in \cite{wang2018} when the noises are serially correlated.  

Next, we discuss the modification of the proposed  matrix-variate factor model with diverging noise effect.  According to Figure 1 of the main text, the white noise effect can be prominent in some economic and financial data. Therefore, we may construct a model in the following way:
\[\textit{Data}=\textit{Dynamically dependent common components}+\textit{Serially correlated errors}+\textit{Diverging white noises}.\]
Mathematically, we may consider the following matrix-variate factor model:
\begin{equation}\label{mf:md}
\bY_t=\bL_1\bF_t\bR_1'+\bfeta_t+\bL_2\bxi_t\bR_2',
\end{equation} 
 where $\bF_t\in R^{r_1\times r_2}$ is a dynamically dependent common factor process, $\bfeta_t$ is a serially correlated error term, and $\bxi_t\in ^{k_1\times k_2}$ is a white noise. In other words, the idiosyncratic term in Model (2.1) can be modified by
 \begin{equation}\label{idi:mod}
 \bE_t=\bfeta_t+\bL_2\bxi_t\bR_2',
 \end{equation}
 which consists of a prominent white noise term  $\bL_2\bfeta_t\bR_2'$ and a serially correlated error term $\bfeta_t$.  For the identification issue, we assume the three terms in (\ref{mf:md}) are independent of each other.  It is not hard to see that the covariance of the idiosyncratic term is
 \begin{equation}\label{cov:wt:sr}
 \Cov(\vc(\bE_t))=\Cov(\vc(\bfeta_t))+\bR_2\otimes\bL_2\Cov(\vc(\bxi_t))\bR_2'\otimes\bL_2,
 \end{equation}
which generalizes the covariance assumption of (2.2) in two ways. First,  the covariance of the vectorized idiosyncratic term is not necessarily a Kronecker structure as that in (2.2).  Even if $\Cov(\vc(\bxi_t))=\bI_{k_1k_2}$,  the covariance $\Cov(\vc(\bE_t))=\Cov(\vc(\bfeta_t))+\bR_2\bR_2'\otimes\bL_2\bL_2'$ is still more general than the one in (2.2). Second,  the idiosyncratic term also consists of a serially correlated term $\bfeta_t$, which makes the idiosyncratic term serially correlated  too. 
 
 In view of the above discussion, we formalize the matrix-variate factor model with diverging noise effect as
 \begin{equation}\label{MF:DV}
 \bY_t=\bL_1\bF_t\bR_1'+\bE_t,\,\,\bE_t=\bfeta_t+\bL_2\bxi_t\bR_2', t=1,...,T,
 \end{equation}
 where $\bfeta_t$ is serially correlated, and $\bxi_t$ is a white noise sequence, making the idiosyncratic term $\bE_t$ a serially correlated one.
 
 Now, we explain why the proposed method still works for extracting the common factors and mitigating the diverging noise effect. The key observations are as follows:
 \begin{enumerate}
 \item According to the discussion in Steps 1--3 above and the proofs in the main article,  the autocovariance-based method can reduce the noise effect and still provides satisfactory convergence rates for the estimated loading matrices following the proofs of Theorem 1 in the main text. In other words, the estimated loading matrices $\wh\bA_1$, $\wh\bP_1$,  and their orthogonal complements are consistent. The error bound is $\|\wh\bA_1-\bA_1\bH_1'\|_2=O_p((p_1p_2)^{-1}+T^{-1/2})$ as explained in the proof of Theorem 5 above, which is outlined in (\ref{a1hat:rate:s}).
 
 \item For the two-way projected principal component analysis in Section 2.2.2, the orthogonal complement spaces of $\bL_2$ can still be estimated using the eigenanalysis on $\bS_1$ in (2.11) of the main text because the diverging effect of the cross-sectional covariance comes from the white noise part in $\wh\bB_1^{'}\bY_t\wh\bQ_1$ because $\wh\bB_1$ and $\wh\bQ_1$ are consistent as discussed above.   Similar argument also applies to $\bR_2$. Therefore, the two-way projected PCA is still able to estimate the loading spaces of $\bL_2$, $\bR_2$, and their orthogonal complements, implying that our proposed method still works for matrix-variate factor models with diverging white noise effect and serially correlated errors.
 \item The white noise testing method in Section 2.2.3 cannot be used to determine the order of the factor matrix since $\bE_t$ is no longer a white noise sequence.  However, we can still adopt  the eigenvalue ratio-based method as that in \cite{lamyao2012} and \cite{wang2018}, although its finite sample properties may deteriorate if the noise effect is very prominent. In fact, we can still show that the ratio-based method is asymptotically valid because the autocovariance can mitigate some noise effect in a  large sample scenario.
 \end{enumerate}
 Therefore, the propose procedure is still able to estimate the matrix-variate factor model with serially correlated errors and diverging white noise effect as the one in (\ref{MF:DV}).  The detailed proof can be established by following the argument in the proof of Theorem 5, by imposing similar conditions as Conditions 3.1-3.4 above.  The proof  is available upon request, and we leave this issue for future research.
 

\end{document}